\documentclass[fleqn,usenatbib]{mnras}

\usepackage{newtxtext,newtxmath}
\usepackage[T1]{fontenc}
\usepackage{graphicx}   
\usepackage{amsmath}    
\usepackage{deluxetable}

\title[OIII in ERQs are compact]{[\ion{O}{III}]\,$\lambda$5007 emissions in extremely red quasars 
(ERQs) are compact}

\author[M. W. Lau et al.]{
Marie Wingyee Lau,$^{1}$\thanks{E-mail: wingyeel@ucr.edu (MWL)}
Serena Perrotta,$^{2}$
Fred Hamann,$^{1}$
Jarred Gillette,$^{1,3}$
\newauthor David S. N. Rupke,$^{4,5}$
Andrey Vayner,$^{6}$
Nadia L. Zakamska$^{6}$
Dominika Wylezalek $^{5}$\\
$^{1}$Department of Physics \& Astronomy, University of California, Riverside, CA 92521, USA\\
$^{2}$Center for Astrophysics and Space Sciences, University of California, San Diego, CA 92093,
USA\\
$^{3}$Eureka Scientific, Oakland, CA 94602, USA\\
$^{4}$Department of Physics, Rhodes College, Memphis, TN 38112, USA\\
$^{5}$Astronomisches Rechen-Institut, Zentrum f\"{u}r Astronomie der Universit\"{a}t Heidelberg, D-69120 Heidelberg, Germany\\
$^{6}$Department of Physics \& Astronomy, Johns Hopkins University, Baltimore, MD 21218, USA\\}
\date{Accepted XXX. Received YYY; in original form ZZZ}

\pubyear{2023}

\begin{document}
\label{firstpage}
\pagerange{\pageref{firstpage}--\pageref{lastpage}}
\maketitle

\begin{abstract}

``Extremely red quasars'' (ERQs) are a non-radio-selected, intrinsically luminous population of 
quasars at cosmic noon selected by their extremely red colour from rest-frame UV to 
mid-IR. ERQs are uniquely associated with exceptionally broad and blueshifted 
[\ion{O}{III}]\,$\lambda$5007 emission reaching speeds $>$6000\,km\,s$^{-1}$. We obtained 
adaptive optics integral-field spectroscopic observations using Keck/OSIRIS and 
Gemini/NIFS of a sample of 10 ERQs with bolometric luminosities 
(10$^{47.0}$\textendash10$^{47.9}$)\,erg\,s$^{-1}$ at $z\sim$~(2.3\textendash3.0). The goal is to 
measure the sizes and spatially-resolved kinematics of the [\ion{O}{III}]-emitting regions. We 
study the surface brightness maps and aperture-extracted spectra and model the 
point-spread functions. We identify signs of merger activities in the continuum 
emissions. We identify physically distinct [\ion{O}{III}] kinematic components that are bimodal 
and respectively trace ERQ-driven outflows of velocity dispersion $\gtrsim$250\,km\,s$^{-1}$ and 
dynamically quiescent interstellar media. We find that the ERQ-driven ionized outflows are 
typically at $\sim$1\,kpc scales whereas the quiescent ionized gas extend to a few kpc. 
Compared to normal quasars the extremely fast ERQ-driven [\ion{O}{III}] outflows tend to 
be more compact, supporting the notion that ERQs are in a young stage of quasar/galaxy evolution 
and represent unique physical conditions beyond orientation differences with normal  
quasar populations. The kinematically quiescent [\ion{O}{III}] emissions in ERQs tend to be 
spatially-resolved but less extended than in normal quasars, which can be explained by 
global and patchy dust obscuration. The hint of ionization cones suggests some of the obscuration 
can be partially explained by a patchy torus. 

\end{abstract}


\section{Introduction}

Quasar outflows have been invoked in galaxy evolution models to regulate or halt both star 
formation in their host galaxies and the accretion of material onto the central supermassive black 
holes. In certain theoretical scenarios, after triggering events for massive starbursts that 
shroud the galaxies in gas and dust, important feedback of quasar outflows is expected to occur 
during a brief dusty ``blowout'' phase of galaxy evolution. During this particular ``blowout'' 
stage outflows remove dust and gas from their hosts, revealing the visibly luminous quasars and 
driving the transition from the initial dusty starbursts to the later quiescent galaxy phases 
\citep[e.g.][]{Sanders+88,Hopkins+06,Hopkins+08a,Hopkins+16}
Feedback from quasar outflows can also aide in establishing the black 
hole{\textendash}bulge correlations in the local universe \citep{Gebhardt+00,Hopkins+06}, and 
prevent the overgrowth of massive galaxies \citep{Croton+06}. 

Adaptive-optics-assisted and spaceborne integral-field spectroscopy have established that ionized 
outflows driven by radio-loud or jetted sources, as well as the much more common radio-quiet 
quasars, can have a major impact on the gas reservoir in their host galaxies especially at high 
redshifts \citep[e.g.][]{Cresci+15,Perna+15,Brusa+16,Nesvadba+17,Williams+17,Kakkad+20,Perna+23,
Vayner+21b,Cresci+23,Veilleux+23}. However, the impact of ionized outflows from obscured, 
dust-reddened radio-quiet quasars on their host galaxies has only recently begun to be explored 
with high-resolution integral-field data. Radiation-hydrodynamic simulations predict that 
reprocessing quasar radiation and trapping the IR radiation by dust can launch fast galactic 
outflows \citep{Costa+18,SolimanHopkins23}. 

Detailed observational studies at high redshift are challenging due to a combination of dust 
reddening and cosmic surface brightness dimming \citep[see e.g.][]{Perna+15,Chen+17}. Obscured, 
dust-reddened quasars are particularly important for studies of quasar feedback because they are 
believed to be still partially embedded in dusty starbursts \citep[e.g.][]{CanalizoStockton01,
UrrutiaLacyBecker08,Assef+15,Banerji+15,Noboriguchi+19,CalistroRivera+21,Glikman+22}. Obscured 
quasars are therefore 
prime candidates for quasars actively clearing their hosts and disrupting star formation. The 
[\ion{O}{III}]\,$\lambda$5007 emission line, being a low density-forbidden transition, is 
particularly important for identifying outflows on galactic, kpc scales. At lower redshifts 
$z\sim0.5$, past integral-field observations of the [\ion{O}{III}]\,$\lambda$5007 emission in 
radio-quiet obscured quasars found outflow signatures of velocities $\sim$800\,km\,s$^{-1}$ in 
essentially all targets examined \citep{Liu+13,Shen+23}. These observations demonstrate that 
ionized outflows in radio-quiet quasars are powerful and ubiquitous. 

However, it is at the peak epoch of galaxy formation and quasar activity, $z\sim$\,2\textendash3, 
that quasar feedback should have the greatest impact on massive galaxy evolution. Recently, a 
remarkable population of ``extremely red quasars'' (ERQs) at $z\sim$\,2\textendash3 have been 
discovered in the Sloan Digital Sky Survey-III Baryon Oscillation Sky Survey 
\citep[BOSS;][]{Ross+15,Hamann+17}. ERQs have extremely red colours in the rest-frame UV to 
mid-IR, high bolometric luminosities $>$10$^{47}$\,erg\,s$^{-1}$, normal to quiet radio 
luminosities $\sim$10$^{41}$\,erg\,s$^{-1}$, and sky densities a few per cent of luminous blue 
quasars consistent with being a short obscured phase of quasar activity. ERQs might also be 
cosmic-noon analogs of ``extremely red objects'' (EROs, likely to be quasars) discovered recently 
at $z\sim$\,5\textendash9 with JWST \citep{Barro+23,Noboriguchi+23}. Importantly, ERQs also have a 
suite of extreme spectral properties unlike any other known quasar populations \citep{MonadiBird22,
Gillette+24}. Near-IR longslit spectroscopy revealed that ERQs systematically have among the 
broadest and most blueshifted [\ion{O}{III}]\,$\lambda$5007 lines ever reported with FWHM 
velocities and blueshifted wings reaching $>$6000\,km\,s$^{-1}$ \citep{Zakamska+16,Perrotta+19}. 
Such extreme [\ion{O}{III}] outflows have only been observed in one other population called the 
hot dust-obscured galaxies (Hot DOGs) which overlap with the ERQ population \citep{Finnerty+20}. 
These velocities are too large to be contained by any realistic galaxy potential and are tell-tale 
signs of powerful outflows. These extreme [\ion{O}{III}] outflows are strongly correlated with the 
extreme red colours of ERQs, and not with radio loudness, high bolometric luminosities, or high 
Eddington ratios \citep{Shen16a,Bischetti+17}. The [\ion{O}{III}] outflows from ERQs are 
typically three times faster than in blue quasars at similar redshifts and luminosities. 
\cite{Perrotta+19} also estimated that the [\ion{O}{III}] outflows from ERQs have kinetic energy 
luminosities at a few per cents of the quasar bolometric luminosities, which should be sufficient 
to drive important feedback in the host galaxies \citep[e.g.][]{HopkinsElvis10,ZubovasKing12}. 

\cite{Lau+22} and \cite{Gillette+23} observed the Ly$\alpha$-emitting haloes around ERQs and 
found they are kinematically quiet down to the spatial resolution $\sim$6\,kpc from the central 
quasars. This implies that the spatial scales of the extremely fast outflows extend out to less 
than $\sim$6\,kpc, although this does not rule out the presence of an undetected underdense phase 
of fast circumgalactic-scale outflows \citep{Nelson+19,Costa+22}. With spaceborne or 
adaptive-optics integral-field observations we can directly measure the extents of the 
[\ion{O}{III}]\,$\lambda$5007 outflows on galactic scales. Detailed study of one ERQ using the 
Near Infrared Spectrograph (NIRSpec) onboard JWST and the Near-Infrared Integral Field 
Spectrograph (NIFS) on the Gemini North telescope (Gemini-N) has been published \citep{Vayner+21a,
Wylezalek+22,Vayner+23,Vayner+24}. Its [\ion{O}{III}] emission traces powerful, extensive 
outflows out to 10\,kpc and clumpy ionized gas in the host out to at least 16\,kpc. 

To determine the characteristic sizes and the spatially resolved kinematics of the [\ion{O}{III}] 
emissions in ERQs as a population, we carried out adaptive-optics observations of nine more ERQs 
using the OH-Suppressing Infrared Integral Field Spectrograph (OSIRIS) on the Keck I telescope, 
forming a total sample of 10. In this paper we test the hypothesis that ERQs are generally young 
objects whose high-speed ionized outflows have not had time to further expand into their host 
interstellar media. In addition to statistically characterizing the extents of the 
[\ion{O}{III}]-traced outflows, we also measure the extents of the more quiescent 
[\ion{O}{III}]-traced interstellar media whenever detected. These measurements of the 
[\ion{O}{III}]-traced interstellar media complement the measurements of continuum-traced starlight 
in \cite{Zakamska+19}. In this work we do not attempt to quantitatively assess dust extinction due 
to the large systematic uncertainties involved. 

This paper is organized as follows. Section 2 describes the observations and data reduction. 
Section 3 describes the data analysis procedures and the results. Section 4  discusses the 
implications of our findings in the context of other quasars in the literature. Section 5 presents 
the conclusions. Throughout this paper we adopt a ${\rm \Lambda}$ cold dark matter cosmology with 
$H_0=69.6$\,km\,s$^{-1}$\,Mpc$^{-1}$, ${\rm \Omega}_{\rm M}=0.286$, 
${\rm \Omega}_{\rm \Lambda}=0.714$, as adopted by the online cosmology calculator developed by 
\citep{Wright06} at the the time of this writing. All distances are proper. For the redshift range 
covered by this sample 2.3\textendash3.0, 1\,arcsec corresponds to (8.4\textendash7.9)\,kpc. We use 
photometric magnitudes in the AB system. We report vacuum wavelengths in the heliocentric frame 
and name the line transitions by their air wavelengths. 

\section{Observations and Data Reduction}

\subsection{Sample Selection}

\begin{figure}
\includegraphics[width=\columnwidth]{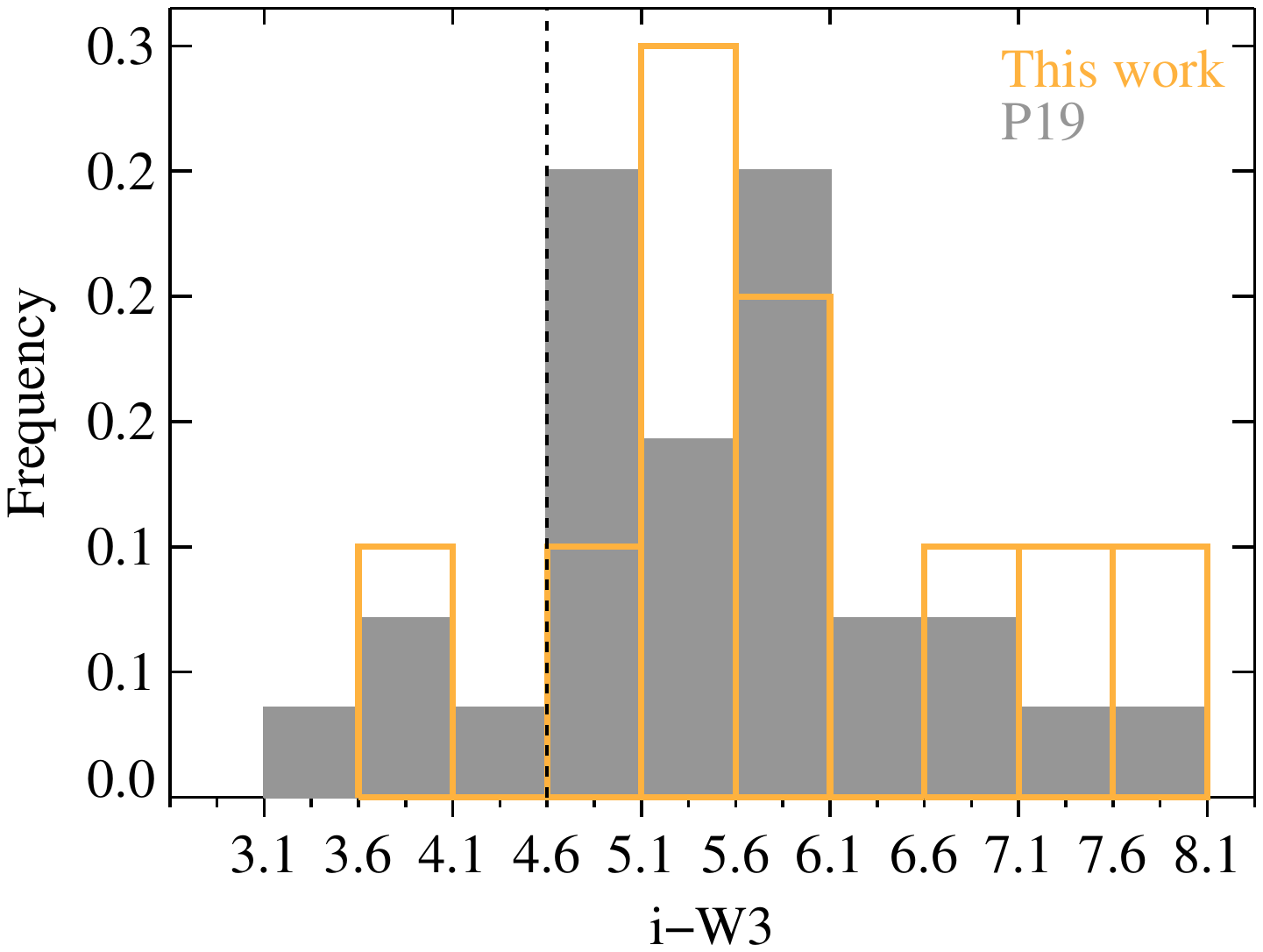}
\caption{Histograms of the $i-W3$ colours of the ERQs and ERQ-like sources in this integral-field 
sample and the \citet{Perrotta+19} longslit sample. The dotted line marks the ERQ definition of 
$i-W3>4.6$.}
\label{fig:imW3histo}
\end{figure}

We select targets with existing longslit [\ion{O}{III}] spectra from \cite{Perrotta+19}, with a 
preference for sources with very red colours and/or very fast [\ion{O}{III}] outflows. The only 
exception is J2223+0857, for which we present the first [\ion{O}{III}] observations here. 
Table~\ref{tab:smpl} lists some previously-reported basic properties of the ERQs in this sample. 
The ``extremely-red'' colour is defined by $i-W3>4.6$ from the Sloan Digital Sky Survey-III 
\citep{Eisenstein+11} and the Wide-field Infrared Survey Explorer \citep{Wright+10}. For 
reference, the $i-W3$ colours of this work range between (3.8\textendash8.0) while that of 
\cite{Perrotta+19} range between (3.3\textendash8.0). The [\ion{O}{III}]\,$\lambda$5007 $w_{90}$ 
velocity widths (defined as encompassing 90\% of the line fluxes) of this work and 
\cite{Perrotta+19} both range between (2450\textendash7230)\,km\,s$^{-1}$. Sources of 
this sample are classified as ERQs except J1550+0806, which is listed as a red ``ERQ-like'' quasar 
by \cite{Hamann+17} based on extreme emission-line properties like ERQs but with a less extreme red 
$i-W3$ colour. J1550+0806 has $i-W3=3.8$ which is still significantly redder the median of  
$W3$-detected BOSS quasars at $i-W3=2.5$. This red ERQ-like quasar overlaps with ERQs in many 
emission-line properties including high rest equivalent widths and extreme [\ion{O}{III}] 
kinematics, and its measurements does not skew any results on the ERQ sample. 
Figure~\ref{fig:imW3histo} shows the $i-W3$ distributions of this sample and the 
\cite{Perrotta+19} sample. For simplicity we refer to all our targets as ERQs. When 
available, we obtain accurate systemic redshifts measured from CO host emission, Ly$\alpha$ halo 
emission, or ``spike''-like narrow Ly$\alpha$ quasar emission as reported in \cite{Gillette+24}. 
The intrinsic bolometric luminosities are estimated from applying a bolometric correction factor 
of 8 to the rest-frame 5\,$\umu$m luminosities extrapolated from the $W3$-band photometry. 

We choose to target the [\ion{O}{III}]\,$\lambda\lambda$4959,5007 emissions over the H$\alpha$ 
emissions. The H$\alpha$-traced outflow and interstellar medium emissions are difficult to extract 
as they are heavily blended with the broad-line region and 
[\ion{N}{II}]\,$\lambda\lambda$6549,6584 emissions \citep[see][]{Vayner+21a,Kakkad+23}.  
This sample is a total of 10 ERQs observed with laser-guided adaptive-optics-assisted near-IR 
integral-field spectrographs. We include ERQ J1652+1728 which has Gemini-N/NIFS observations and 
good adaptive optics correction published in \cite{Vayner+21a}. One more ERQ 
J2323$-$0100 has shallower Keck-I/OSIRIS observations briefly described in \cite{Vayner+21a}. We 
obtained deeper OSIRIS observations of this source plus new observations of other ERQs. 

We select targets based on availability of nearby bright stars, existence of multiwavelength data, 
observation scheduling constraints due to object visibility, object brightness, and weather 
conditions. To enable tip/tilt correction by the Keck I laser guide 
star adaptive optics system, we require availability of an $r<17$\,mag star within 
$\sim$50\,arcsec of the science target. All but one ERQ, J2223+0857, in this sample have existing 
near IR longslit spectra where [\ion{O}{III}] kinematics and energetics are already analyzed 
\citep{Perrotta+19}, and six of them have Atacama Large Millimeter Array Band 6 observations for 
measuring molecular gas and dust (F.\ Hamann et al., in preparation). The long requisite exposure 
times and complex mechanics of the laser guide star adaptive optics system limit observation 
scheduling to at most one target per half good night. 

\subsection{Observing and Reducing Data}

The published NIFS observations of J1652+1728 has a wavelength range of the full {\it K}-band and 
a field of view of 3$\times$3\,arcsec$^2$, and has an absolute flux uncertainty at 10\textendash15 
per cent. OSIRIS provides spectroscopy with median resolution 
$R\sim3800$ with a variable field-of-view utilizing a lenslet design. The OSIRIS observations were 
conducted in the Hn3, Hn4, and Hn5 narrowband filters using the 50\,milliarcsec plate scale mode. 
The Hn3 filter allows extracting a wavelength range of (15940\textendash16760)\,\AA and a field of 
view of 2.4$\times$3.2\,arcsec$^2$ per frame. The Hn4 filter has a wavelength range of 
(16520\textendash17370)\,\AA and a field of view of 2.1$\times$3.2\,arcsec$^2$. The Hn5 filter has 
a wavelength range of (17210\textendash1808)\,\AA and a field of view of 
1.6$\times$3.2\,arcsec$^2$. We observed the tip/tilt stars immediately before and in between their 
science target frames. During science exposures we dithered along the short axis to enlarge the 
field of view, yielding a final co-added field-of-view ranging from 1.7$\times$2.4\,arcsec$^2$ to 
2.8$\times$2.8\,arcsec$^2$ with the edges trimmed. The chosen field of view was informed by the 
extent of the [\ion{O}{III}] emission 
in the NIFS observations of J1652+1728. To enable sky subtraction we nodded by 5\,arcsec from the 
science target to blank sky in the repeating object-sky-object sequences. The exposure time of 
each frame was 900\,s and the total on-source exposure times ranged from 2700\,s to 18000\,s. To 
correct for atmospheric absorption and to flux calibrate the final co-added science cube, telluric 
standard stars were observed with the same setup as the science observations within 0.2 airmass of 
the average of the science observations before or after the science observations. Spectral types 
of the standard stars are close to A0. In Table~\ref{tab:obslogs} we present the logs of the 
OSIRIS observations. The reference adaptive-optics-corrected point spread function (PSF) sizes are 
measured from fitting Gaussian models to the azimuthally-averaged surface brightness radial 
profiles of the paired tip/tilt stars or the H$\beta$ broad line region of the ERQs, which range 
from 0.15\,arcsec to 0.23\,arcsec.

For each science target, we reduced the OSIRIS observations using the Keck OSIRIS Spectroscopic 
Data Reduction Pipeline \citep{Lyke+17,Lockhart+19} and the general-purpose IDL library ISFRED 
\citep{Rupke14a}. The OSIRIS pipeline first creates a master dark by median-combining several dark 
observations taken in the afternoon. After dark subtraction, the OSIRIS pipeline extracts the 
spectra using a Lucy-Richardson deconvolution using a known point spread function for each lenslet 
stored in the rectification matrix. The OSIRIS pipeline then applies wavelength calibration and 
combines the spectra into a three-dimensional data cube. The OSIRIS pipeline 
performs scaled sky subtraction by scaling families of OH emission lines in the sky cubes to match 
the science observations. The OSIRIS pipeline then removes the telluric absorption features by 
removing hydrogen features from the telluric star spectrum, dividing by a blackbody, normalizing 
it, and then dividing this response function into the science frames. Using routines of the IFSRED 
library we performed further sky subtraction by selecting a sky aperture in each science frame and 
subtracting the spatial median of it across the entire field. We aligned the individual reduced 
data cubes using the position of the quasar centroid and median-combined them. We extracted the 
standard star flux and compared to existing {\it H}-band photometry to obtain the absolute flux per 
unit count per second, and absolute-flux-calibrated the mosaicked science data cube. We estimate 
the absolute fluxing accuracy at 10\textendash15 per cent. Finally we 
visually examined the reduced, mosaicked, flux-calibrated science data cube for any unaccounted 
background artifacts caused by the complex optics and manually subtracted them. 

\begin{table*}
\caption{Basic Properties of ERQs Observed with Adaptive Optics}
\label{tab:smpl}
\setlength{\tabcolsep}{0.1in}
\begin{tabular}{lcccccccc}
\hline
Name & R.A. & Dec. & Systemic Redshift (Indicator) \tablenotemark{a} & {\it i}$-${\it W}3\tablenotemark{b} & Bolometric Luminosity (erg s$^{-1}$)\tablenotemark{c} \\
\hline
J0006$+$1215 & 00:06:10.67 & $+$12:15:01.2 & 2.3183 (CO) & 8.0 & 7.58$\times10^{47}$ \\
J0209$+$3122 & 02:09:32.15 & $+$31:22:02.7 & 2.3595 (H$\beta$) & 5.1 & 1.02$\times10^{47}$ \\
J1031$+$2903 & 10:31:46.53 & $+$29:03:24.1 & 2.2956 ([\ion{O}{III}]) & 6.0 & 1.91$\times10^{47}$ \\
J1217$+$0234 & 12:17:04.70 & $+$02:34:17.1 & 2.4280 (CO) & 5.7 & 2.49$\times10^{47}$ \\
J1232$+$0912 & 12:32:41.73 & $+$09:12:09.3 & 2.4050 (CO) & 5.6 & 6.34$\times10^{47}$ \\
J1550$+$0806 & 15:50:57.71 & $+$08:06:52.1 & 2.5088 (H$\beta$) & 6.8 & 1.16$\times10^{47}$ \\
J1604$+$5633 & 16:04:31.55 & $+$56:33:54.2 & 2.4914 (Low ions) & 3.8 & 1.96$\times10^{47}$ \\
J1652$+$1728 & 16:52:02.64 & $+$17:28:52.3 & 2.9548 (Ly$\alpha$ halo) & 5.7 & 6.80$\times10^{47}$ \\
J2223$+$0857 & 22:23:07.12 & $+$08:57:01.7 & 2.2902 (CO) & 5.4 & 1.73$\times10^{47}$ \\
J2323$-$0100 & 23:23:26.17 & $-$01:00:33.1 & 2.3805 (CO) & 5.6 & 2.87$\times10^{47}$ \\
\hline
\end{tabular}
\tablenotetext{a}{\raggedright Obtained from \cite{Perrotta+19} and \cite{Gillette+24}.}
\tablenotetext{b}{\raggedright Obtained from SDSS and WISE photometry.}
\tablenotetext{c}{\raggedright Estimated based on $W3$-band photometry.}
\end{table*}

\begin{table*}
\caption{Observing Logs of ERQs}
\label{tab:obslogs}
\setlength{\tabcolsep}{0.03in}
\begin{tabular}{lccccccccc}
\hline
Name & Date (UT) & Instrument & Band & Field of View\tablenotemark{a} & Integration Time & Airmass & Seeing & Corrected PSF (Indicator) \tablenotemark{b} \\
 & & & (\AA) & (arcsec$\times$arcsec) & (s) & & (arsec) & (arcsec/kpc) \\
\hline
J0006$+$1215 & 2020 Oct 3 & OSIRIS & Hn3/(15940\textendash16760) & 2.8$\times$2.8 & 8100 & 1.1 & 0.9 & 0.17/1.4 (tip/tilt star) \\
J0209$+$3122 & 2020 Oct 3 & OSIRIS & Hn4/(16520\textendash17370) & 2.8$\times$2.8 & 9000 & 1.1 & 0.9 & 0.19/1.6 (tip/tilt star) \\
J1031$+$2903 & 2021 May 3 & OSIRIS & Hn3/(15940\textendash16760) & 2.4$\times$2.8 & 2700 & 1.1 & 0.7 & 0.19/1.6 (tip/tilt star) \\
J1217$+$0234 & 2021 May 4 & OSIRIS & Hn4/(16520\textendash17370) & 2.4$\times$2.6 & 6300 & 1.1 & 1.2 & 0.17/1.4 (H$\beta$) \\
J1232$+$0912 & 2020 Jun 5 & OSIRIS & Hn4/(16520\textendash17370) & 2.8$\times$2.8 & 5400 & 1.2 & 1.0 & 0.22/1.8 (tip/tilt star) \\
J1550$+$0806 & 2021 May 3 & OSIRIS & Hn5/(17210\textendash18080) & 1.7$\times$2.4 & 8100 & 1.2 & 0.4 & 0.23/1.9 (tip/tilt star) \\
J1604$+$5633 & 2021 May 4 & OSIRIS & Hn5/(17210\textendash18080) & 1.7$\times$2.4 & 3600 & 1.3 & 1.2 & 0.22/1.8 (tip/tilt star) \\
J1652$+$1728 & 2019 May 7; 2019 May 8 & NIFS & K/(19900\textendash24000) & 2.6$\times$2.8 & 6000 & 1.1; 1.1 & 0.4 & 0.20/1.6 (continuum) \\
J2223$+$0857 & 2018 Sep 1 & OSIRIS & Hn3/(15940\textendash16760) & 2.5$\times$2.8 & 7200 & 1.2 & 0.6 & 0.19/1.6 (tip/tilt star) \\
J2323$-$0100& 2017 Aug 13; 2018 Aug 11 & OSIRIS & Hn4(16520\textendash17370) & 2.6$\times$2.8 & 18000 & 1.1; 1.1 & 0.5; 0.6 & 0.15/1.3 (tip/tilt star) \\
\hline
\end{tabular}
\tablenotetext{a}{\raggedright The effective useful co-added field of view.}
\tablenotetext{b}{\raggedright FWHM obtained from fitting the Gaussian core of the PSF. The corresponding physical scale at the redshift of the target is also reported.}
\end{table*}

\section{Analysis and results}

As the spatially-integrated [\ion{O}{III}] kinematics have been analyzed in \cite{Perrotta+19}, 
this work focuses on the spatial extents and the spatially resolved kinematics. Due to limited 
signal-to-noise (S/N) we do not attempt to examine the spectra spaxel by spaxel, but rather we 
examine narrowband images and aperture spectra. We first create narrowband continuum and 
[\ion{O}{III}] images and spatially-integrated spectra for quicklook visualization of the 
adaptive optics data. Then we use two techniques, namely the radial profile method and the PSF 
subtraction method, to carefully assess if the continuum and [\ion{O}{III}] emissions are 
extended. We consider emission to be extended if either technique reveals it to be extended. 
After that we assess the physical interpretation of all the [\ion{O}{III}] kinematic components as 
a sample. 

\subsection{Narrowband Images and Integrated Spectra}

While the NIFS data are broadband, the OSIRIS targets were observed through narrowband filters of 
widths (83\textendash88)\,\AA, which do not cover the entire H$\beta$-[\ion{O}{III}] region. We 
therefore utilize the broadband longslit data and the best fits from \cite{Perrotta+19} to inform 
de-belending of the data into different emission components and therefore generation of narrowband 
continuum and [\ion{O}{III}] images from the OSIRIS data cubes. For J2223+0857 which is not part 
of the longslit sample, the bulk of H$\beta$ emission is fortunately cleanly separated from 
[\ion{O}{III}] and we can therefore perform a fit to the OSIRIS spatially-integrated spectrum. We 
fit a constant to the continuum, a single Gaussian to H$\beta$, and two double Gaussians that are 
tied in kinematics and 1:3 amptlitude ratios to [\ion{O}{III}]\,$\lambda\lambda$4959,5007. These 
are the least number of components that best fits the spatially-integrated spectrum of J2223+0857. 

From the fits to the longslit spectra or the spatially-integrated OSIRIS spectrum we can then 
identify wavelength channels that are blend of (H$\beta$+continuum), blend of 
(H$\beta$+[\ion{O}{III}]+continuum), blend of ([\ion{O}{III}]+continuum), and pure continuum. 
From each data cube, we create a continuum flux map by summing wavelength channels that are 
identified to contain purely continuum and channels that contain H$\beta$ plus continuum 
emissions, as shown in the first column of Fig.~\ref{fig:nbimgsintspec}. This is done to increase 
the S/N of this narrowband image by relying on the assumption that H$\beta$ is spatially 
unresolved, justified on the measured sub-parcsec sizes of quasar broad-line regions 
\citep[see][]{Williams+17} while neglecting the narrow-line region component by necessity. The 
maps are spatially smoothed with a Gaussian kernel of FWHM 0.1\,arcsec and we mark contours of S/N 
of two to aid visualization of the ERQ emission. We determine S/N from the data and variance 
channels containing purely continuum/H$\beta$ as delineated in the integrated spectra shown in the 
third column of Fig.~\ref{fig:nbimgsintspec}, ignoring strong sky emissions and filter edges. 
From each data cube, we create an [\ion{O}{III}] map from summing wavelength channels that contain 
[\ion{O}{III}]\,$\lambda\lambda$4959,5007 emission and subtract the continuum and any blending 
H$\beta$ emission, as shown in the second column of Fig.~\ref{fig:nbimgsintspec}. The continuum 
level of each spaxel is approximately constant across the narrowband and is determined by the 
two-sigma-clipped mean of the channels used to generate the continuum map. The blending H$\beta$ 
emission flux relative to continuum level is determined from the longslit best-fit, relying on the 
assumption that H$\beta$ is spatially unresolved. The maps are spatially smoothed and we mark 
contours of S/N of two to aid visualization. We determine S/N from channels containing 
[\ion{O}{III}] as delineated in the integrated spectra, ignoring strong sky emissions and filter 
edges. 

From each data cube, we create a spatially-integrated spectrum by summing all spaxels of S/N 
greater than two over the full wavelength range, as shown in the third column of 
Fig.~\ref{fig:nbimgsintspec}. We mark wavelength boundaries between (H$\beta$+continuum), 
(H$\beta$+[\ion{O}{III}]+continuum), ([\ion{O}{III}]+continuum), and continuum-only emissions. 
With these clearly delineated wavelength regions, we ensure that no [\ion{O}{III}]-containing 
channels contribute to the continuum flux maps in the first column of Fig.~\ref{fig:nbimgsintspec}. 
Where a longslit spectrum exists, we overplot it to give context to the narrowband spectrum. The 
small discrepancy between the OSIRIS spatially-integrated spectra and the longslit spectra comes 
from the different physical regions extracted. The OSIRIS spatially-integrated spectra only 
include S/N\,$>$\,2 spaxels while the longslit spectra are 0.7-arcsec extractions. For J2223+0857 
which does not have a longslit spectrum we overplot our best fit and separately show the H$\beta$ 
and the two individual [\ion{O}{III}] components. 

Since a longslit spectrum of J2223+0857 has not been previously obtained, we measure its 
[\ion{O}{III}]\,$\lambda$5007 emission-line properties from the OSIRIS spatially-integrated 
spectrum to put it in context with the other ERQs. We calculate the ${\rm v}_{98}$ and the 
$w_{80}$ values to quantify the kinematics as in \cite{Perrotta+19}. We define ${\rm v}_{98}$ as 
the velocity that encompasses 98\% of the line flux integrating from the red end, which is a 
measure of the maximum outflow velocity traced by an emission line. We define ${\rm v}_{50}$ as 
the line centroid, and is a measure of the bulk velocity of the line-emitting gas. We define 
$w_{80}$ as the velocity width that encompasses 80\% of the line flux around the centroid and 
excludes 10\% of the red and blue wings, which is often used to find signature of AGN-driven 
outflows, and similarly we define $w_{90}$. The non-parametric ${\rm v}_{98}$, ${\rm v}_{50}$, 
$w_{80}$, and $w_{90}$ values of the total [\ion{O}{III}]\,$\lambda$5007 emission derived from our 
best-fit are respectively $(-7690\pm300)$\,km\,s$^{-1}$, $(-5510\pm200)$\,km\,s$^{-1}$, 
$(6030\pm170)$\,km\,s$^{-1}$ and $(7760\pm320)$\,km\,s$^{-1}$. These values make the 
[\ion{O}{III}] kinematics of J2223+0857 typical of the ERQs measured in the longslit sample. The 
rest equivalent width and the line luminosity of the total [\ion{O}{III}]\,$\lambda$5007 emission 
derived from our best-fit are respectively $(139\pm5)$\,\AA\ and 
$(4.2\pm0.1)\times10^{43}$\,erg\,s$^{-1}$, again typical of other measured ERQs. 

Fig.~\ref{fig:nbimgsintspec} reveals that in J0006+1215, J1217+0234, J1232+0912, and J2323$-$0100 
there exists continuum emission not consistent with a point-spread function. Four sources 
of this sample, namely J1217+0234, J1232+0912, J1652+1728, and J2323$-$0100, overlap with the 
\cite{Zakamska+19} sample of Hubble Space Telescope (HST) near-IR images that correspond to the 
rest-frame {\it B}-band and do not cover [\ion{O}{III}]\,$\lambda\lambda$4959,5007. For 
J0006+1215, the OSIRIS narrowband images reveal very extended continuum and [\ion{O}{III}] 
emissions. The analysis of Section~3.2 will show that the extended [\ion{O}{III}] likely traces 
dynamically quiescent interstellar media, supporting the extended continuum being diffuse 
starlight. For J1217+0234, in the OSIRIS images we find spatially-resolved, clumpy, disturbed 
continuum and [\ion{O}{III}] emissions. The HST image reveals extended continuum emission around 
this source out to 3\,arcsec, with the larger detectable extent due to higher surface brightness 
sensitivity. Section~3.2 will show that the extended [\ion{O}{III}] is likely interstellar media. 
The HST image and the extended quiescent [\ion{O}{III}] both support the extended continuum 
detected in OSIRIS data being diffuse starlight. For J1232+0912, in the OSIRIS image we find a 
hint of extended continuum. The HST image reveals extended continuum emission around this source 
out to 3\,arcsec at higher sensitivity. The HST image independently confirms the extended 
continuum detected in OSIRIS data being diffuse starlight. For J2323$-$0100, a second nucleus is 
marginally discernible 1.3\,arcsec away in the north-northeast direction. The HST image reveals 
the ERQ is in an ongoing major merger with a second nucleus at that position. We note that for 
J1652+1728, although extended continuum is not detected in the NIFS data, the HST image finds 
extended continuum around the source out to 6\,arcsec. Overall we identify signs of merger 
activities in ERQs corroborated by findings from HST images, although we do not attempt to explore 
systematic uncertainties and observational biases to quantify a merger fraction 
\cite[see][]{Villforth23}.

\subsection{Surface brightness radial profiles}

The first technique of assessing extended emissions consists of comparing the surface brightness 
radial profiles of the continuum and [\ion{O}{III}] emissions of the ERQ with a reference 
PSF. For each ERQ, we extract a nuclear spectrum from an aperture of radius 0.1\,arcsec centred at 
the centroid of the ERQ emission, and extract from annular apertures of progressively larger radii 
till reaching the edges of the field of view. The annular apertures have variable widths of 
0.1\,arcsec to 0.4\,arcsec, allowing finer spatial sampling in the inner regions and increasing 
S/N of low surface brightness emissions in the outer regions. 

We first perform continuum and line fitting to the nuclear spectrum. Due to the limited wavelength 
range of the OSIRIS data, in most ERQs either the continuum redward of [\ion{O}{III}] or the 
H$\beta$ emission is well covered, but not both. We therefore employ the best-fit parameters to 
the broadband longslit data of \cite{Perrotta+19} to inform our line fitting. We do not force our 
line decomposition to precisely follow the longslit results, as the precise spectral profile 
depends on the regions included in the extraction aperture \citep{Law+18}. We use a constant to 
fit the continuum, as the continuum is approximately flat within a limited waevelength range and 
linear function would introduce too many degrees of freedom for a fit to be successful. Where the 
H$\beta$ peak is not covered we fix the H$\beta$-to-continuum amplitude ratio to the longslit 
value and allow the flat continuum level to vary in the fitting process, relying on the assumption 
that H$\beta$ is spatially unresolved. Where the H$\beta$ peak is covered we allow both the flat 
continuum level and the H$\beta$ amplitude to vary. We use one to two Gaussians to fit H$\beta$, 
and two to three double Gaussians tied in kinematics and amplitude ratios of 1:3 to fit 
[\ion{O}{III}]\,$\lambda\lambda$4959,5007. The number of components follows the longslit results 
and is also motivated by their distinct spatial distributions which give rise to physical meaning, 
which is further described in Section 3.4. The change in $\chi^2$ is minimal if we increase the 
number of fitted components and we prefer the least number of [\ion{O}{III}] components per ERQ 
system that best fits the data. We initialize the free [\ion{O}{III}] kinematic parameters to the 
longslit values, and fix the H$\beta$ kinematic parameters to the longslit values. It has been 
found from the longslit sample that \ion{Fe}{II} contribution does not significantly change the 
[\ion{O}{III}] measurements \citep{Perrotta+19}. We then fit other annular aperture spectra. We 
determine the H$\beta$-to-continuum amplitude and fix the H$\beta$ kinematic parameters in the same 
way as in the nuclear spectrum. With the exception of J1031+2903, we fix the [\ion{O}{III}] 
components' kinematics to the nuclear spectrum's results freeing only their tied amplitudes. For 
J1031+2903 whose second [\ion{O}{III}] component is most apparent in the 
(0.2\textendash0.4)\,arcsec annulus, we fix the second [\ion{O}{III}] component's kinematic 
parameters in all apertures to this annulus's results (we note that the reduced $\chi^2$ values of 
including or excluding the second [\ion{O}{III}] component in the nuclear spectrum are nearly 
equivalent). This is done to reduce the degrees of freedom in fitting low S/N data. For each 
extracted aperture spectrum we keep a component as a detection if its fitted flux is above 
2$\sigma$. We find that the same two to three [\ion{O}{III}] kinematic components can be employed 
to produced reasonable fits across all apertures. We order the components by their velocity 
dispersion values from high to low and name them ``component 1'', ``component 2'', and ``component 
3'' accordingly. Table~\ref{tab:kinematics} presents the velocity centroids and dispersions of 
these fitted [\ion{O}{III}] components, and their formal $\chi^2$ minimization errors. All ERQs but 
J1652+1728 are fitted with two [\ion{O}{III}] components while J1652+1728 is fitted with three. A 
commonly adopted threshold for ionized outflows that require active galactic nucleus (AGN) driving 
is velocity dispersion $\gtrsim$250\,km\,s$^{-1}$, as these velocities are ten times more prevalent 
in AGN-host galaxies than star-forming galaxies \citep{Harrison+16,Kakkad+20}. With respect to this 
reference value, components 1 of all 10 ERQs are broad, tracing AGN-driven outflows, while 
components 2 and 3 are a mix of six broad and five narrow widths, with narrow widths tracing the 
more quiescent interstellar media. The five fitted components that are below the AGN-driven 
outflow cut are the component 2 of J0006+1215, J0209+3122, J1031+2903, J1217+0234 and the 
component 3 of J1652+1728. Among them, J0006+1215 and J1217+0234 have independent accurate 
systemic redshift measurements from CO host emission \citep{Gillette+24}. The narrow component 2
of both of them are fitted with zero velocity shift from the systemic, indicating that they trace 
quiescent interstellar media likely in dynamical equilibrium with their host galaxies. 
For a non-parametric reference, Table~\ref{tab:annuw80} presents the $w_{80}$ value of the total 
[\ion{O}{III}\,$\lambda$5007 line profile in all extracted aperture spectra. 

With the continuum and [\ion{O}{III}] line fluxes derived from these spectral fits, we construct 
their surface brightness radial profiles as functions of the annular aperture radii. We determine 
errors on the radial profiles as propagated from the formal $\chi^2$ minimization errors of the 
fitting the components. We also 
construct surface brightness radial profiles of the tip/tilt stars. We generally use tip/tilt 
stars rather than the extracted ERQ H$\beta$ emission as the PSF model even though the broad-line 
region is spatially unresolved because tip/tilt stars have higher S/N. Exceptions are J1217+0234 
where the tip/tilt star observations are noisy and we instead use the ERQ H$\beta$ for describing 
the PSF, and J1652+1728 where we use the ERQ's spatially-unresolved continuum in the NIFS data for 
consistency with \cite{Vayner+21a}. 
Fig.~\ref{fig:annuprofspecs} presents these nuclear and annular aperture spectra, their best fits, 
and the surface brightness radial profiles. We qualitatively assess whether the continuum 
and the [\ion{O}{III}] components are spatially resolved by comparing with their reference PSF 
profiles. We also quantitatively assess spatial extents by calculating the half-light radii and 
our defined maximum radii that respectively encompass 50\% and 90\% of the fluxes of the different 
emission components. We determine their errors as propagated from the formal errors of the fitted 
components. Table~\ref{tab:COGmeasures} presents these 
characteristic $R_{\rm half}$ and $R_{\rm  max}$ sizes of the continuum, total [\ion{O}{III}], and 
the individual [\ion{O}{III}] components. We report radii that are deconvolved with their 
reference PSFs, and for unresolved emission components we report their upper limit at the PSF 
sizes. As the adaptive-optics-corrected PSF involves a Gaussian-like core and a weak halo 
comparable to the natural seeing \citep{Law+18}, the half-light radii better characterize 
the emission component sizes. The maximum radii, on the other hand, are more commonly reported in 
the literature, hence calculating them enables comparison with measurements on other luminous 
quasars. 

The continuum emissions of J0006+1215, J1232+0912, J1217+0234, and J1604+5633 are spatially 
resolved, although we caution the suboptimal weather of the J1604+5633 observations. In particular 
the continuum emission of J0006+1215 remains detectable out to edge of the field at 1.4\,arcsec. 

Majority of the [\ion{O}{III}] components that exceed the AGN-outflow threshold are spatially 
unresolved. Specifically, among all [\ion{O}{III}] components of the sample that meet the 
AGN-outflow cut, only five out of 16 of them are spatially resolved, namely the J1550+0806 
component 2, the J1652+1728 components 1 and 2, and the J2323$-$0100 components 1 and 2. 
These five components are spatially compact with maximum radii $\lesssim$4\,kpc. 
On the contrary, all five components of the sample that are below the AGN-outflow cut are 
spatially resolved, namely the component 2 of J0006+1215, J0209+3122, J1031+2903, J1217+0234 and 
the component 3 of J1652+1728. The narrow component 2 of J1031+2903 is not centrally concentrated 
but displays a local dip in its surface brightness radial profile at (0.2\textendash0.4)\,arcsec, 
implying clumpiness in the interstellar media. While the detected extent of an individual emission 
component can be moderated by dust extinction, we observe a general trend for the sample that 
narrower, more extended components become more prominent in outer annuli relative to broader, more 
compact components. 

\subsection{Residuals of PSF subtraction}

The second technique of assessing extended emissions consists of generating the residual data 
after subtracting the nuclear spectrum spaxel by spaxel following a two-dimensional empirical PSF 
profile. The PSF on the OSIRIS detector varies by only a few tenths of a pixel as a function of 
lenslet or wavelength \citep{Lockhart+19}. For a spatially unresolved emission component, e.g.\ 
the quasar broad line region or continuum, the spectral shape anywhere in the data cube would be 
about the same as that at the quasar centroid modulo a flux scaling factor. With this, an emission 
component that does not follow the nuclear spectral shape is revealed to be spatially resolved. We 
use the nuclear spectrum of radius 0.1\,arcsec to be the spatially-unresolved quasar spectral 
template, and determine the continuum or fitted H$\beta$ flux level where it is well covered. For 
every spaxel in the data cube, we re-scale the normalization of this unresolved template to match 
the continuum or H$\beta$ flux and subtract it. We conservatively err on the side of 
oversubtraction to ensure that residual emission is not the PSF wings. This method is not 
sensitive to revealing extended continuum emissions and we only assess extended line emissions. 
Due to the limited wavelength coverage and data sensitivity, we do not attempt a more complex PSF 
decomposition treatment. 

To generate the residual surface brightness map of each ERQ, we collapse the PSF-subtracted data 
cube along the wavelength dimension and spatially smooth with a Gaussian kernel of FWHM 
0.1\,arcsec and mark contours of S/N of two. We determine S/N from the PSF-subtracted data 
channels and their corresponding variance channels and the variance channels of the nuclear 
spectra that form the PSF templates. From the PSF-subtracted data cube we extract from 
annular apertures of progressively larger radii till the edges of the field. We fit two to three 
[\ion{O}{III}] components to the residual annular spectra with the same kinematic parameters as 
before PSF subtraction and allowing the ampltiudes to vary. We then calculate the half-light and 
maximum radii from the fitted [\ion{O}{III}] fluxes in the annular bins. Because the residual 
emission tends to be either dominated by one component where the individual components are 
physically distinct, or have very similar contribution from two components where they are broad 
and blended, we only report the total [\ion{O}{III}] profile's characteristic sizes. We determine 
their errors as propagated from the formal $\chi^2$ minimization errors of the components fitted 
to the PSF-subtracted residual data. 
Fig.~\ref{fig:psfres} presents these residual surface brightness maps, residual annular aperture 
spectra, and their best fits. Brightness flares near the edges of the field of view should be 
considered together with the presence of emission-line shapes in the outer annular spectra to 
assess detection.  
Table~\ref{tab:resmeasures} presents the half-light and maximum residual [\ion{O}{III}] radii. 
With the PSF subtraction method we reveal spatially resolved [\ion{O}{III}] emission in 
J0006+1215, J0209+3122, J1031+2903, J1217+0234, J1550+0806, J1652+1728, and J2323$-$0100, 
totalling  seven out of 10 ERQs. In J0006+1215, J0209+3122, J1031+2903, and J1217+0234, the 
residual emission is dominated by a narrow component. In J1550+0806, the residual emission 
is dominated by a component that has the lowest velocity dispersion of all ERQ fitted components 
that meet the AGN-outflow cut. In J1652+1728, the residual emission is dominated by a narrow 
component and a component that has the second lowest velocity dispersion of all components that 
meet the AGN-outflow cut. In J2323$-$0100, where we detect residual emission contributed by both 
components 1 and 2 which both meet the AGN-outflow cut, we have the best corrected angular 
resolution and the deepest observations. 
We note the hints of one or two ionization cones in J0006+1215, J1217+0234, J1550+0806, and 
J1652+1728. 

\subsection{Physical Motivation of the [\ion{O}{III}] Components}

For a spatially-integrated spectrum, fitting Gaussians provides a noiseless approximation to the 
line profile and the individual components are not physically motivated models. The situation is 
however different for integral-field spectroscopy at high spatial resolution.  
Despite the arbitrary nature of Gaussian line profiles, when the individual kinematic components 
of a source can be coherently fitted across many spatial resolutions and have different spatial 
distributions, the individual components can be ascribed to distinct dynamical mechanisms. 
Nonetheless, considering the results from sensitive JWST NIRSpec integral-field spectra of 
J1652+1728, interpreting individual kinematic components averaged over large apertures especially 
when galaxies are in mergers and have various clumps is nontrivial \citep{Wylezalek+22}. 
We therefore also examine the velocity centroid distribution, velocity dispersion distribution, 
and spatial extent distribution of all components. We then consider components 1 and 2 physically 
distinct in J0006+1215, J0209+3122, J1031+2903, J1217+0234, and J1550+0806, and components 1, 2, 
and 3 distinct in J1652+1728. On the other hand, when two components of a source are both broad, 
blended, and have very similar spatial distributions, often times both spatially unresolved, only 
the sum of the Gaussians is meaningful as a quantification of the total outflow line profile 
\citep{ZakamskaGreene14}. We therefore do not consider components 1 and 2 physically distinct in 
J1232+0912, J1604+5633, J2223+0857, and J2323$-$0100. 

\begin{figure}
\includegraphics[width=\columnwidth]{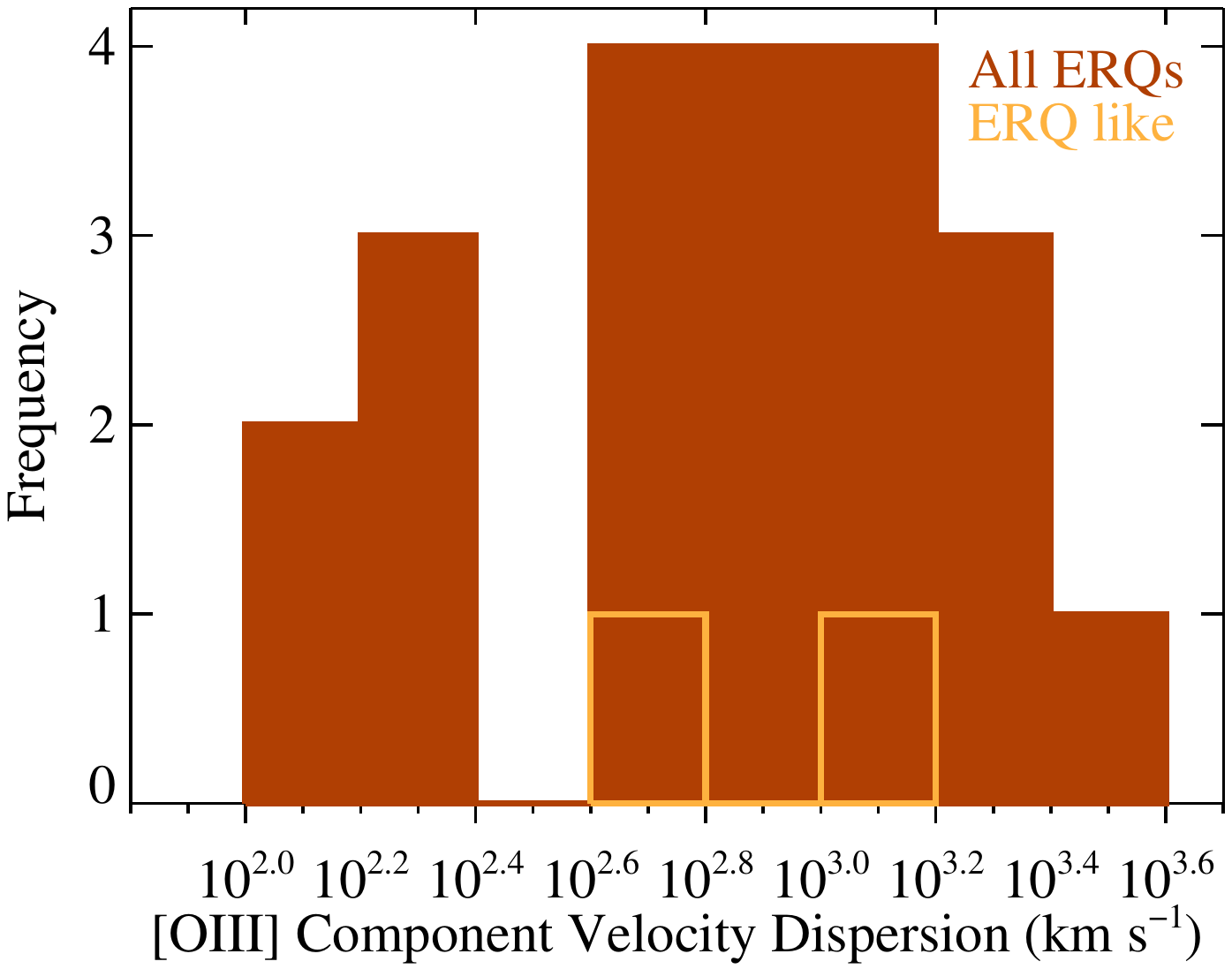}
\caption{A histogram of the velocity dispersion of all [\ion{O}{III}] components fitted to the ERQ 
sample. The full sample in shown in brown solid bars. Components of the one red ERQ-like quasar, 
J1550+0806, is separately shown in orange hollow bars. 
}
\label{fig:disphisto}
\end{figure}

From our radial profile analysis and PSF subtraction analysis, we find that all narrow 
[\ion{O}{III}] components of velocity dispersions below 250\,km\,s$^{-1}$ are spatially resolved, 
while broad [\ion{O}{III}] components are a mix of spatially resolved and unresolved. We can 
demonstrate this finding is not sensitive to the defined AGN-driven outflow threshold. 
Fig.~\ref{fig:disphisto} presents the histogram of velocity dispersions of all fitted 
[\ion{O}{III}] components. A trough in the distribution of velocity dispersions is present at 
$\sim$(250\textendash\,400)\,km\,s$^{-1}$. Thanks to the adaptive optics observations we are able 
to ascribe physical meaning to the narrow versus broad components, which was previously not 
possible with longslit data whose line profile decomposition may be arbitrary. The bimodal 
distribution justifies the adopted AGN-outflow threshold, and the precise velocity cutoff on 
AGN-driving does not affect the discussion in Section~4 comparing the [\ion{O}{III}] spatial 
extents with other AGN-driven outflows in the literature. Further, the bimodal distribution 
implies that the narrow and broad components likely have distinct dynamical mechanisms, with 
narrow components possibly tracing the more quiescent interstellar media. We note that  
in ERQs where both fitted components are broad and blended, the individual components are not 
physically distinct but represent lower limits to the velocity widths of the fast outflowing 
features. The histogram of kinematic components thus underestimates the true velocity distribution 
of fast outflowing materials, and the actual distinction between narrow versus broad 
[\ion{O}{III}]-emitting gas should be even more prominent. 

In the analysis of JWST NIRSpec integral-field spectra of J1652+1728 in \cite{Vayner+23} that 
measure diffuse [\ion{O}{III}] out to 2\,arcsec from the central quasar, they find a similar 
bimodal distribution in the velocity dispersions of components fitted to this single system. This  
further attests to the physical motivation of the [\ion{O}{III}] components when the data has 
sufficient angular resolution. 

\begin{figure*}
\includegraphics[width=0.33\textwidth]{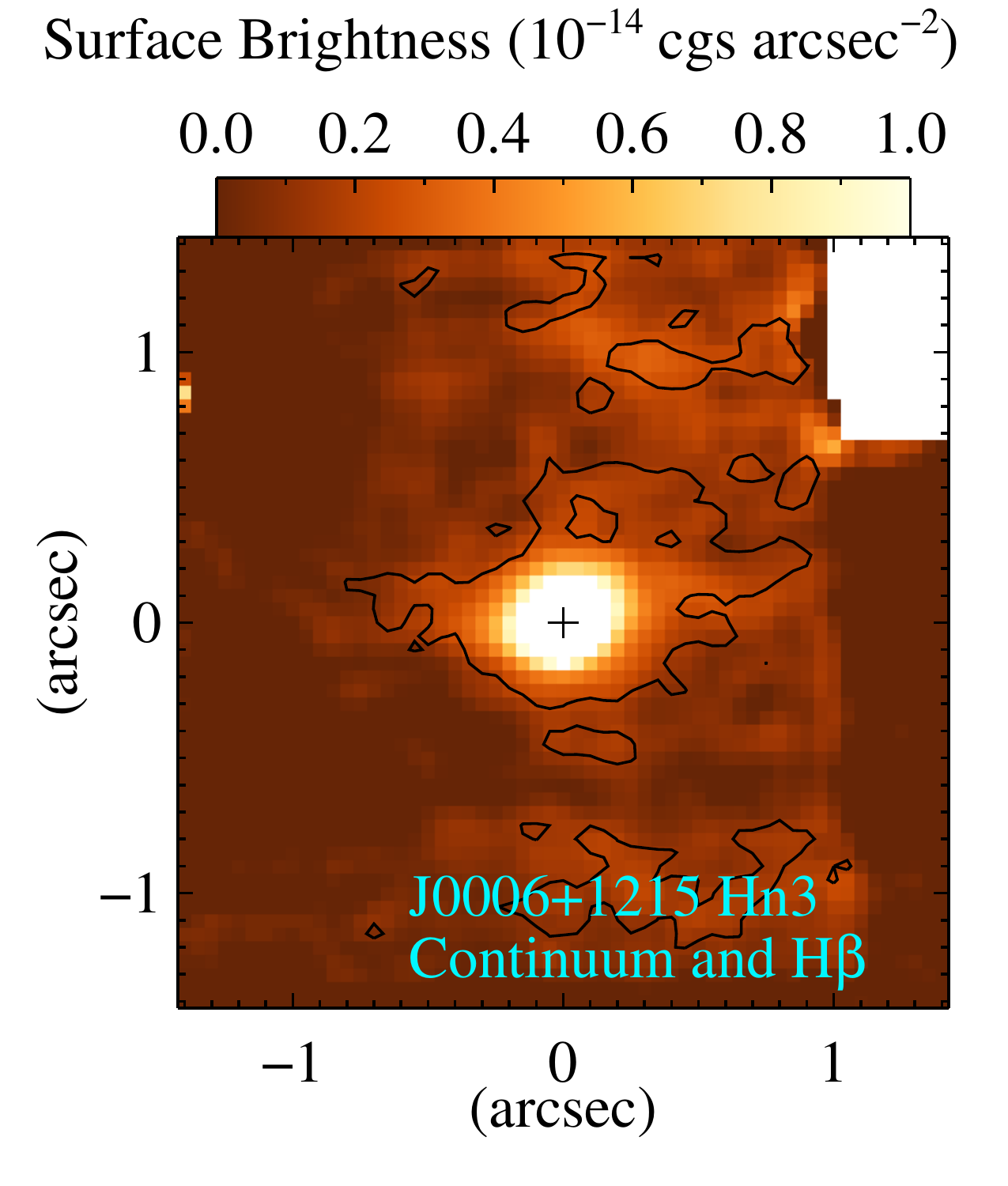}
\includegraphics[width=0.33\textwidth]{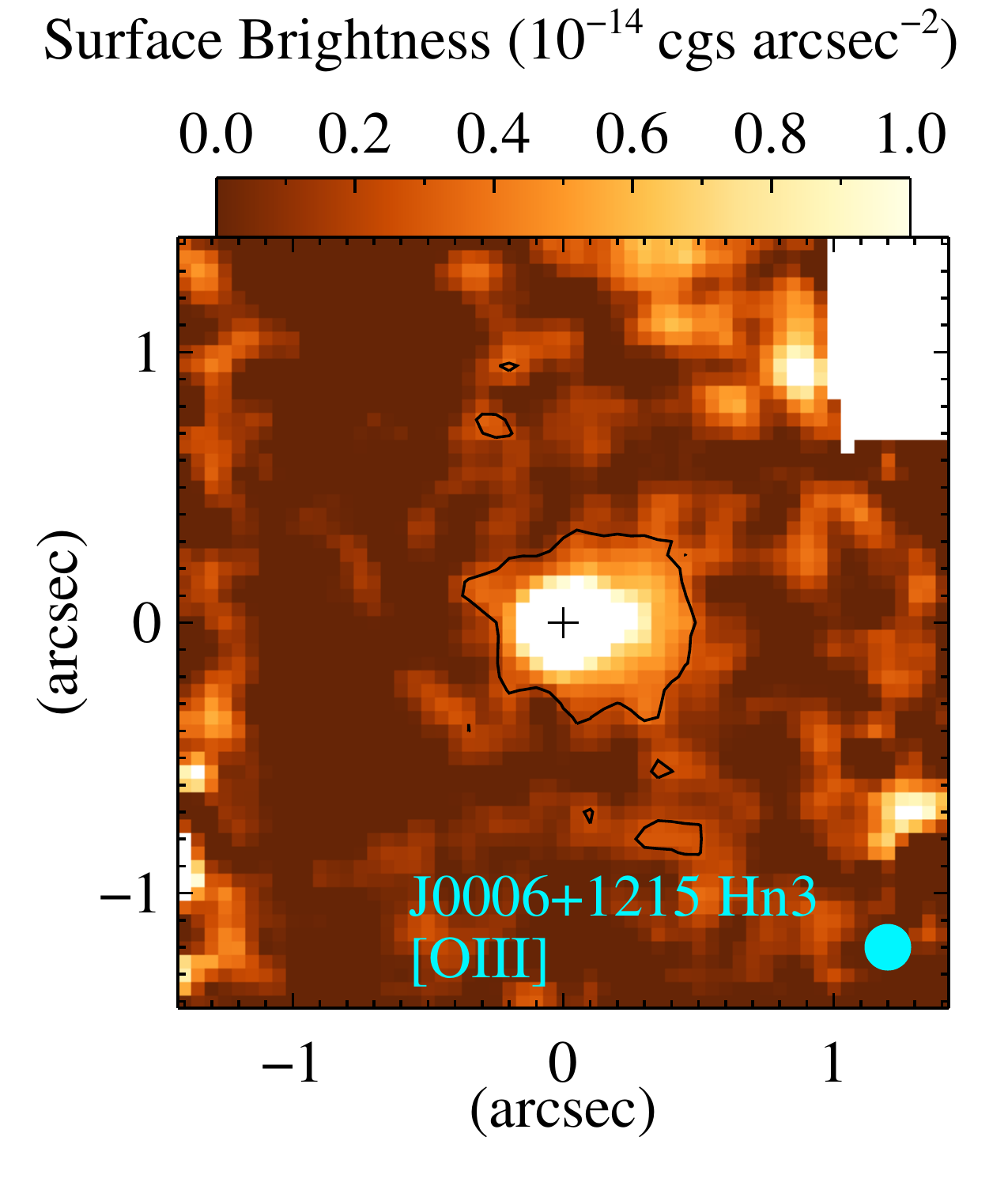}
\includegraphics[width=0.33\textwidth]{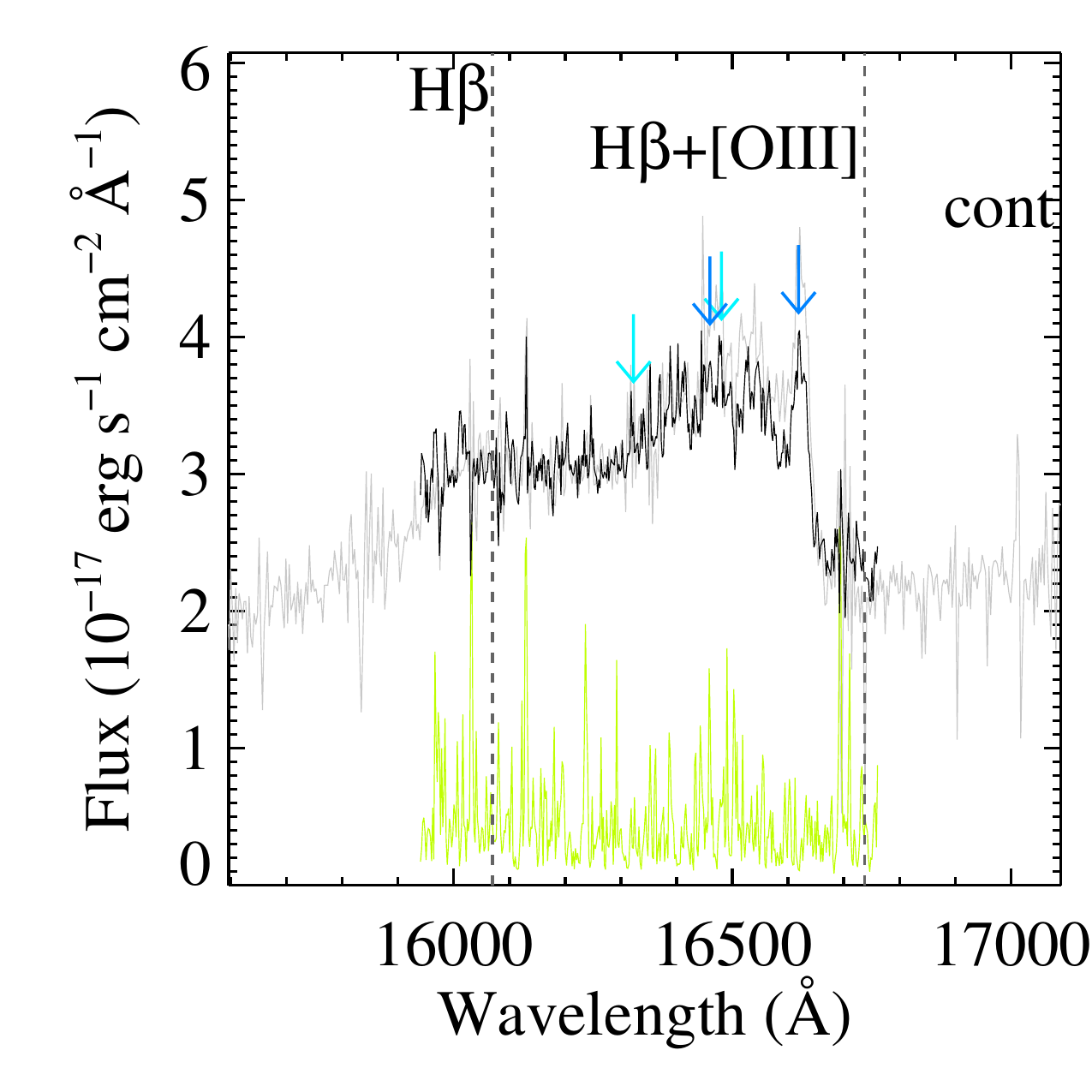}
\includegraphics[width=0.33\textwidth]{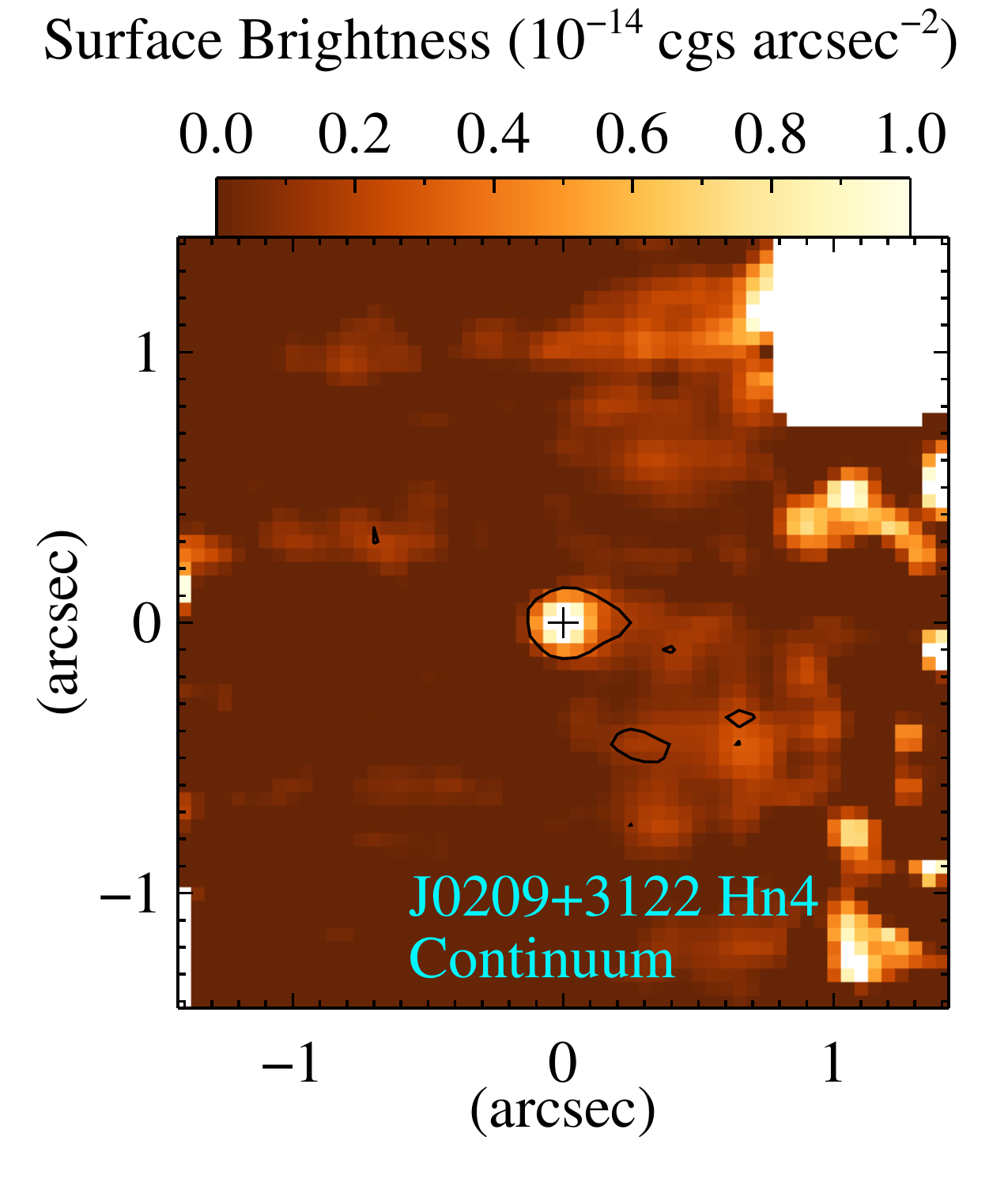}
\includegraphics[width=0.33\textwidth]{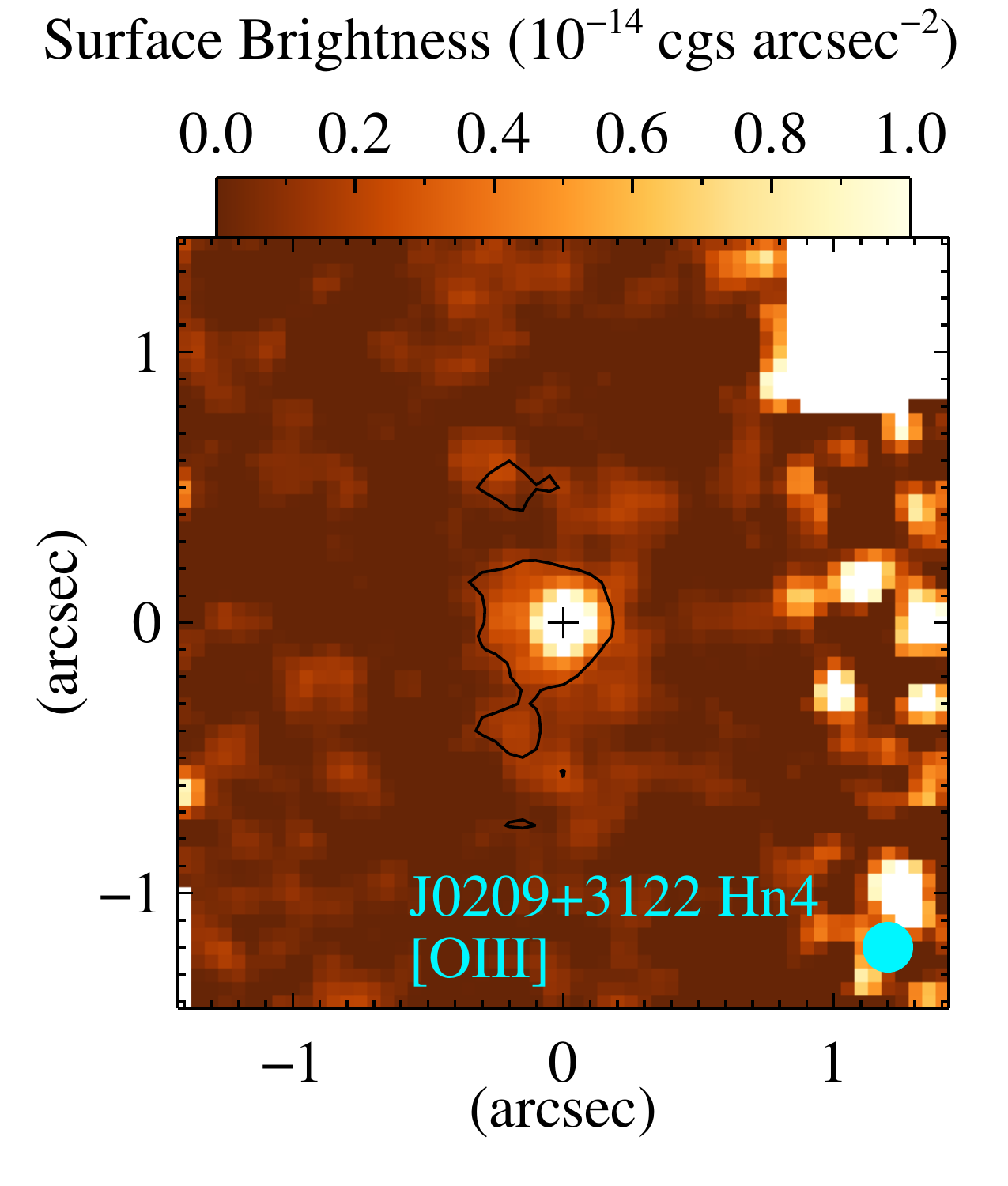}
\includegraphics[width=0.33\textwidth]{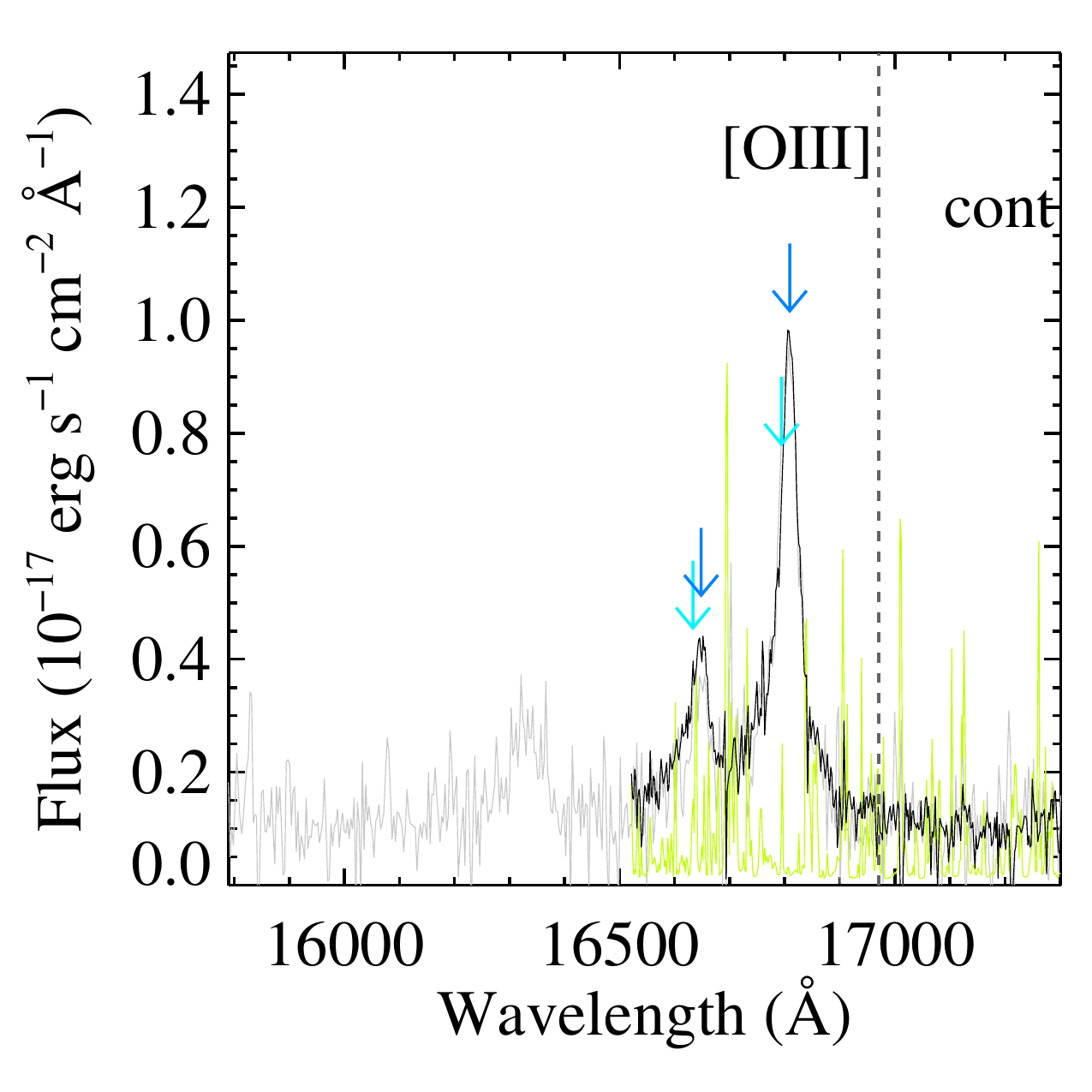}
\includegraphics[width=0.33\textwidth]{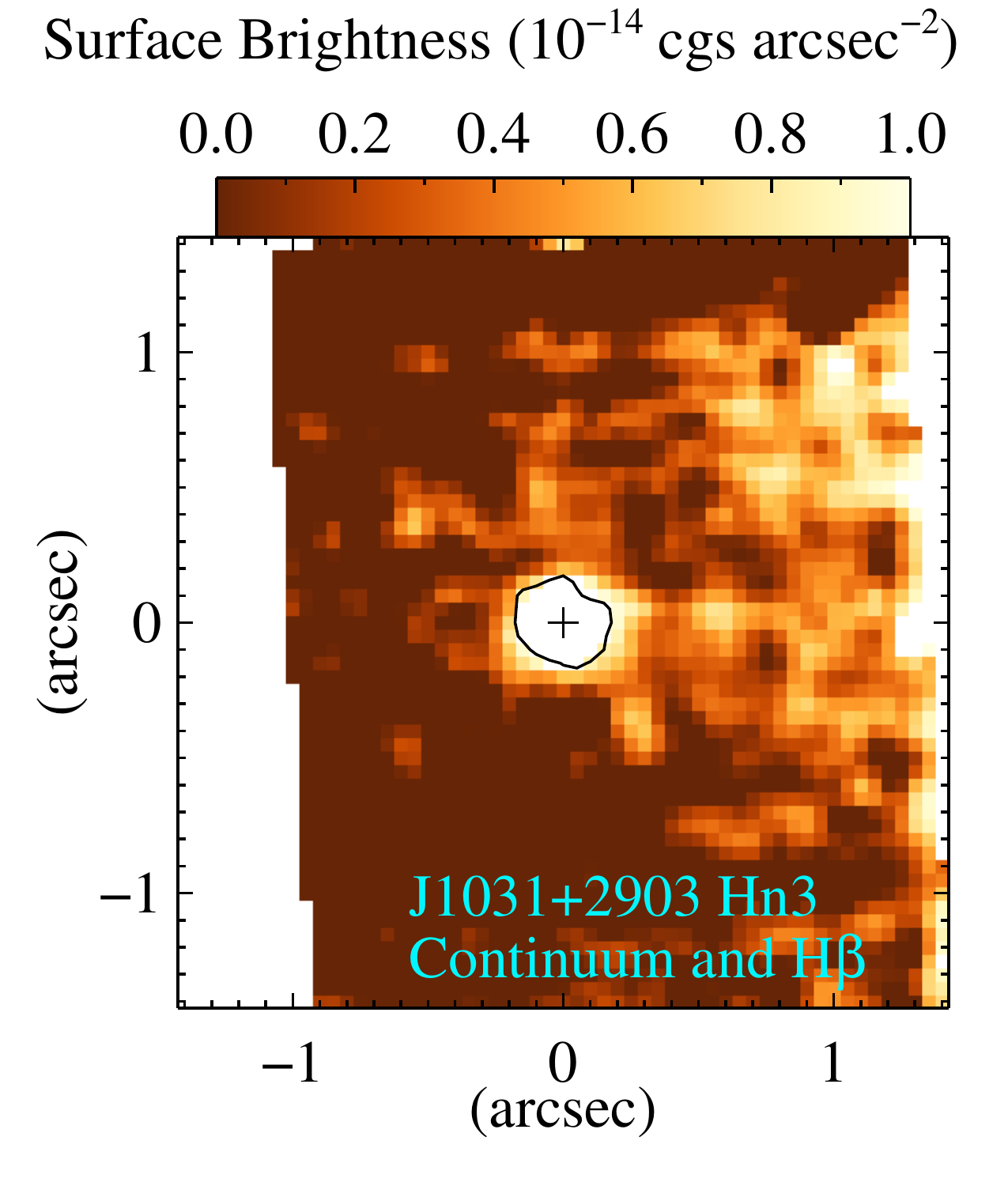}
\includegraphics[width=0.33\textwidth]{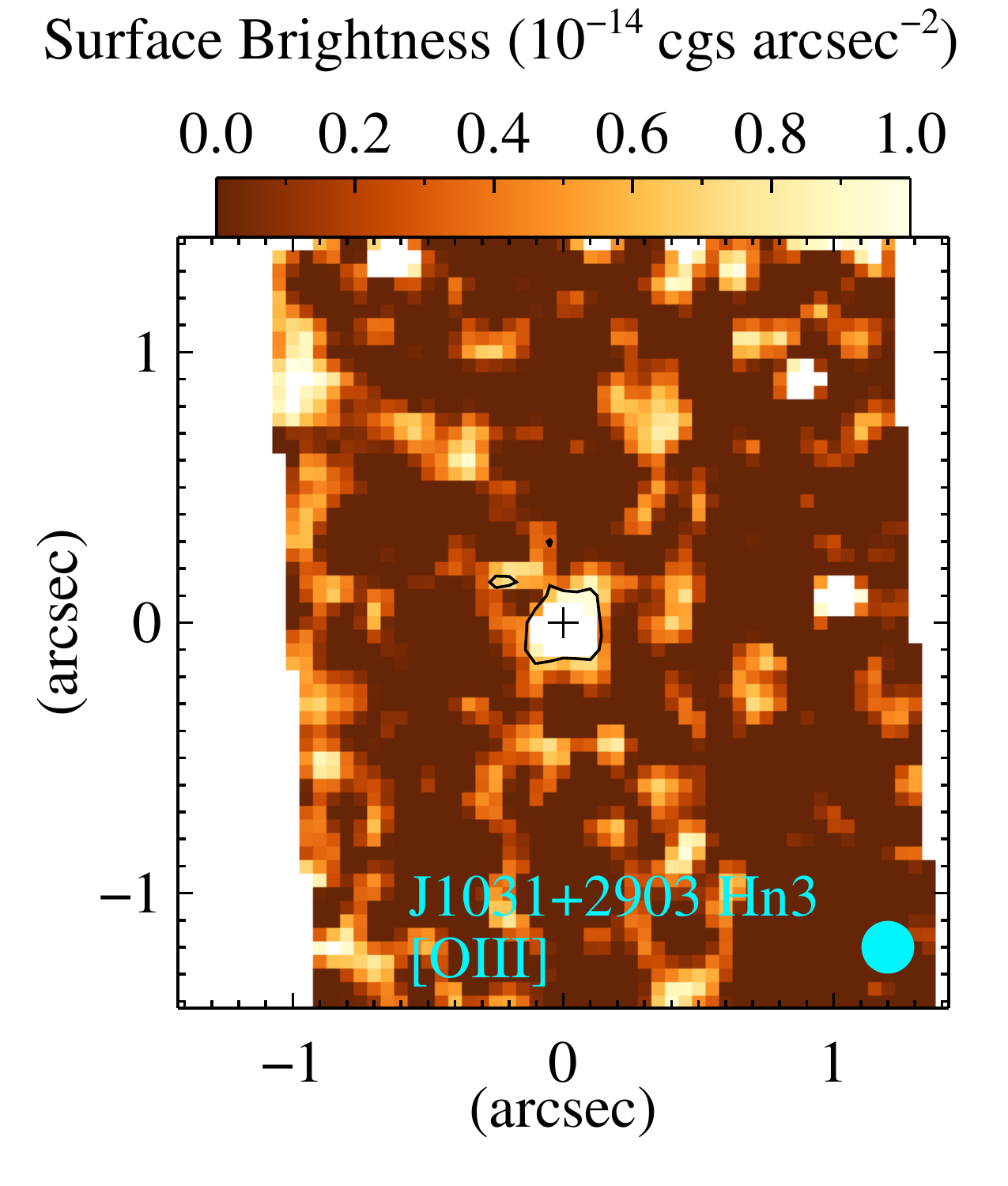}
\includegraphics[width=0.33\textwidth]{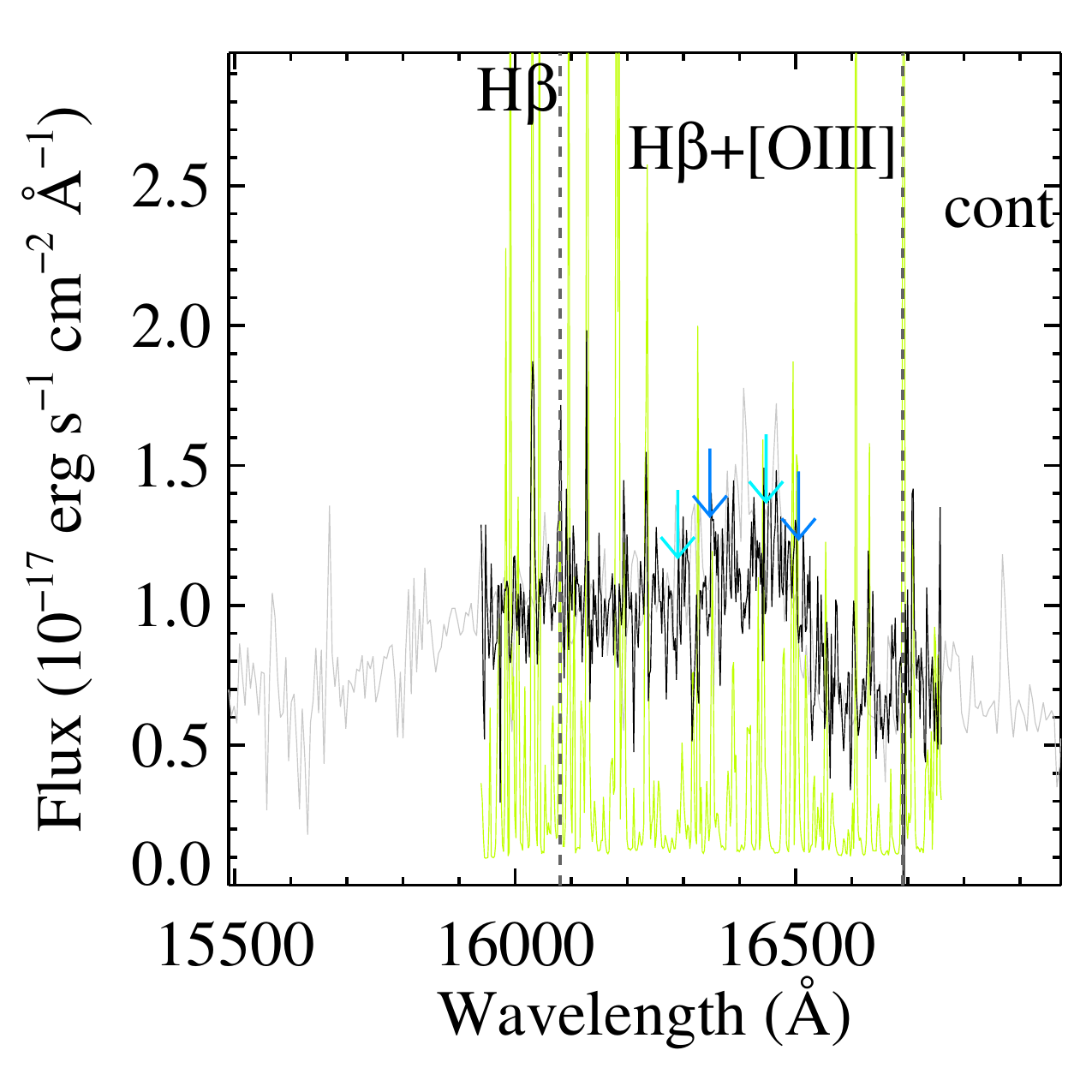}
\caption{Narrowband images and integrated spectra for visualizing adaptive optics observations of 
the ERQ sample. The first column is surface brightness maps summing channels of continuum and 
H$\beta$ emissions. The second column is surface brightness maps summing channels of [\ion{O}{III}] 
emission and subtracting continuum and any blending H$\beta$ emissions. Contours of S/N of two are 
marked on the maps. The Gaussian FWHM size of the PSF is shown as a circle on the [\ion{O}{III}] 
map. The third column is spatially-integrated spectra summing spaxels of S/N greater than two 
shown in observed wavelengths. The flux is in black and the 1-$\sigma$ error is in lime. 
Boundaries of H$\beta$, [\ion{O}{III}], continuum emissions and their blends are marked on the 
spectra. We overplot a previously published longslit spectrum in grey where it exists. For 
J2223+0857 which does not have a longslit spectrum, we overplot our best fit and the H$\beta$ and 
the two [\ion{O}{III}] components in orange, lime, cyan, and blue, and show the residual spectrum.}
\label{fig:nbimgsintspec}
\end{figure*}
\begin{figure*}
\includegraphics[width=0.33\textwidth]{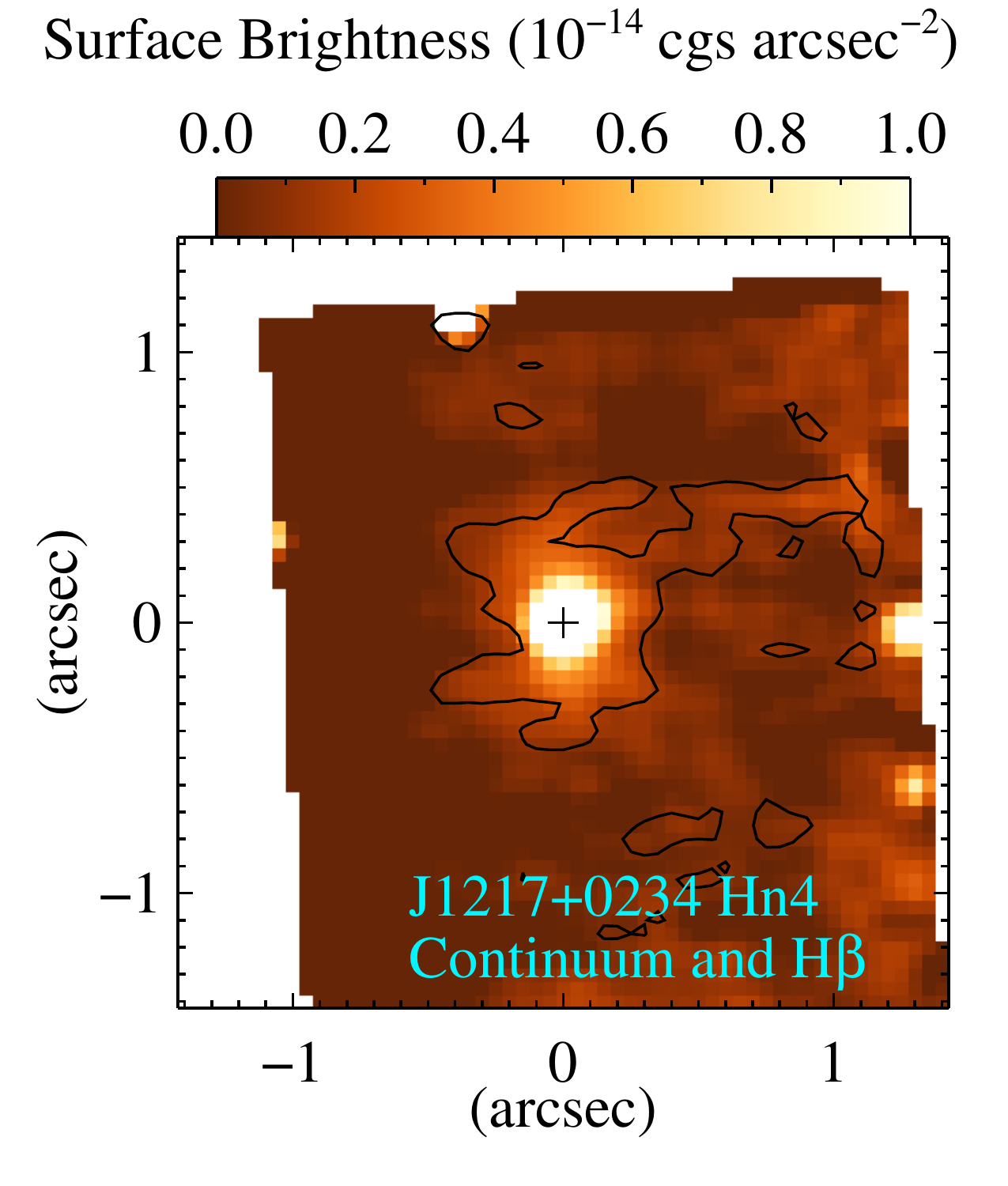}
\includegraphics[width=0.33\textwidth]{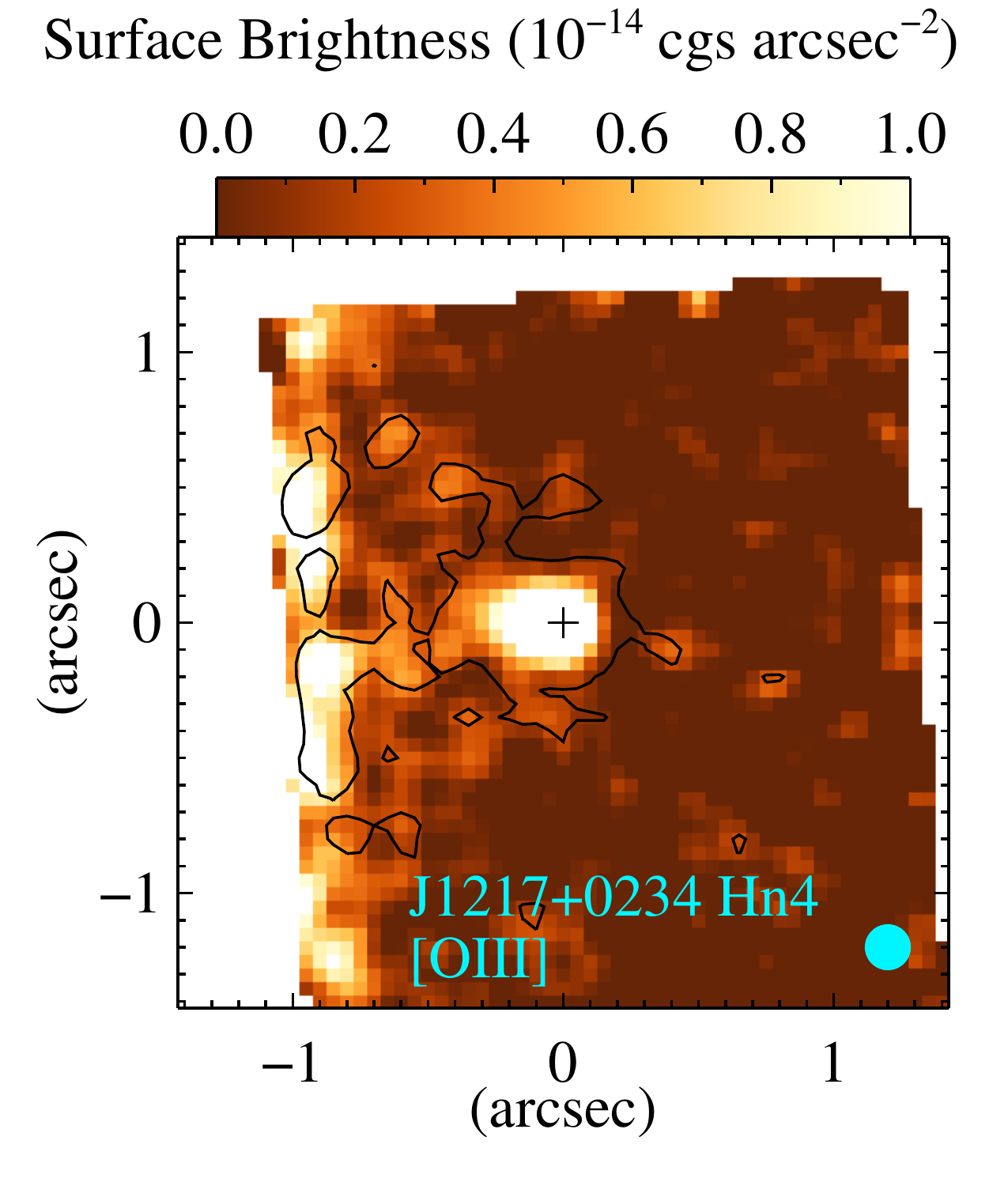}
\includegraphics[width=0.33\textwidth]{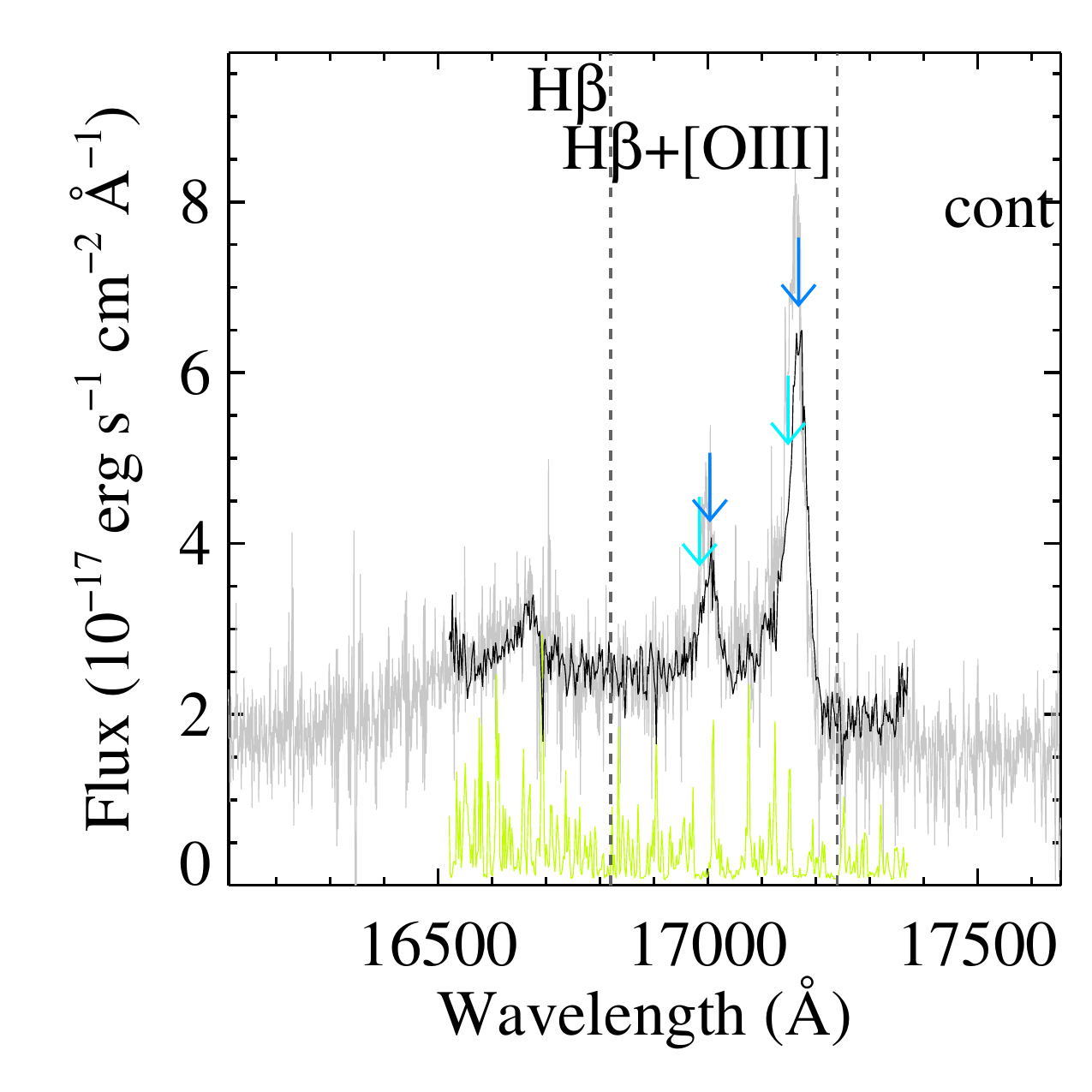}
\includegraphics[width=0.33\textwidth]{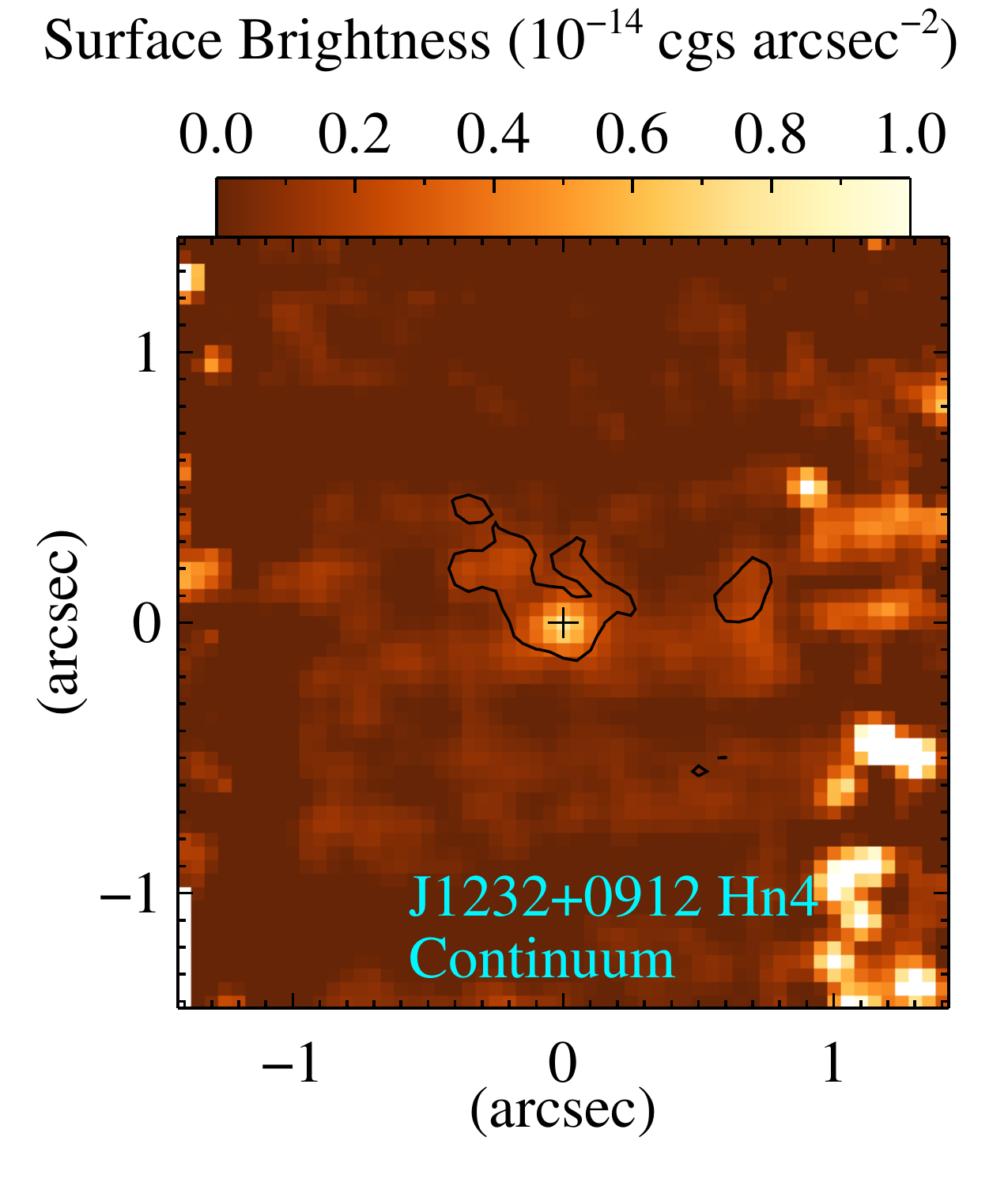}
\includegraphics[width=0.33\textwidth]{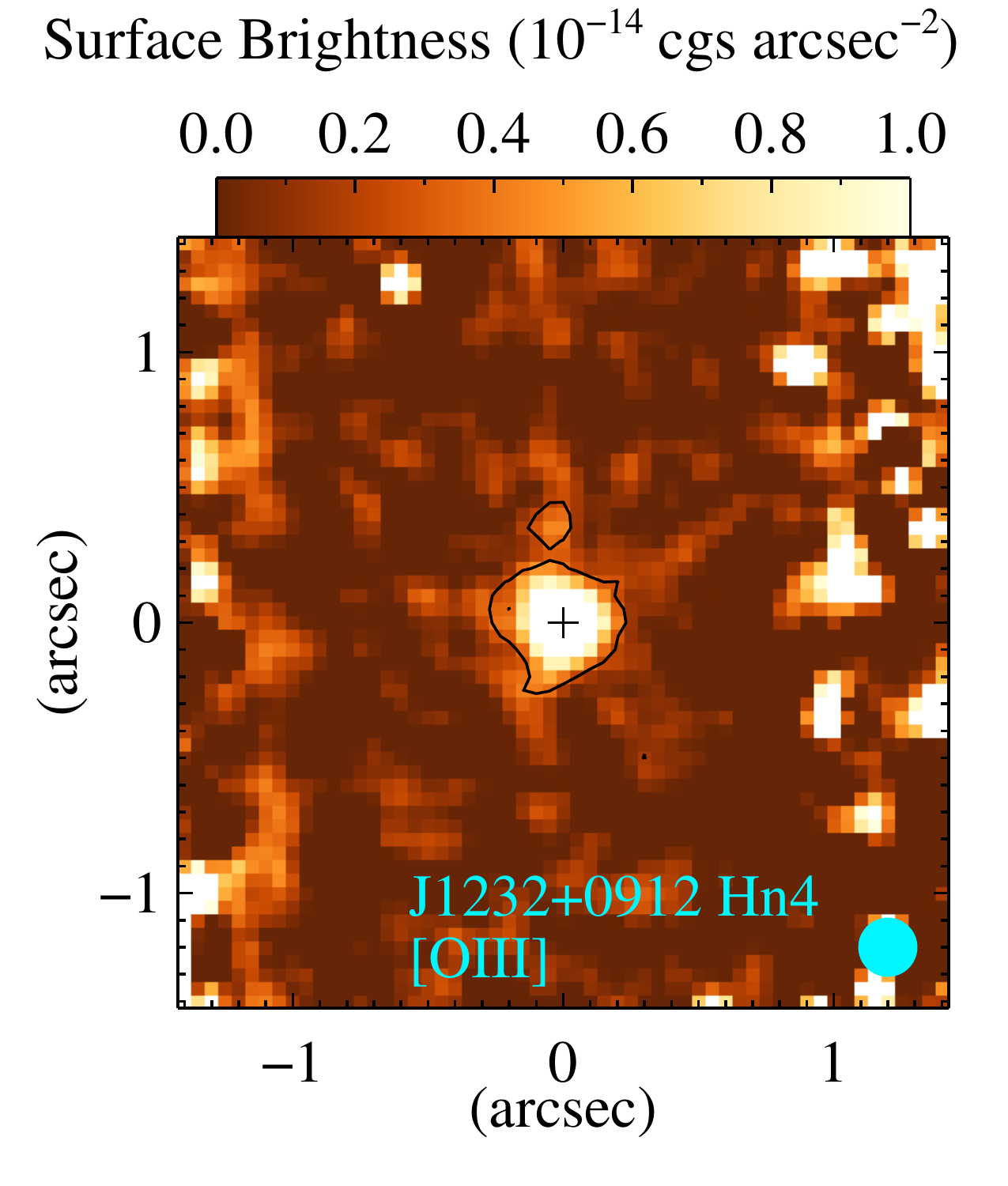}
\includegraphics[width=0.33\textwidth]{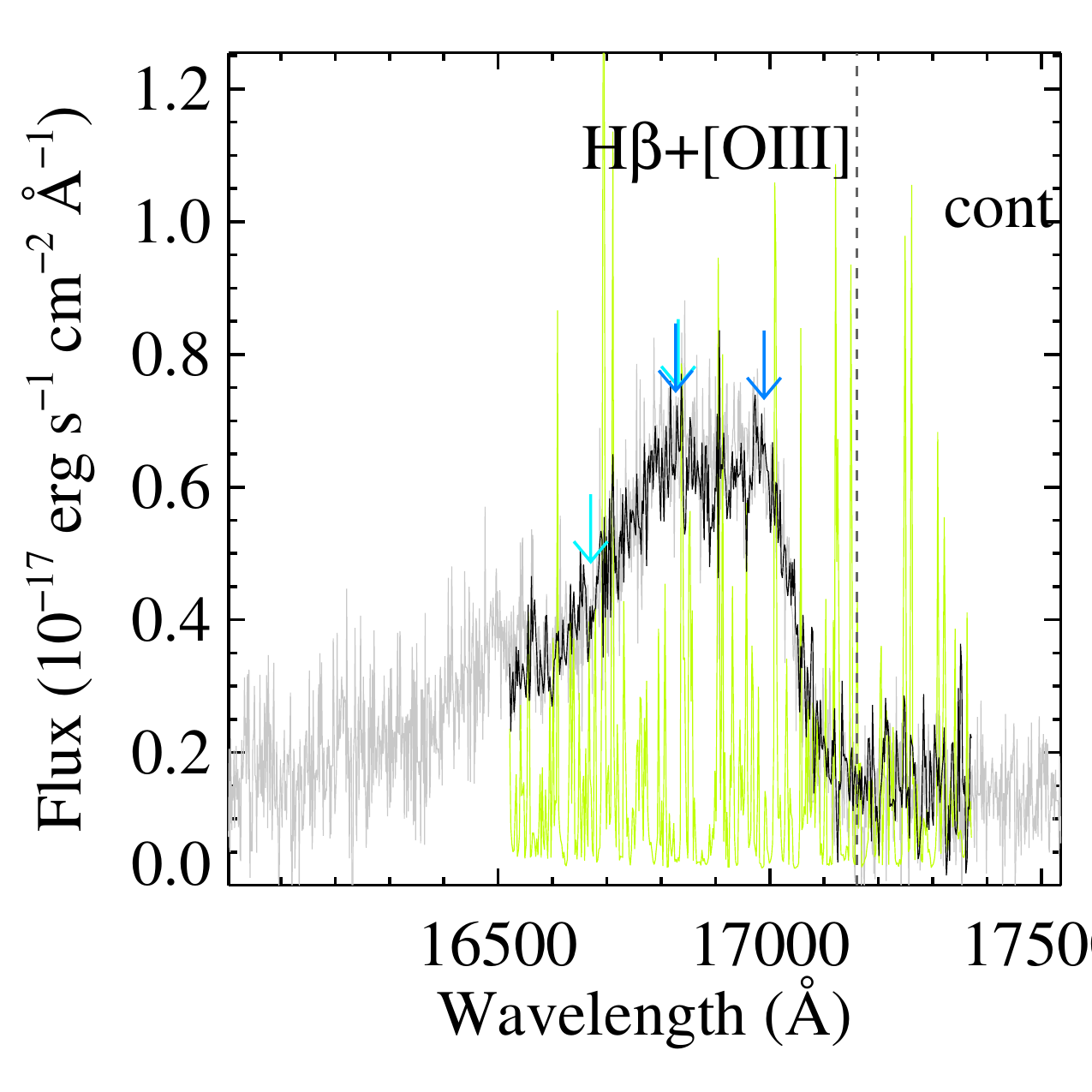}
\includegraphics[width=0.33\textwidth]{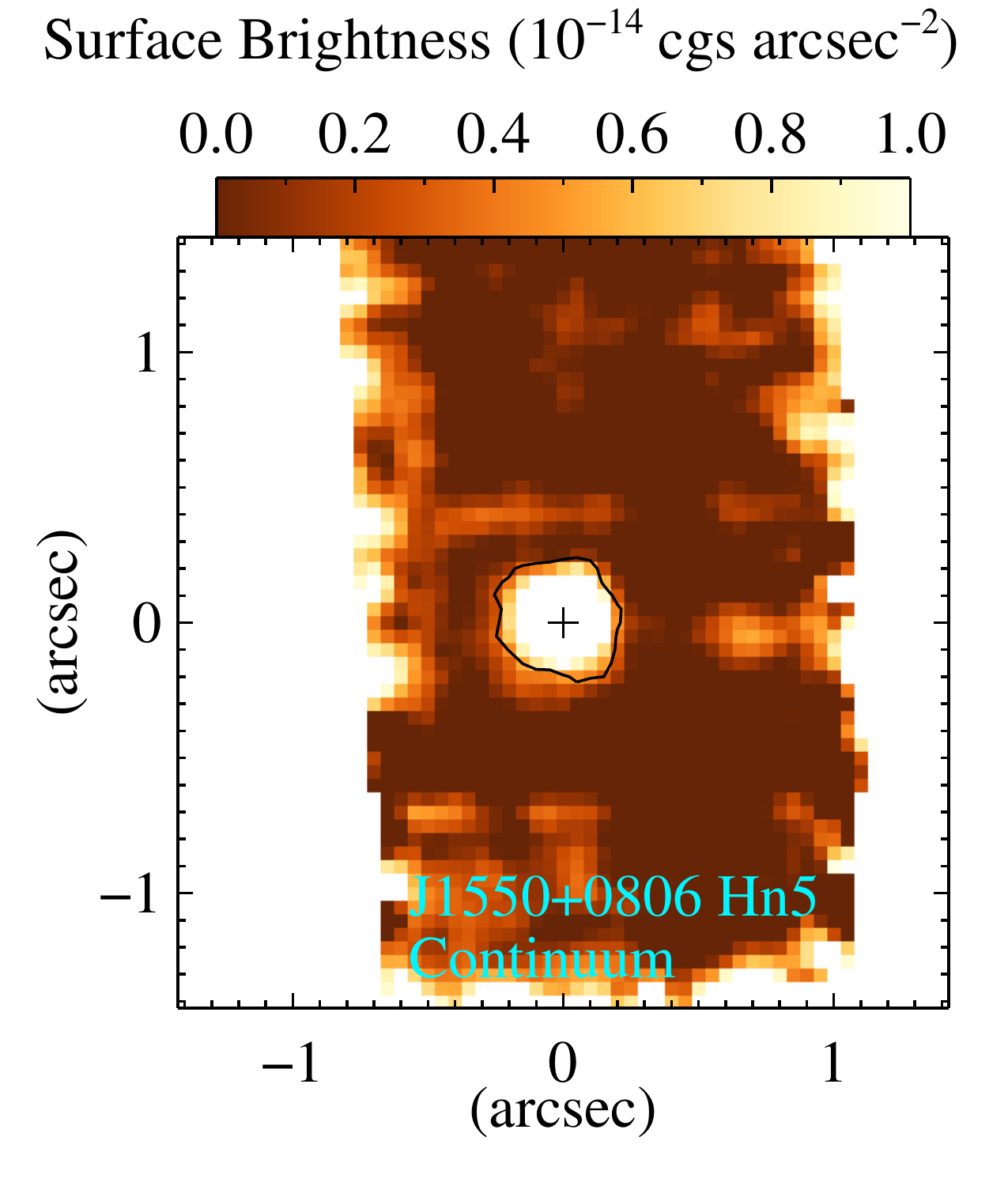}
\includegraphics[width=0.33\textwidth]{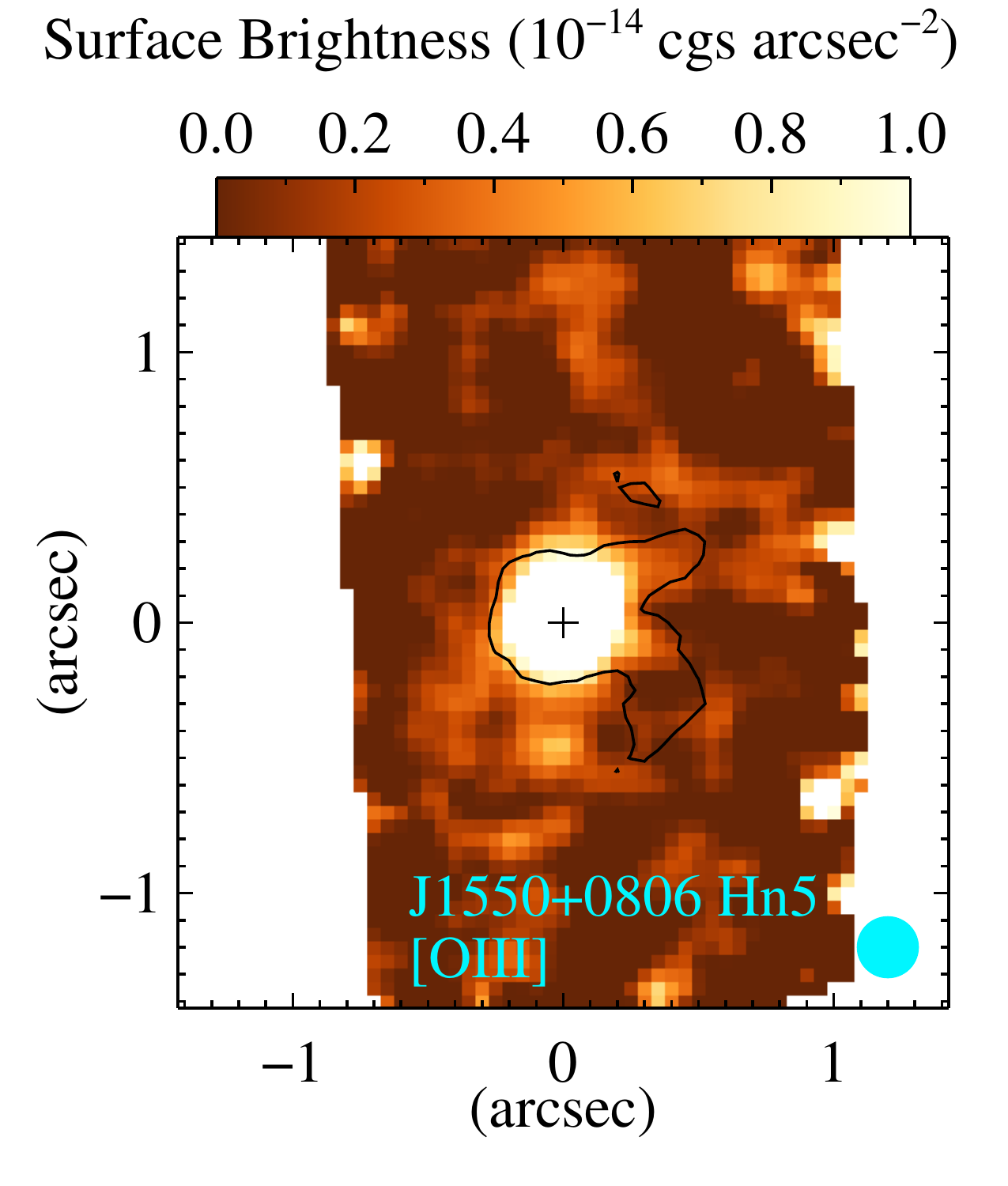}
\includegraphics[width=0.33\textwidth]{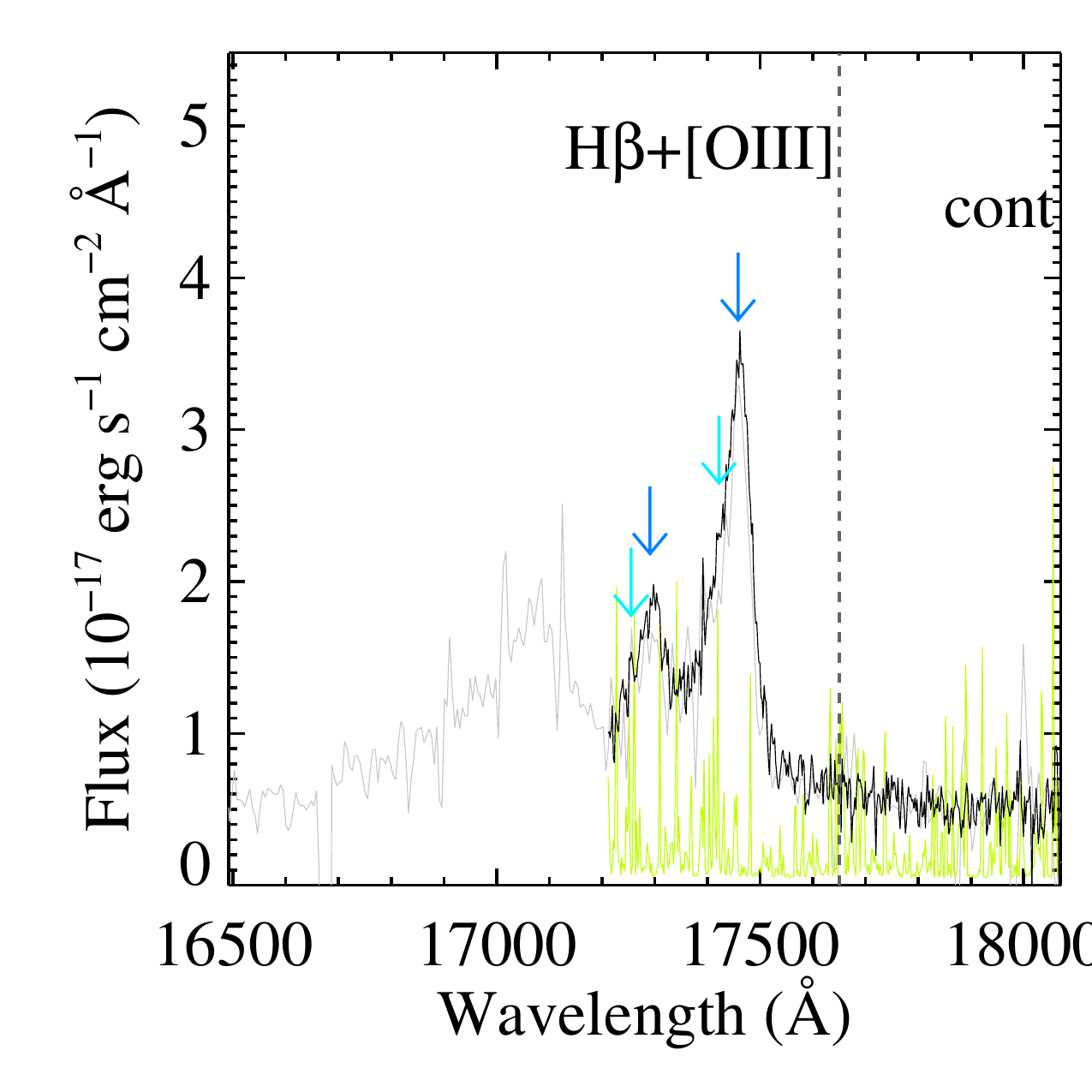}
\contcaption{}
\end{figure*}
\begin{figure*}
\includegraphics[width=0.33\textwidth]{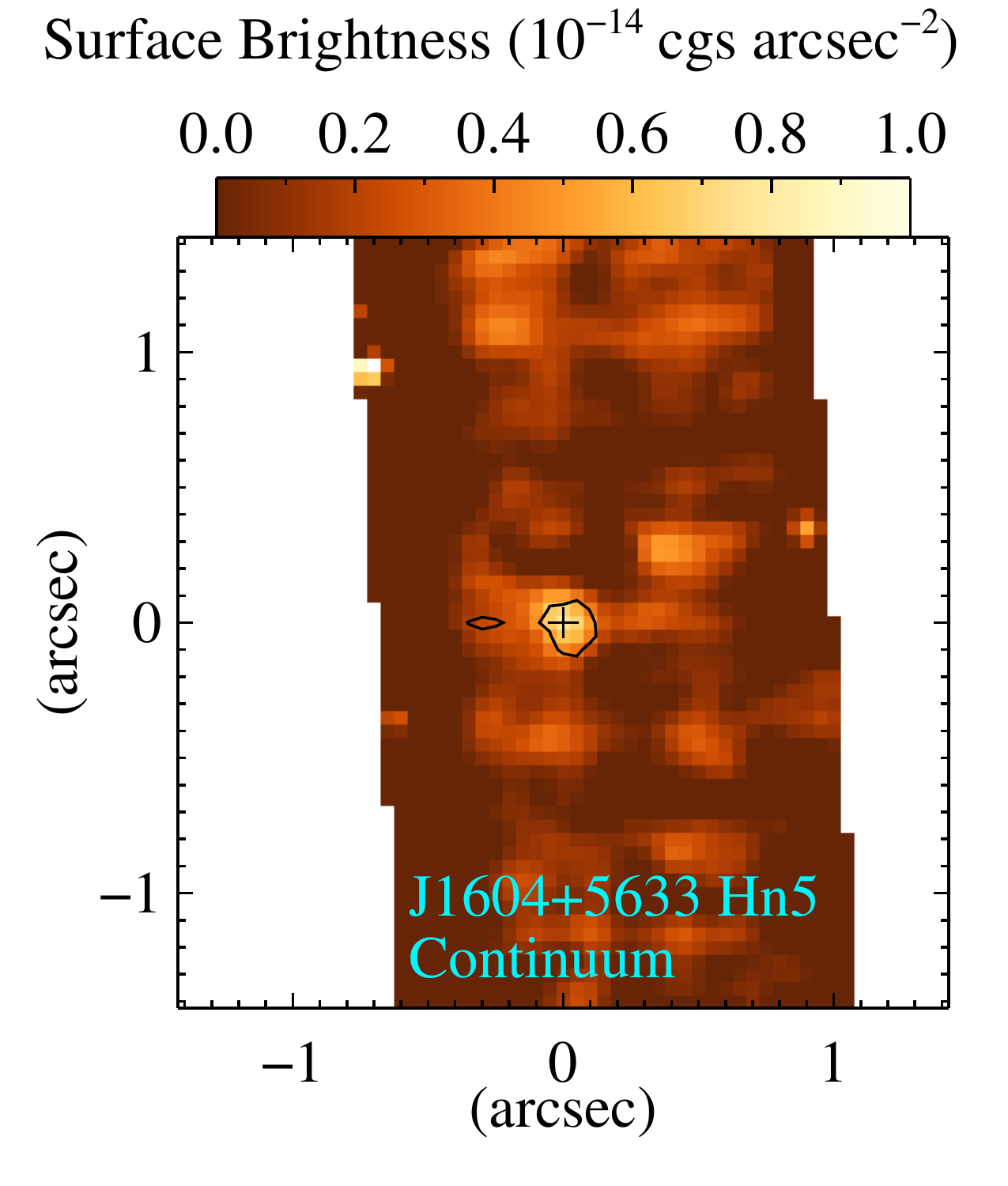}
\includegraphics[width=0.33\textwidth]{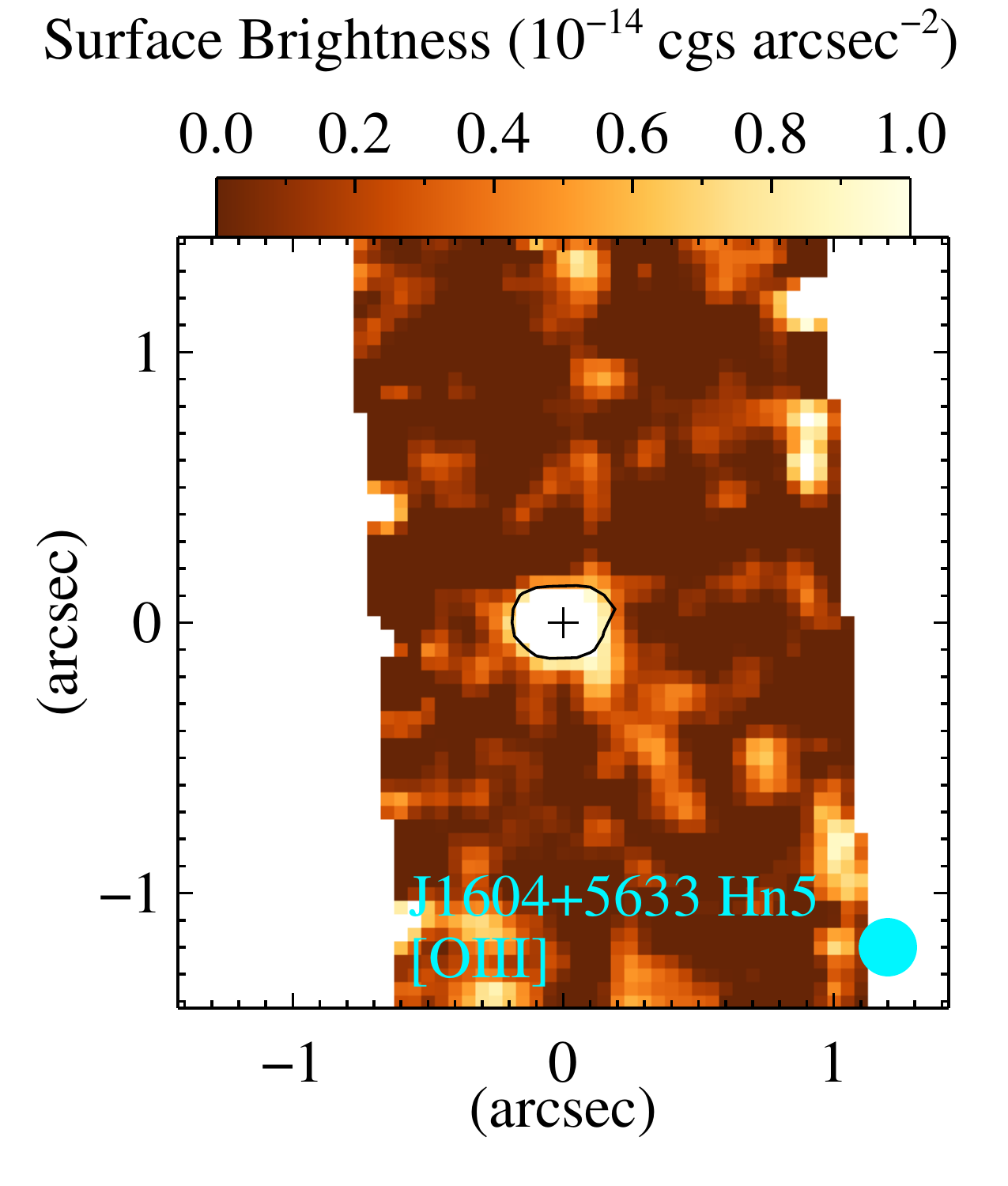}
\includegraphics[width=0.33\textwidth]{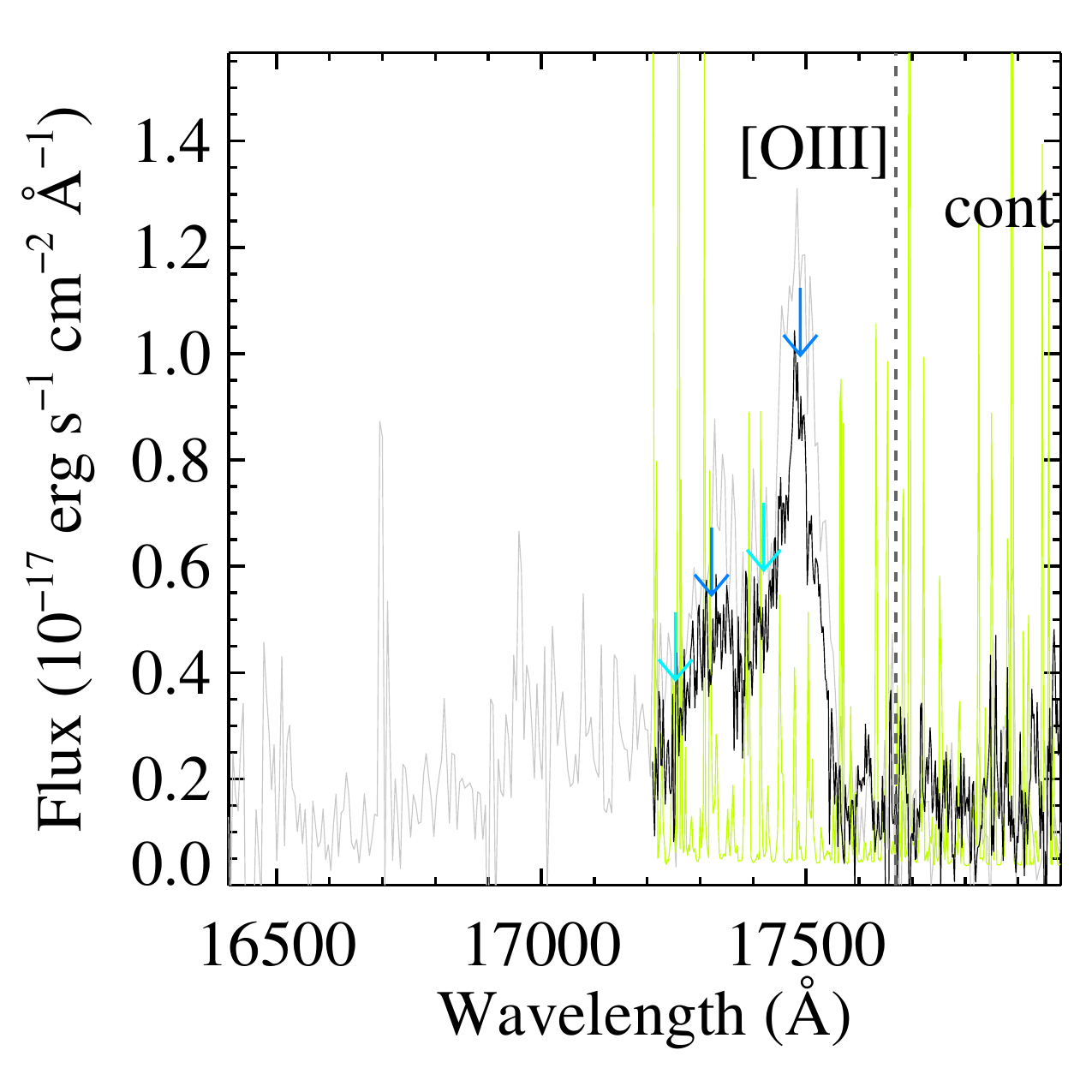}
\includegraphics[width=0.33\textwidth]{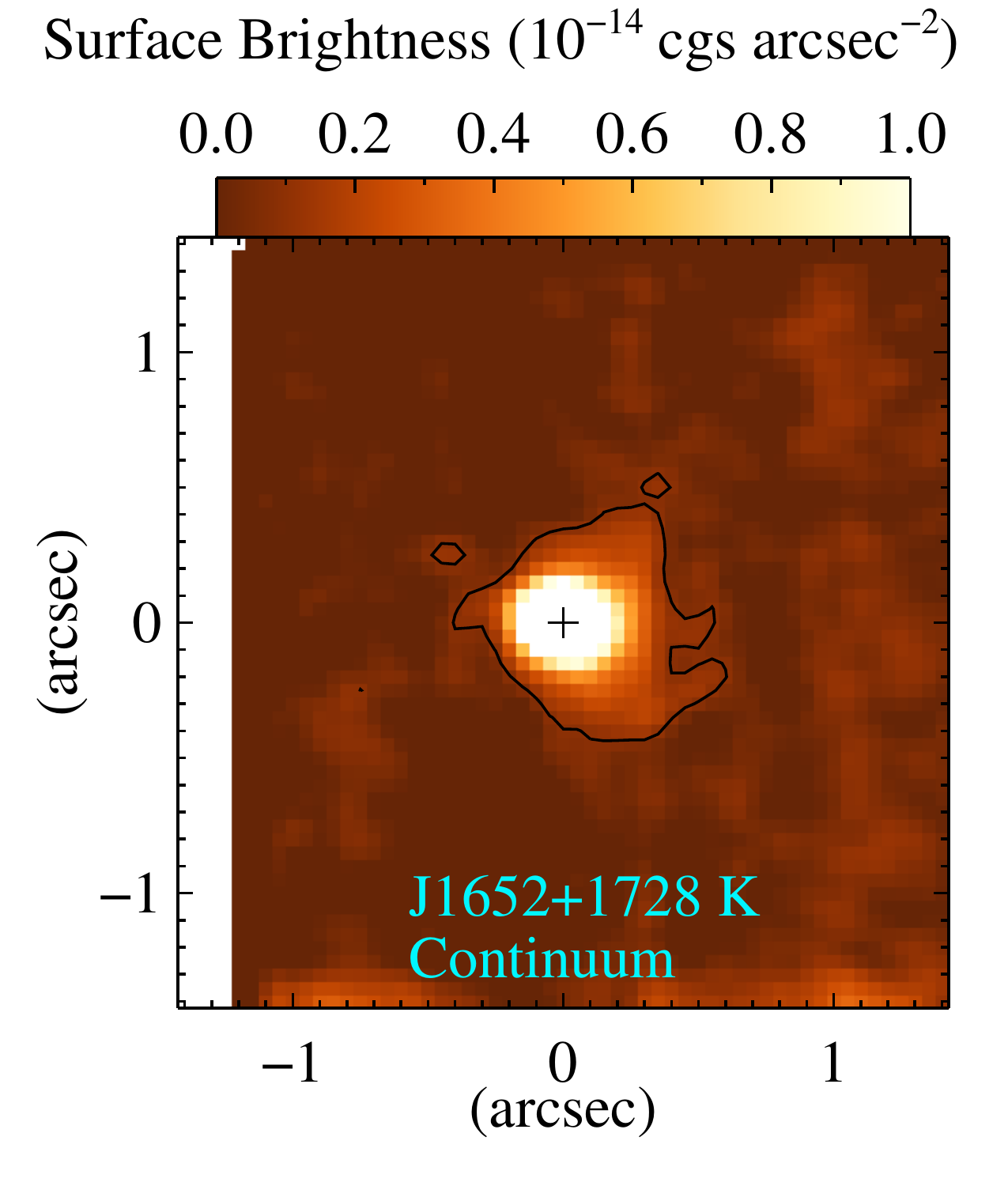}
\includegraphics[width=0.33\textwidth]{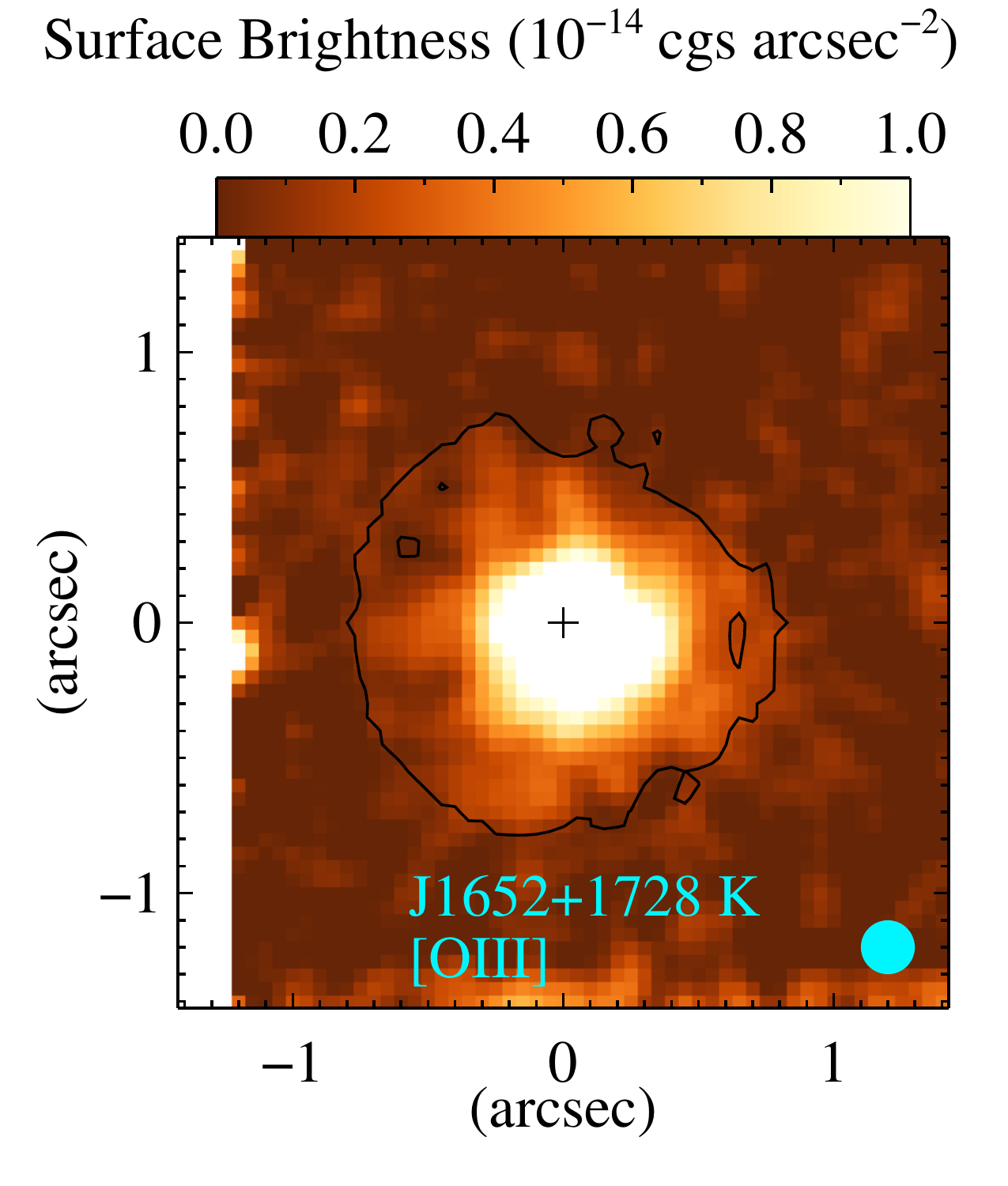}
\includegraphics[width=0.33\textwidth]{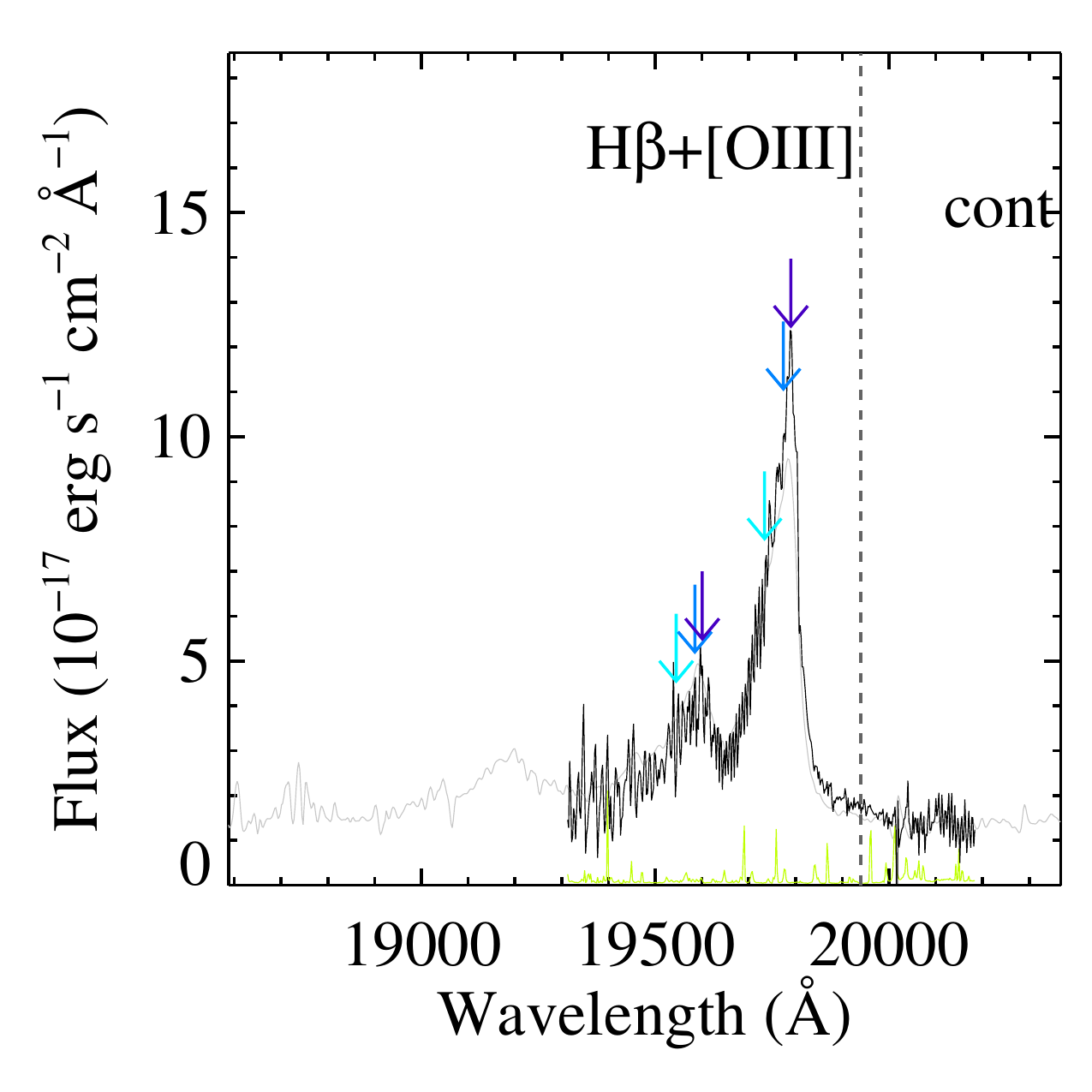}
\includegraphics[width=0.33\textwidth]{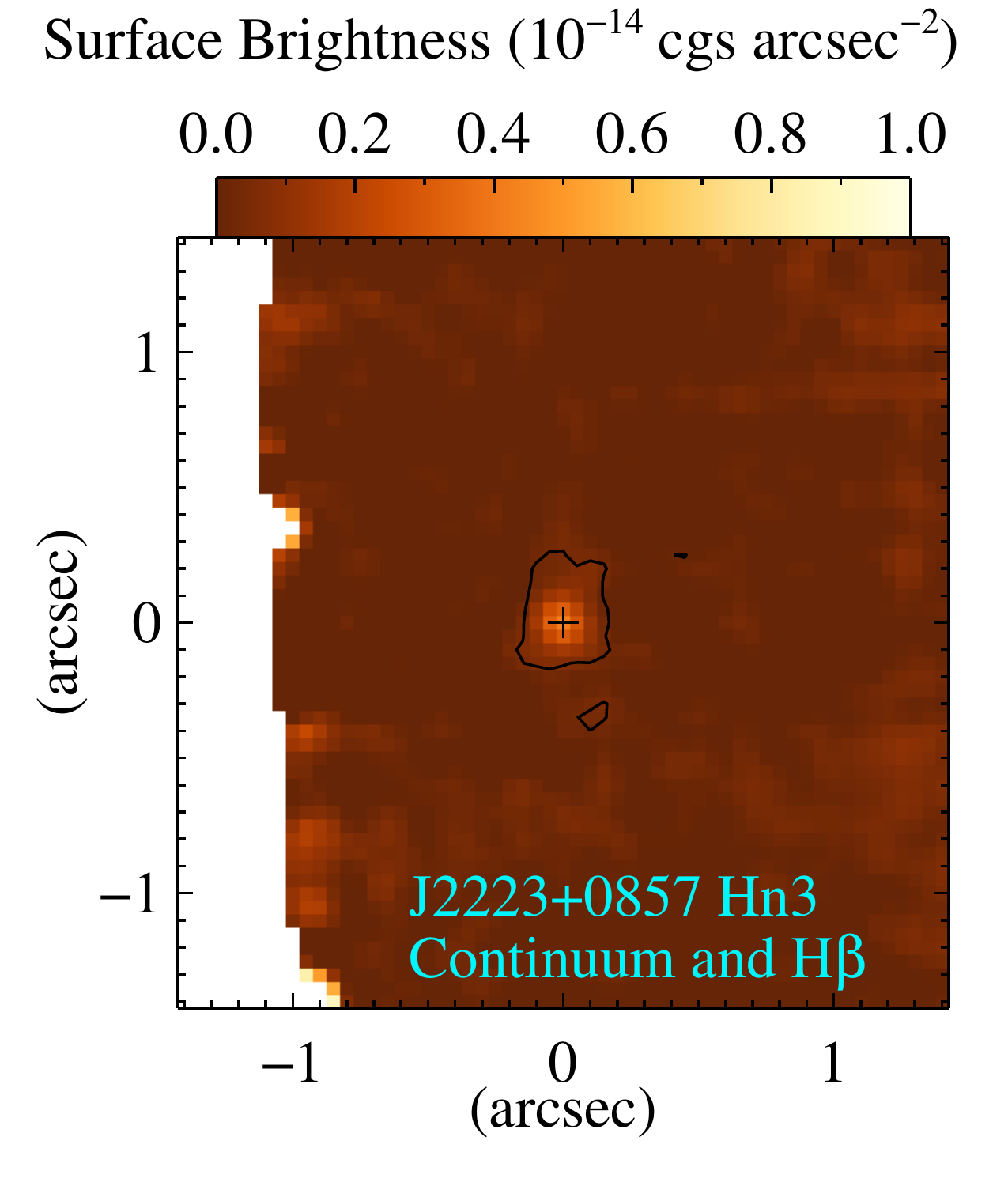}
\includegraphics[width=0.33\textwidth]{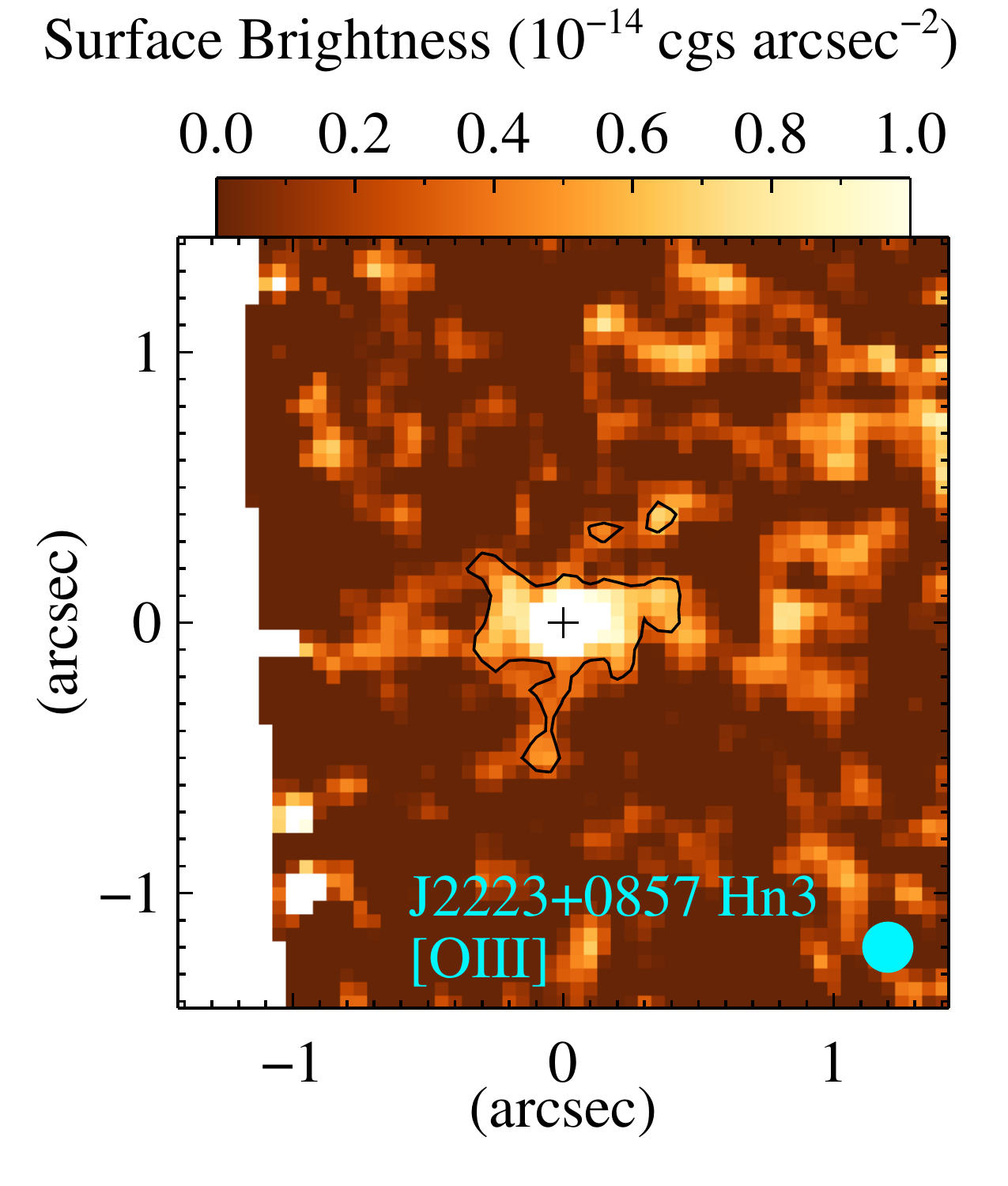}
\includegraphics[width=0.33\textwidth]{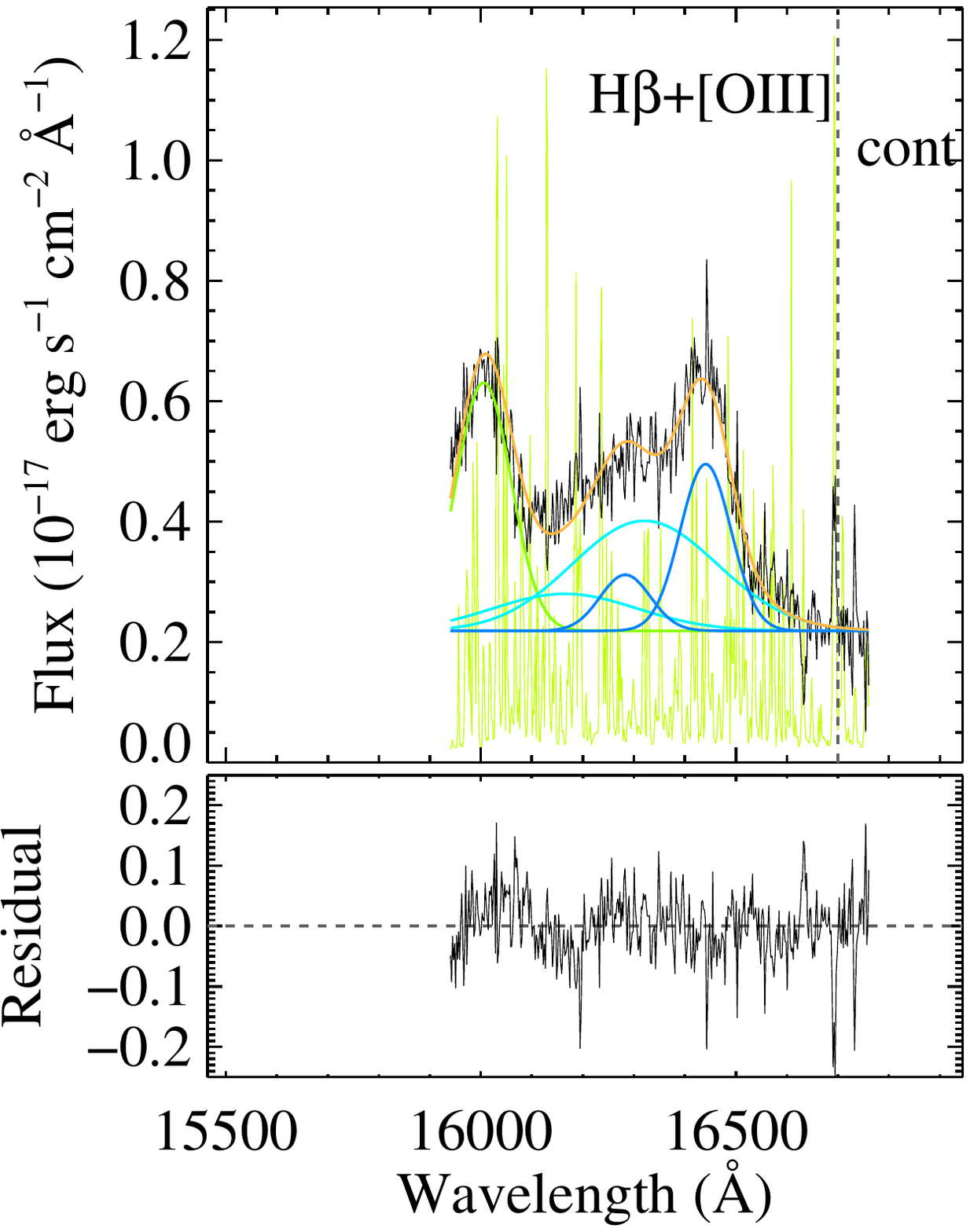}
\contcaption{}
\end{figure*}
\begin{figure*}
\includegraphics[width=0.33\textwidth]{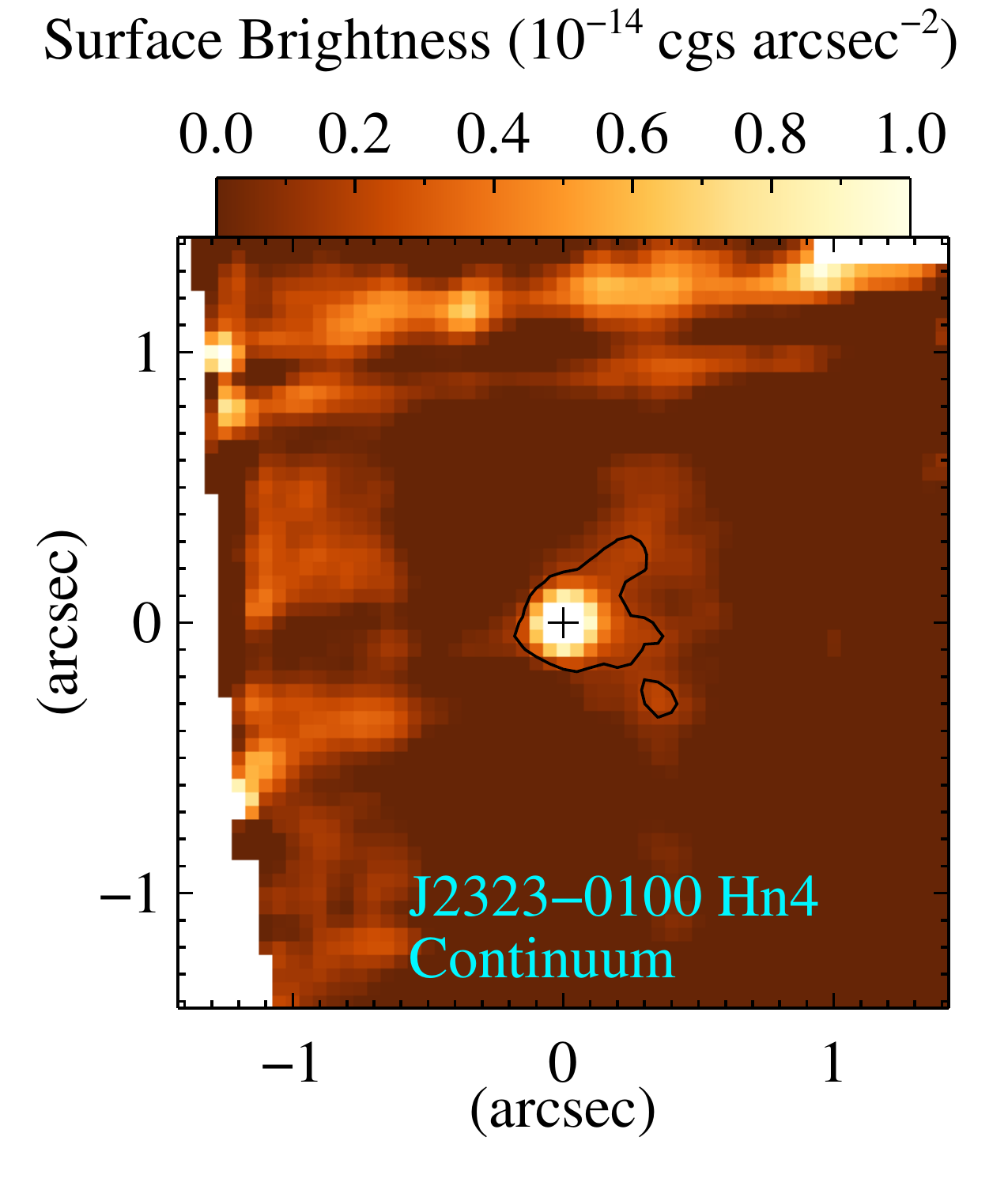}
\includegraphics[width=0.33\textwidth]{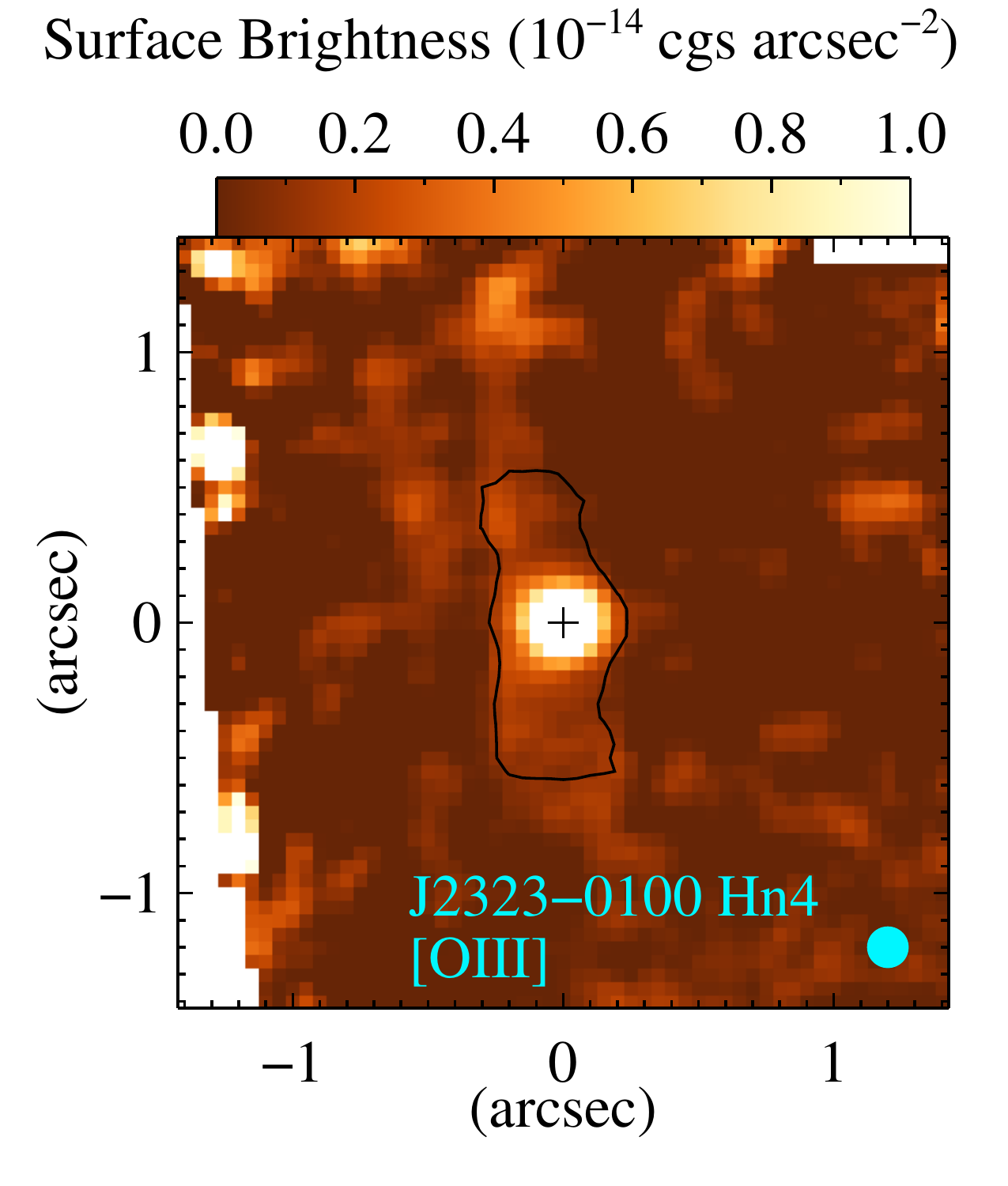}
\includegraphics[width=0.33\textwidth]{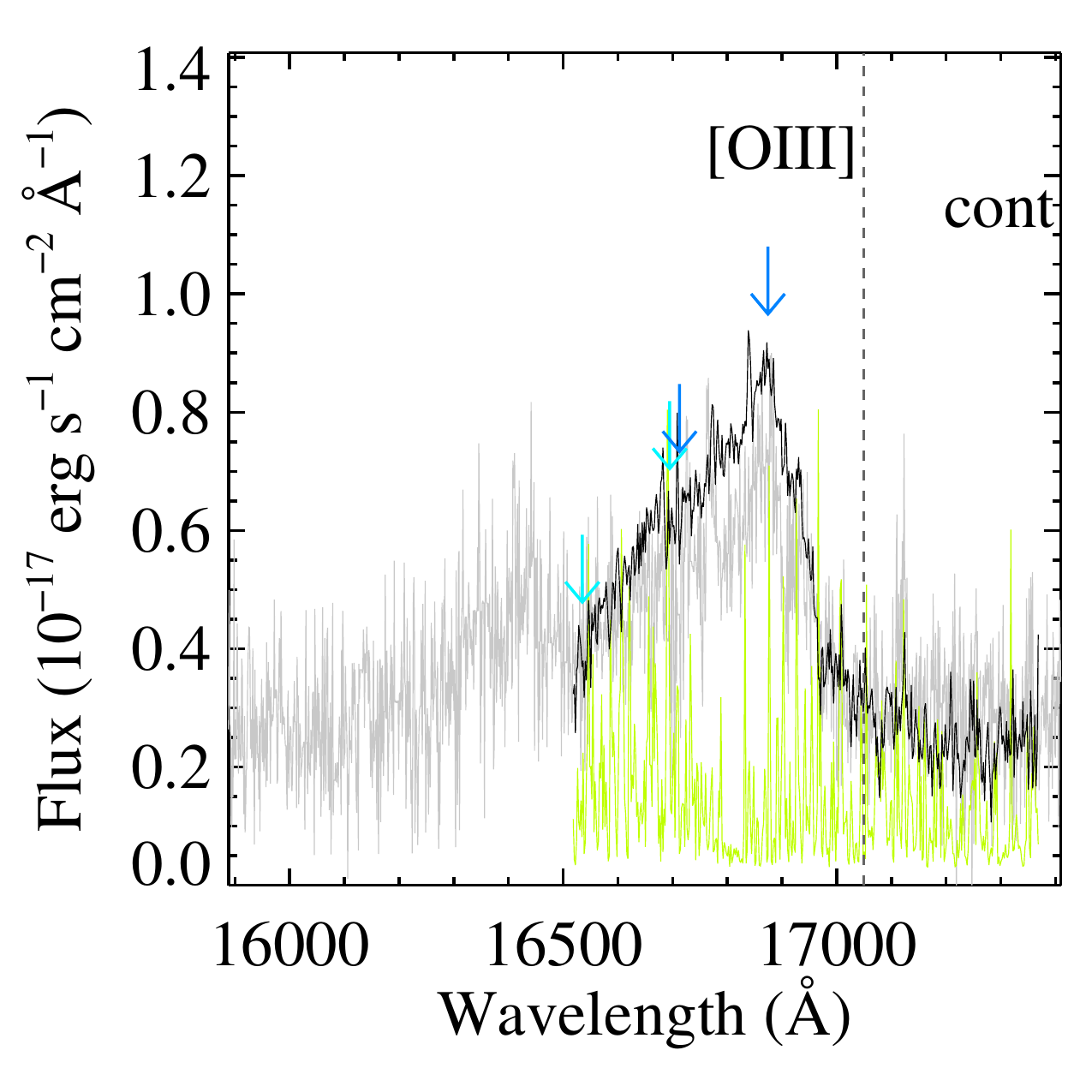}
\contcaption{}
\end{figure*}

\begin{figure*}
\includegraphics[width=0.66\textwidth]{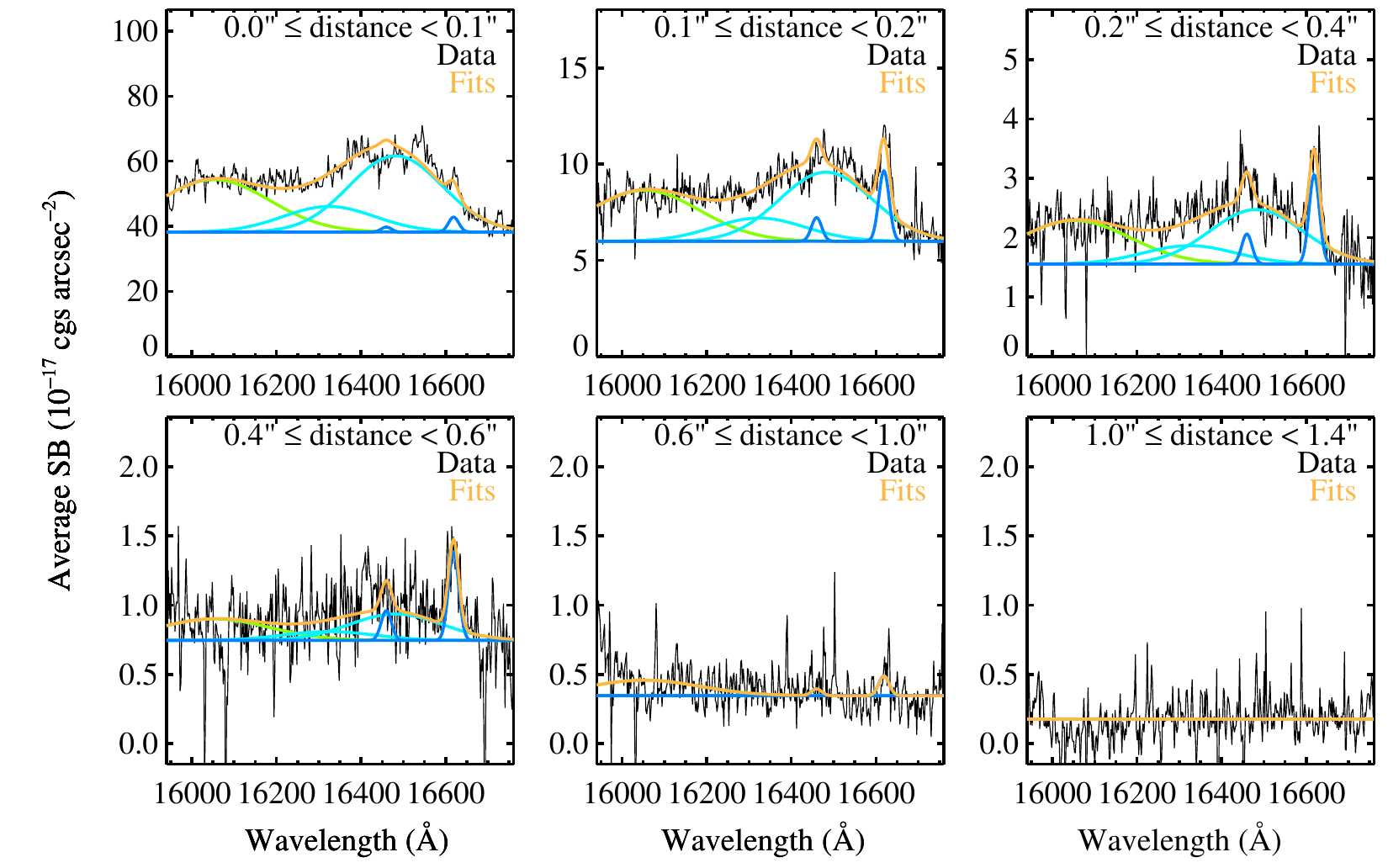}
\hspace{0.2in}
\includegraphics[width=0.30\textwidth]{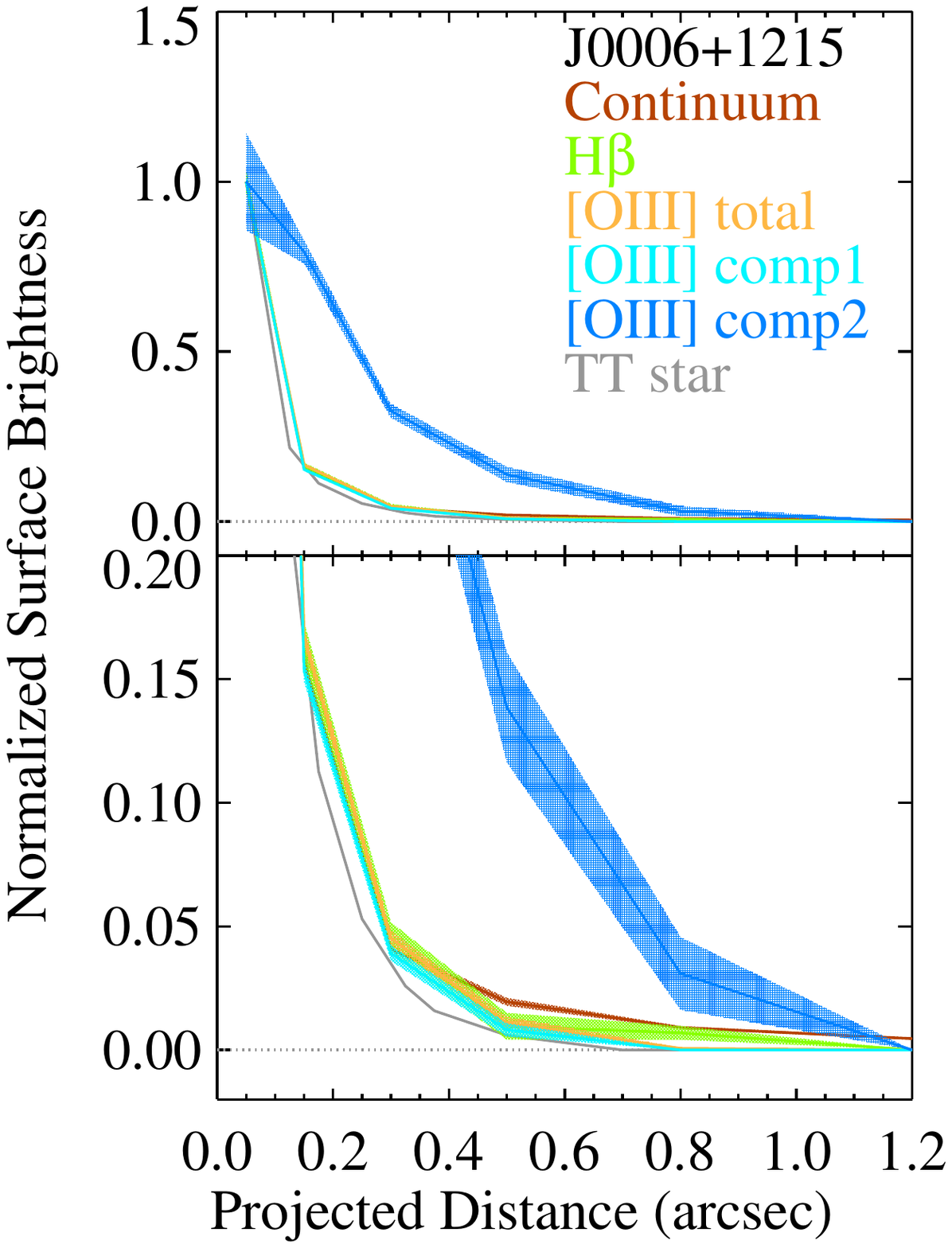}
\includegraphics[width=0.66\textwidth]{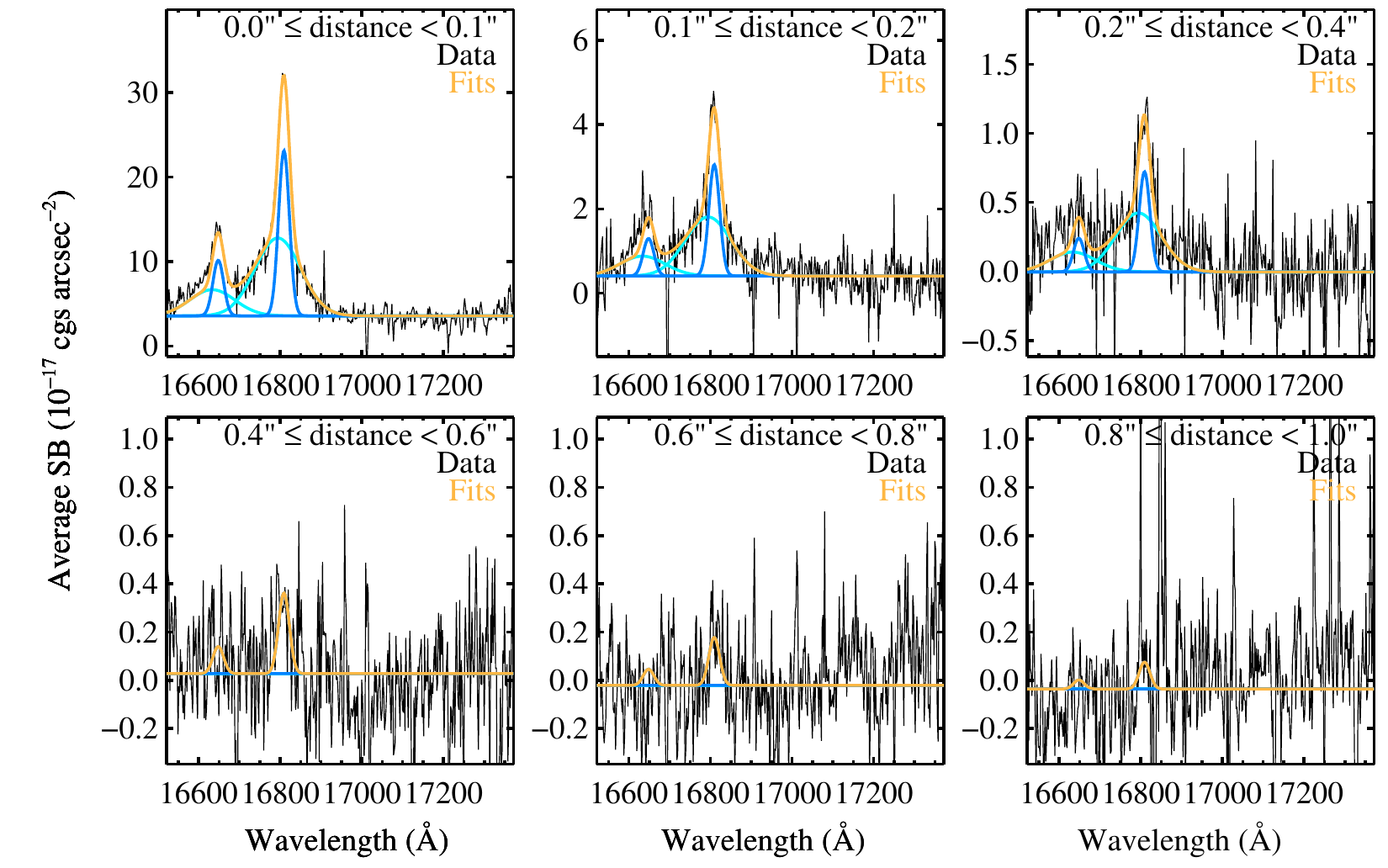}
\hspace{0.2in}
\includegraphics[width=0.30\textwidth]{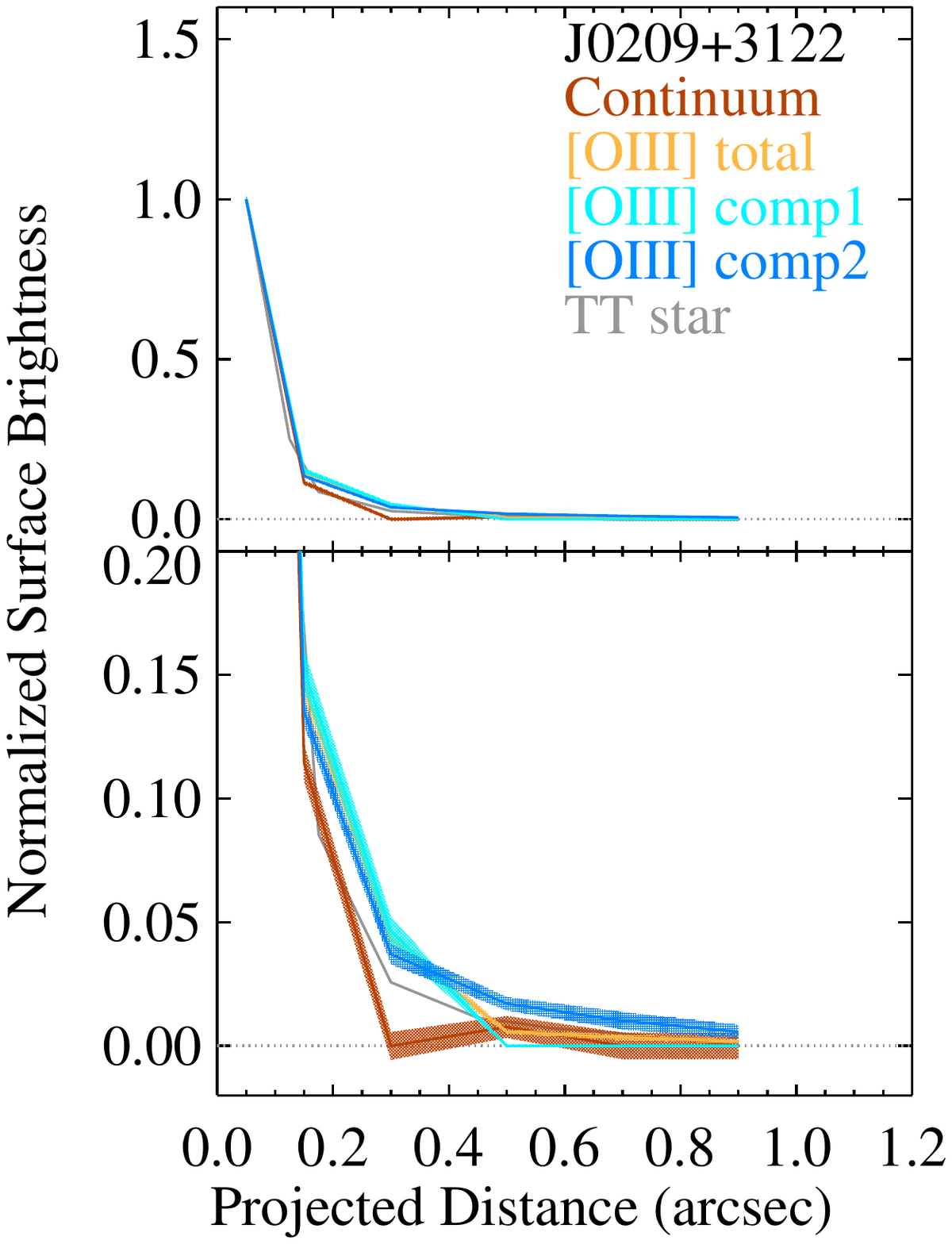}
\includegraphics[width=0.66\textwidth]{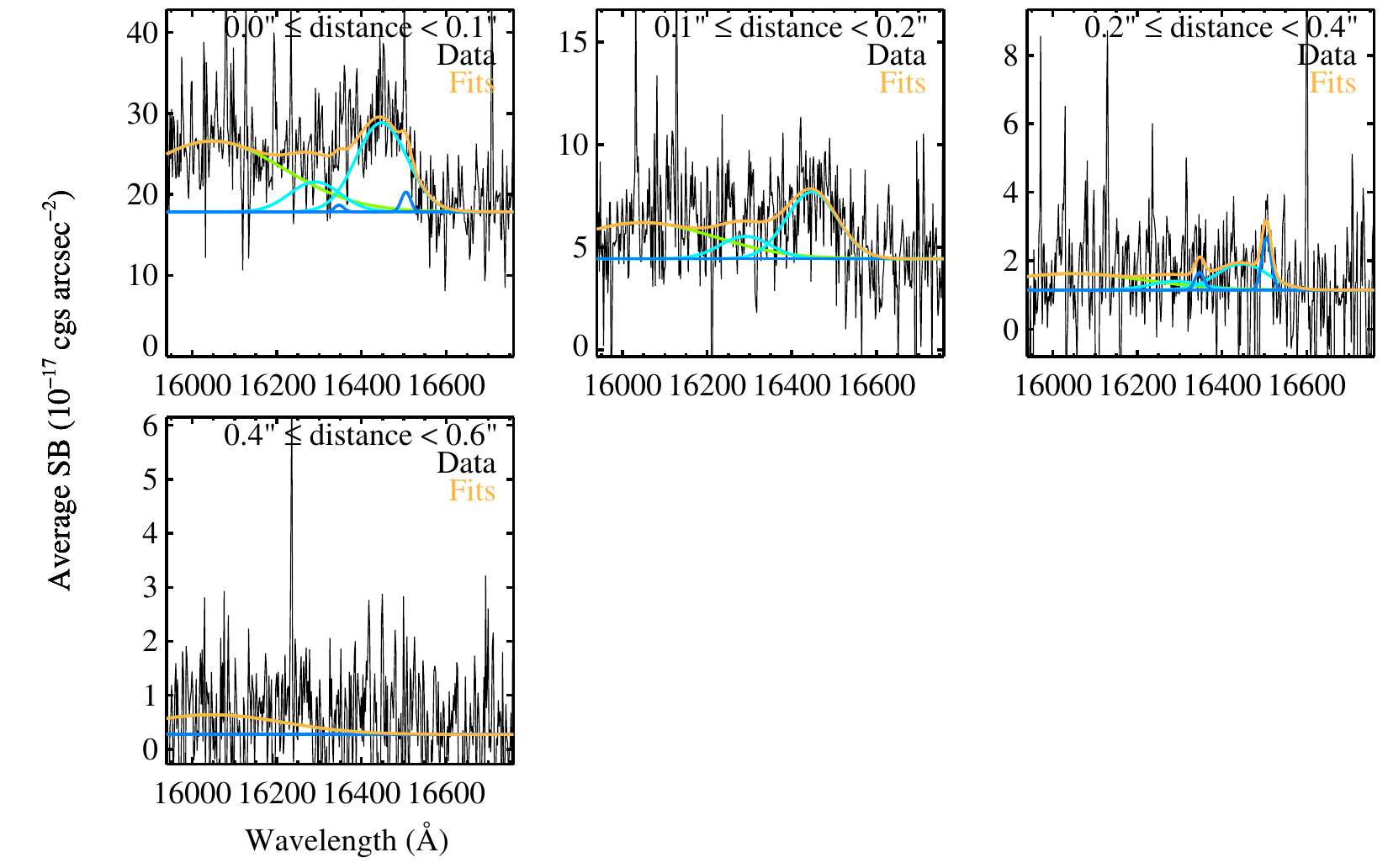}
\hspace{0.2in}
\includegraphics[width=0.30\textwidth]{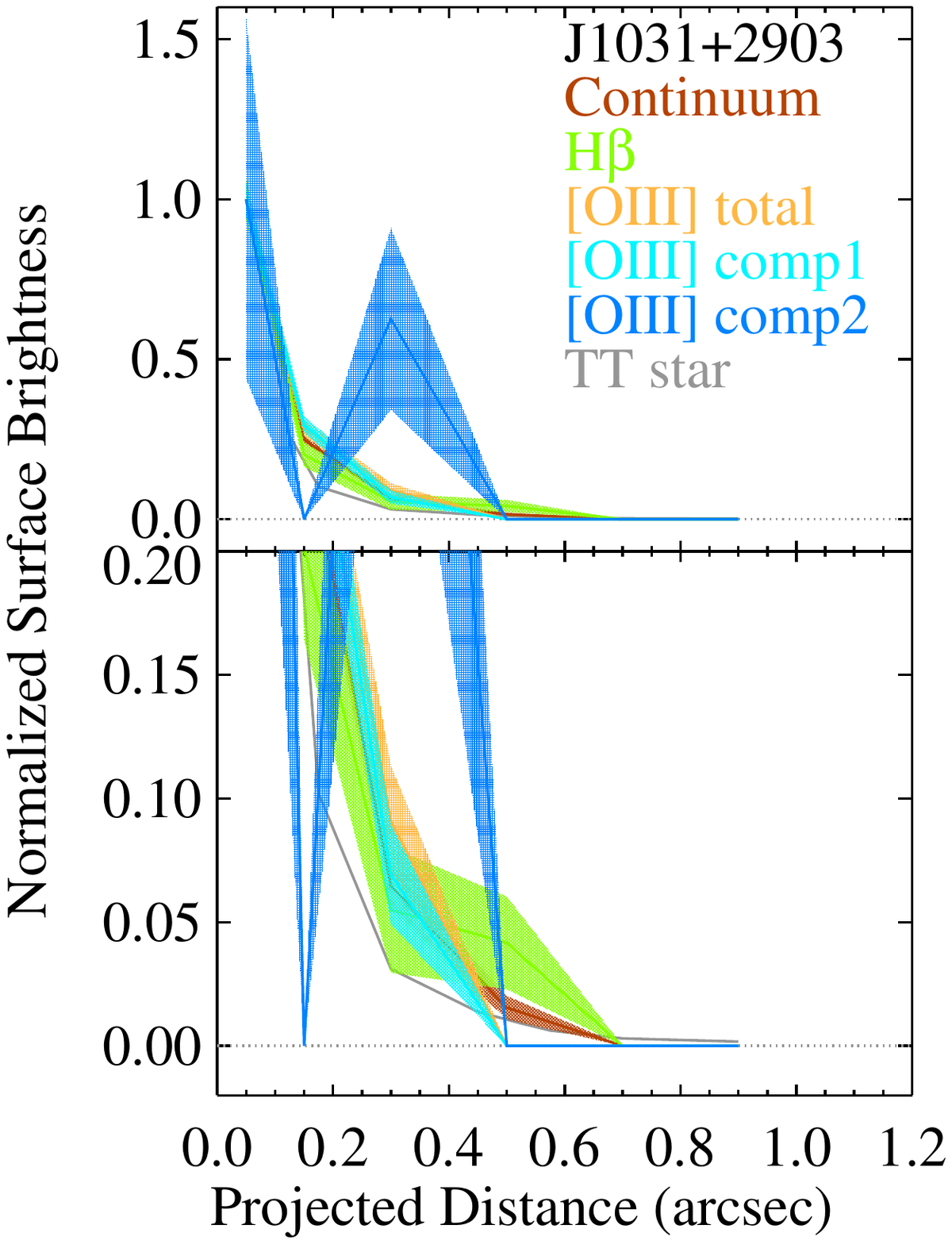}
\caption{The left-hand panels show nuclear and annular spectra generated from OSIRIS and NIFS data 
of ERQs. We overplot the best-fits to the aperture spectra in orange. On each spectrum we overplot 
the H$\beta$ component in lime and the [\ion{O}{III}]\,$\lambda\lambda$4959,5007 components 1, 2, 
and 3 in cyan, blue, and indigo at the continuum level. The right-hand panels show the surface 
brightness radial profiles of the different emission components normalized to the brightest 
aperture, in two different zoom scales. Shown on the panels are the continuum 
emission in brown, H$\beta$ emission in lime, total [\ion{O}{III}] emission in orange, 
[\ion{O}{III}] components 1, 2, and 3 in cyan, blue, and indigo, and the tip/tilt star in 
grey. The shaded regions are the 1-$\sigma$ uncertainties in the surface brightnesses.}
\label{fig:annuprofspecs}
\end{figure*}
\begin{figure*}
\includegraphics[width=0.66\textwidth]{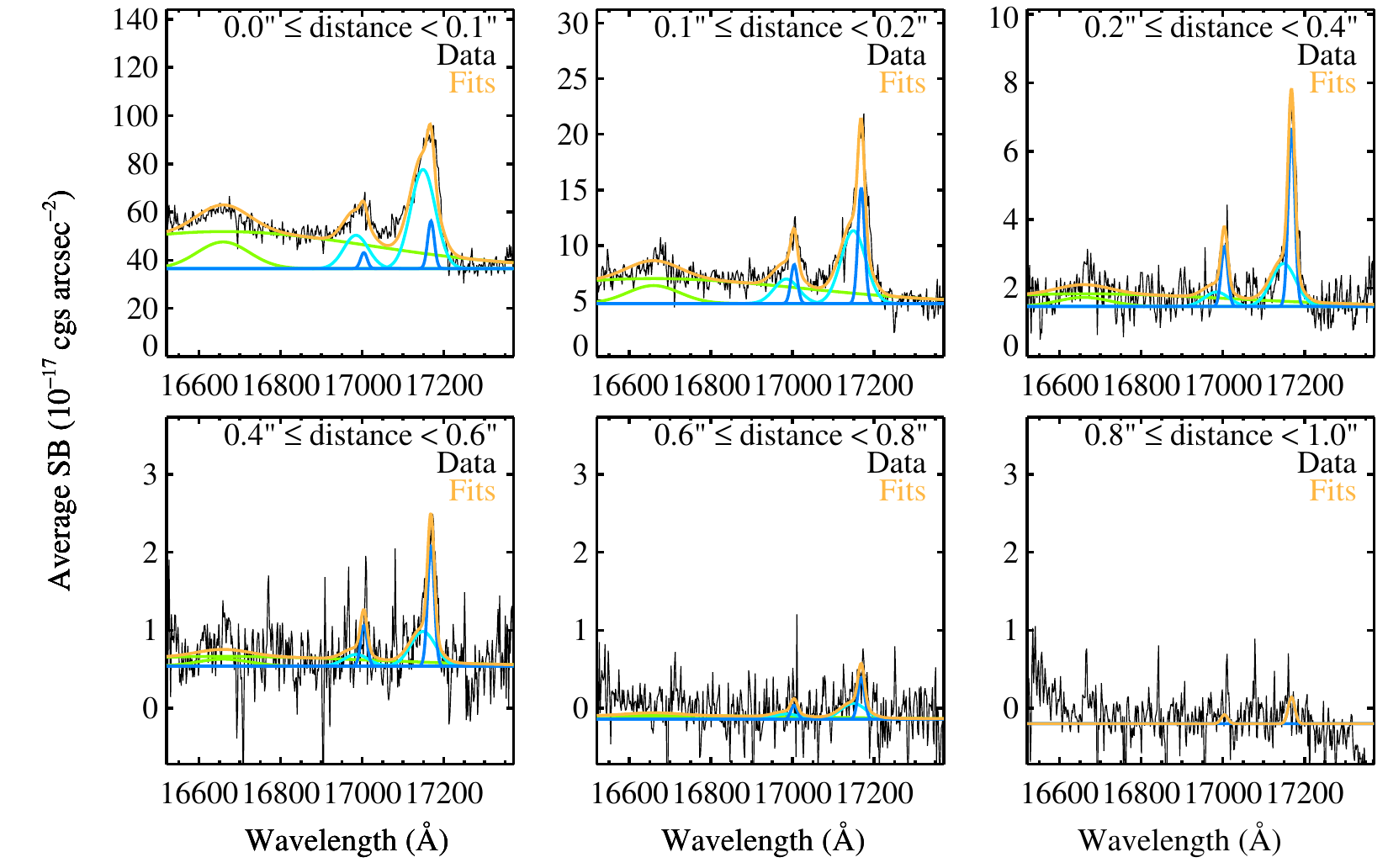}
\hspace{0.2in}
\includegraphics[width=0.30\textwidth]{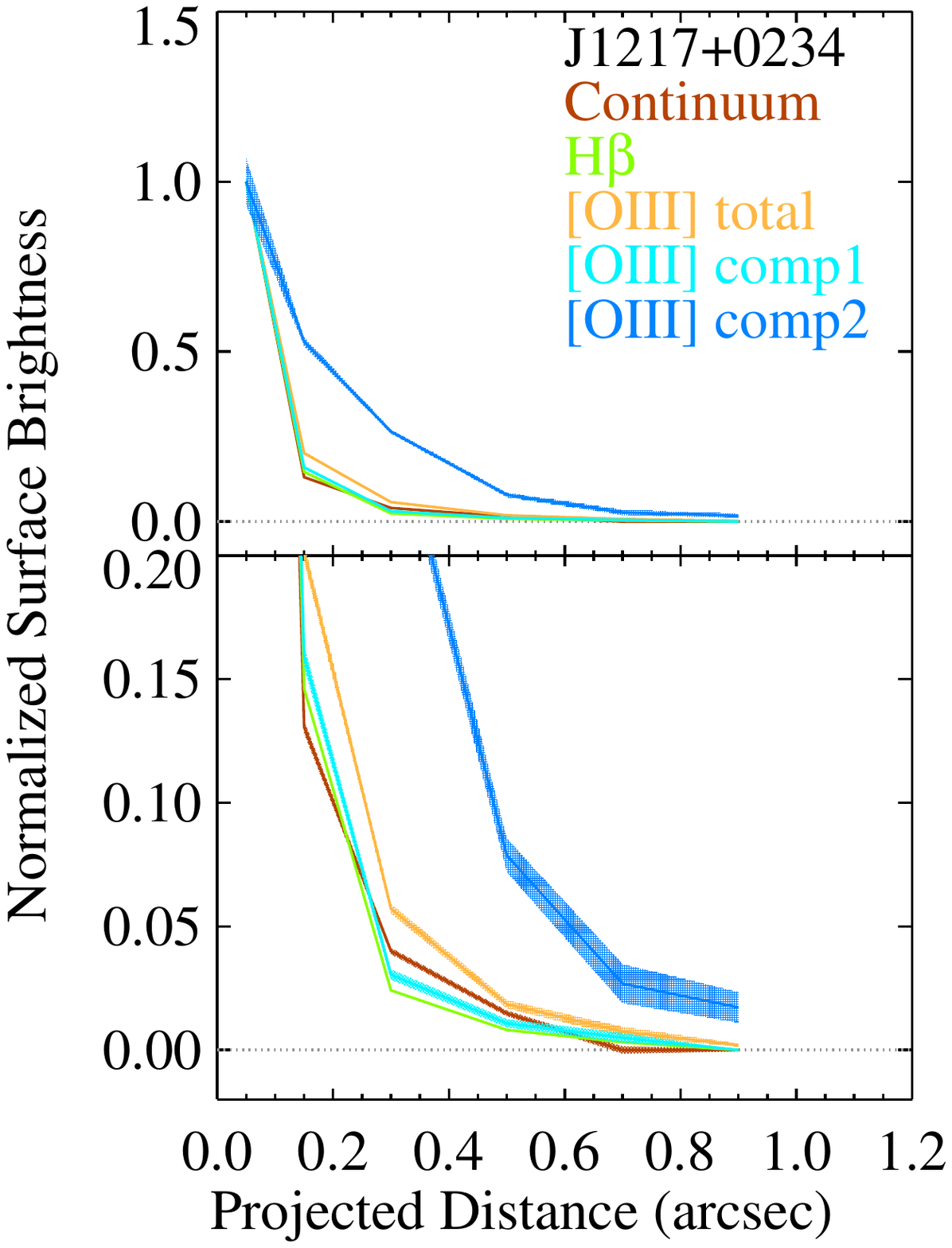}
\includegraphics[width=0.66\textwidth]{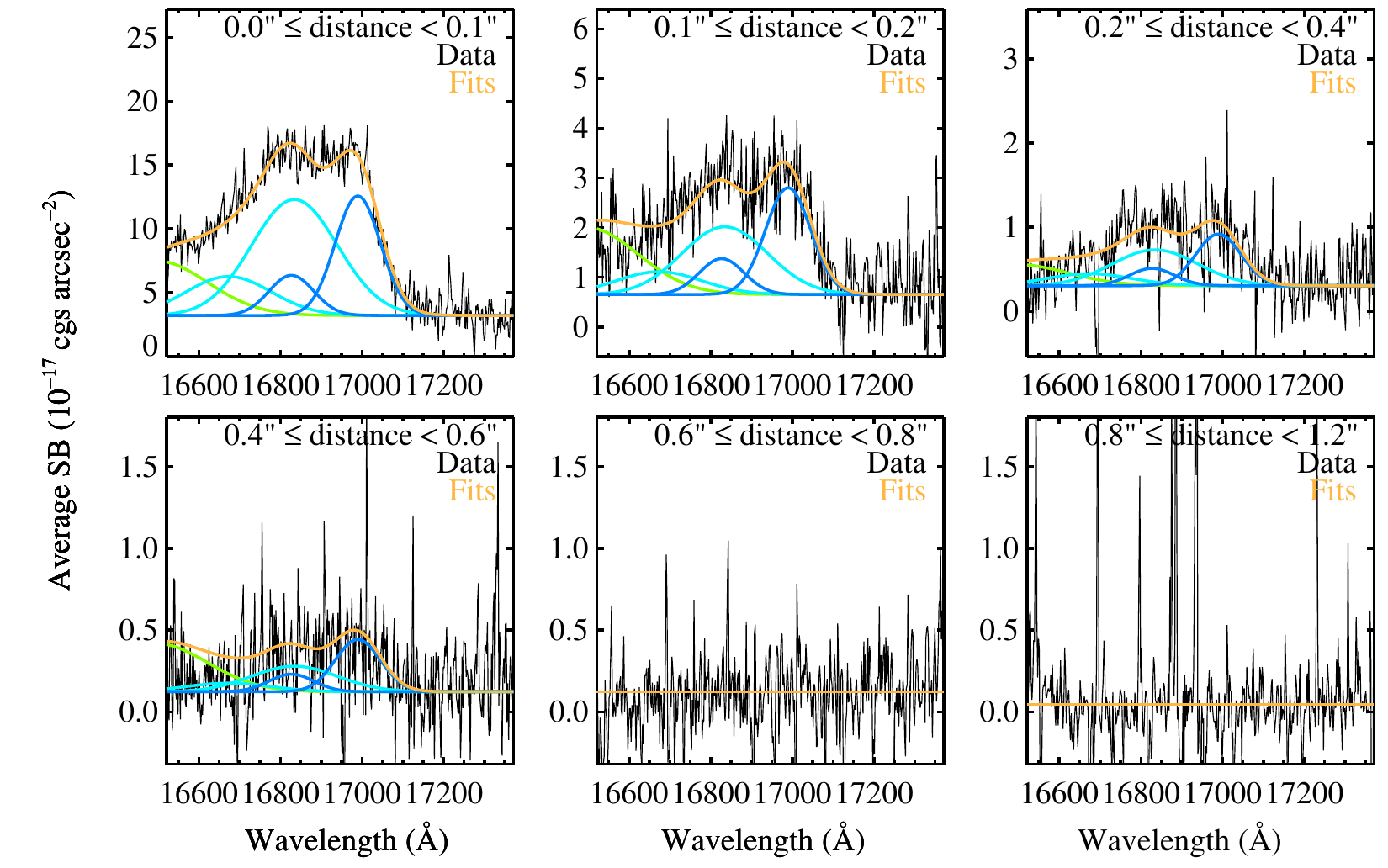}
\hspace{0.2in}
\includegraphics[width=0.30\textwidth]{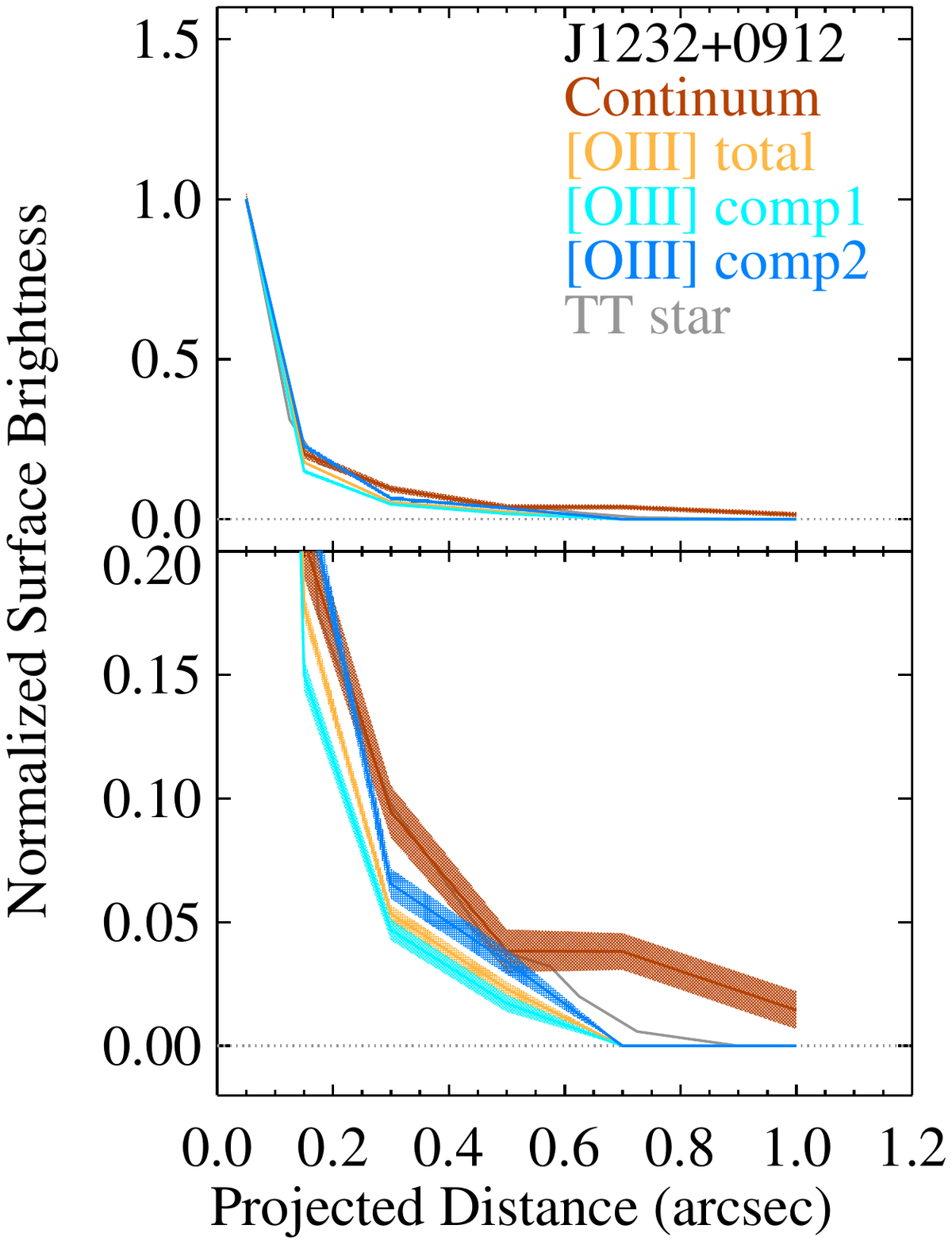}
\includegraphics[width=0.66\textwidth]{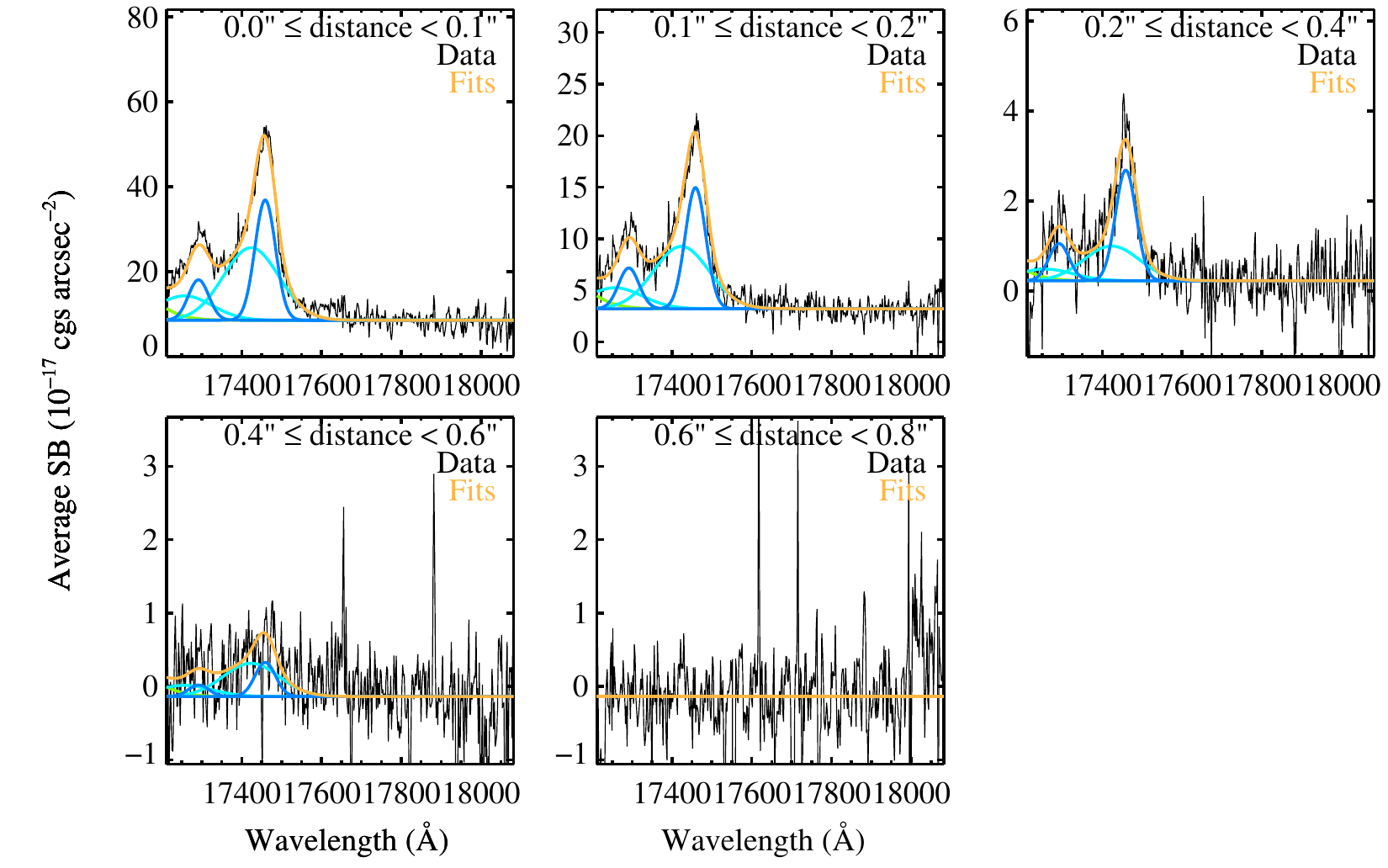}
\hspace{0.2in}
\includegraphics[width=0.30\textwidth]{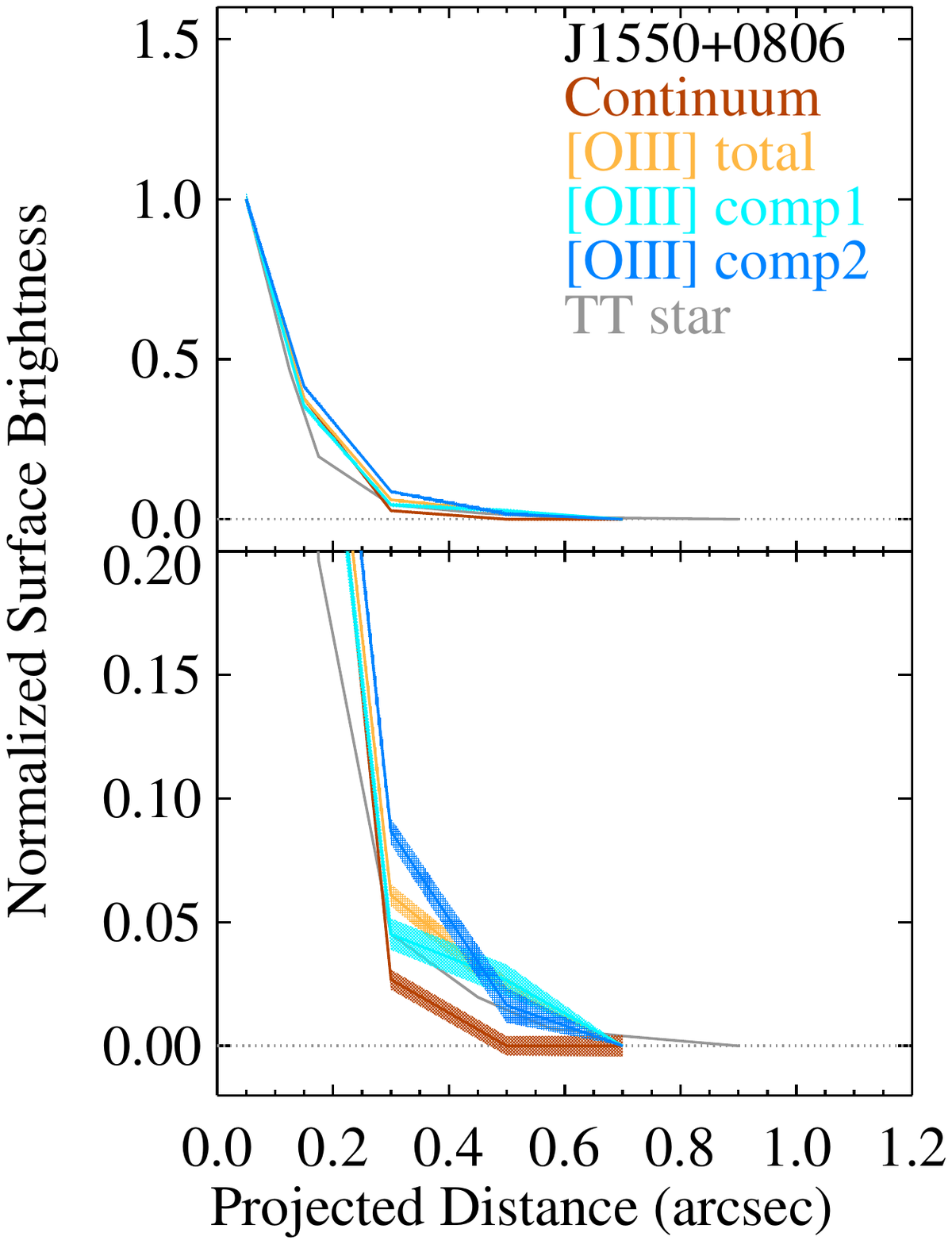}
\contcaption{}
\end{figure*}
\begin{figure*}
\includegraphics[width=0.66\textwidth]{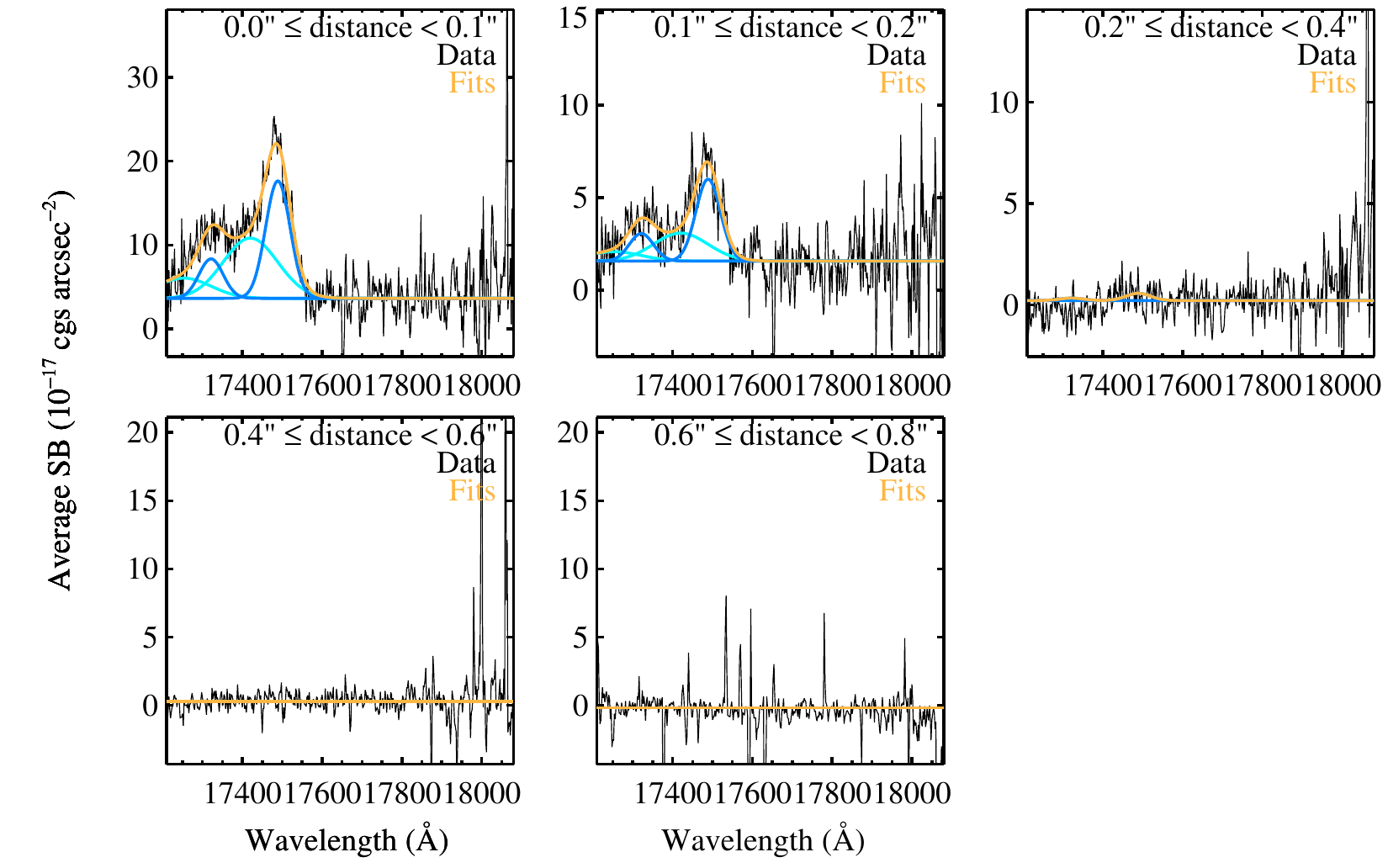}
\hspace{0.2in}
\includegraphics[width=0.30\textwidth]{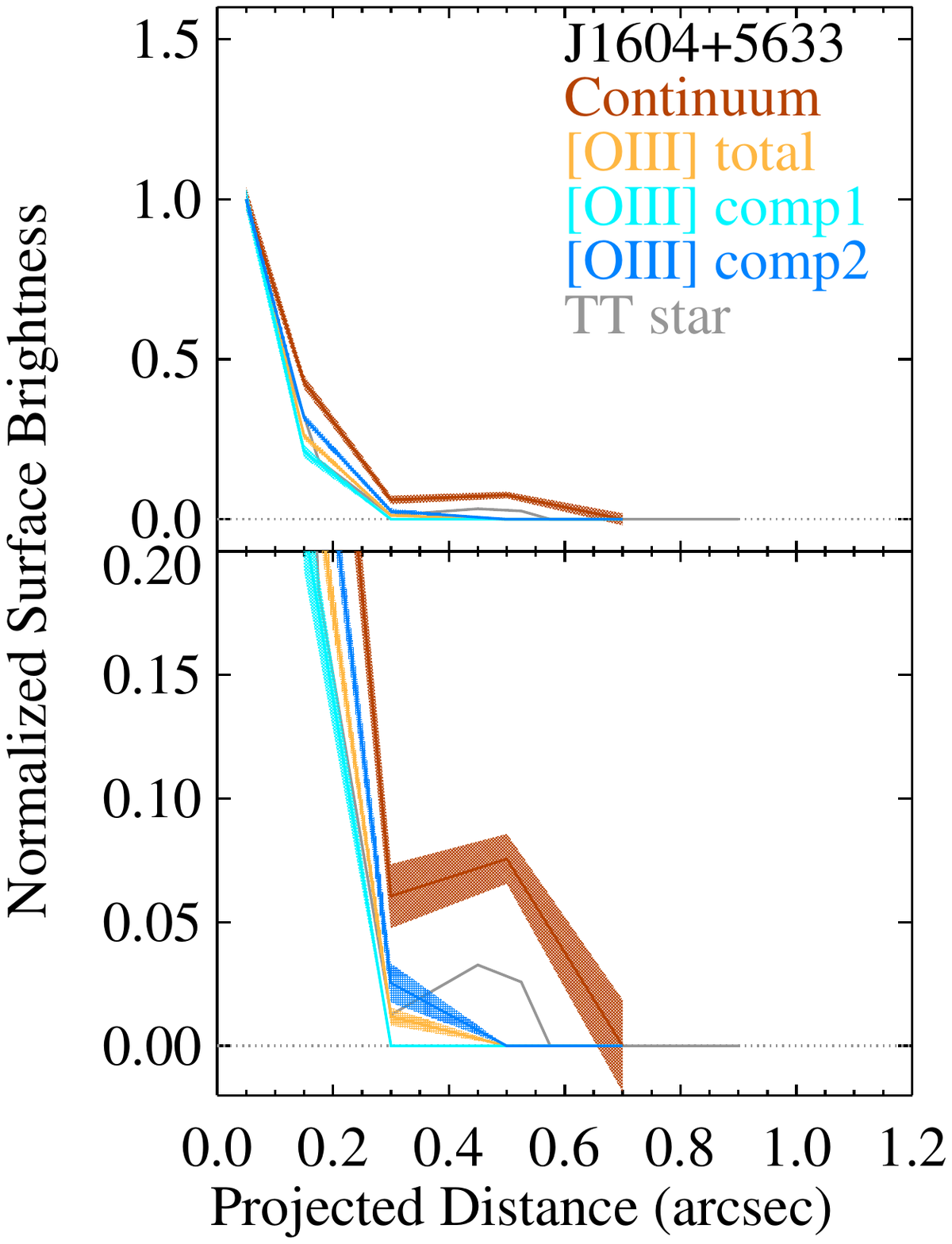}
\includegraphics[width=0.66\textwidth]{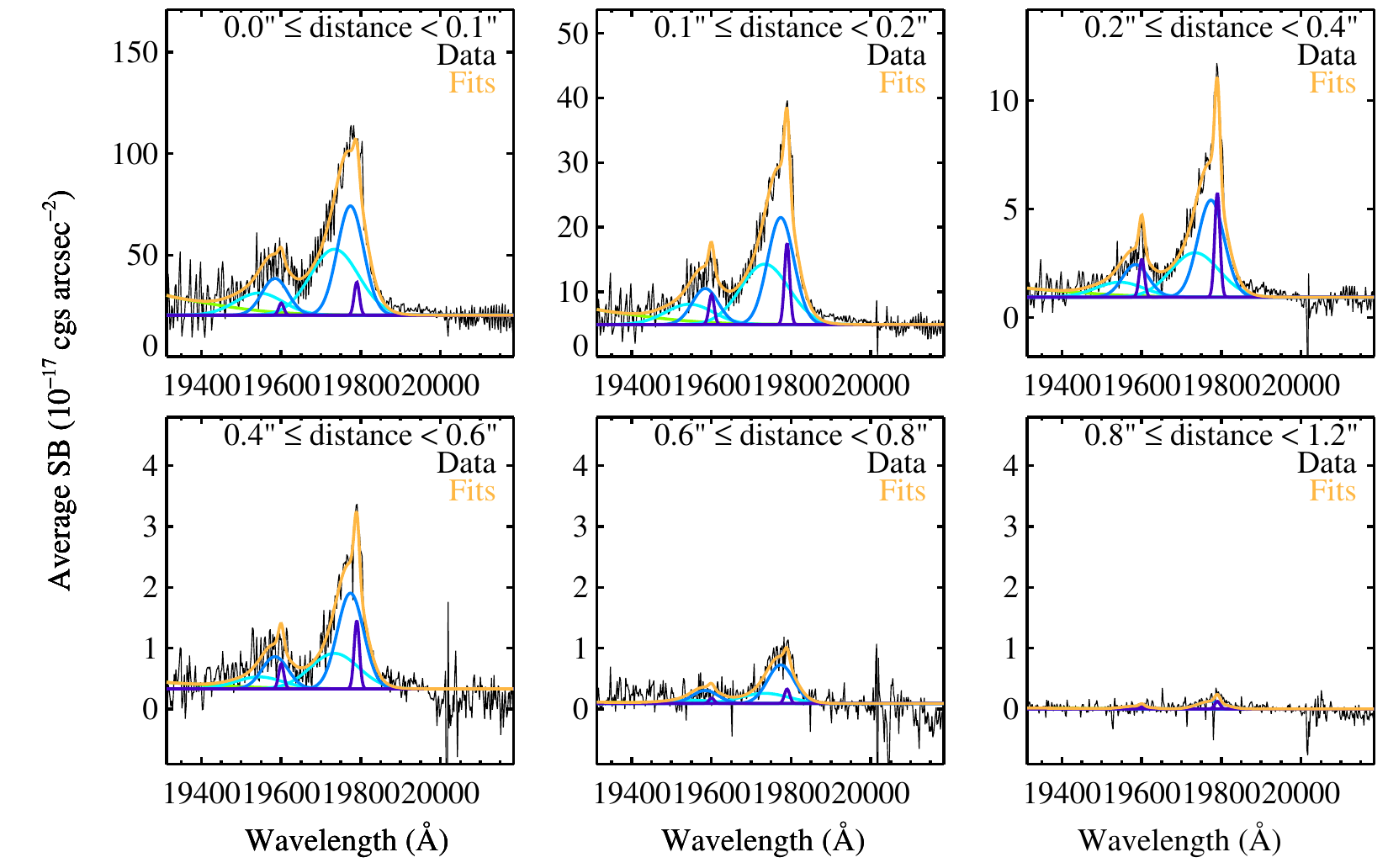}
\hspace{0.2in}
\includegraphics[width=0.30\textwidth]{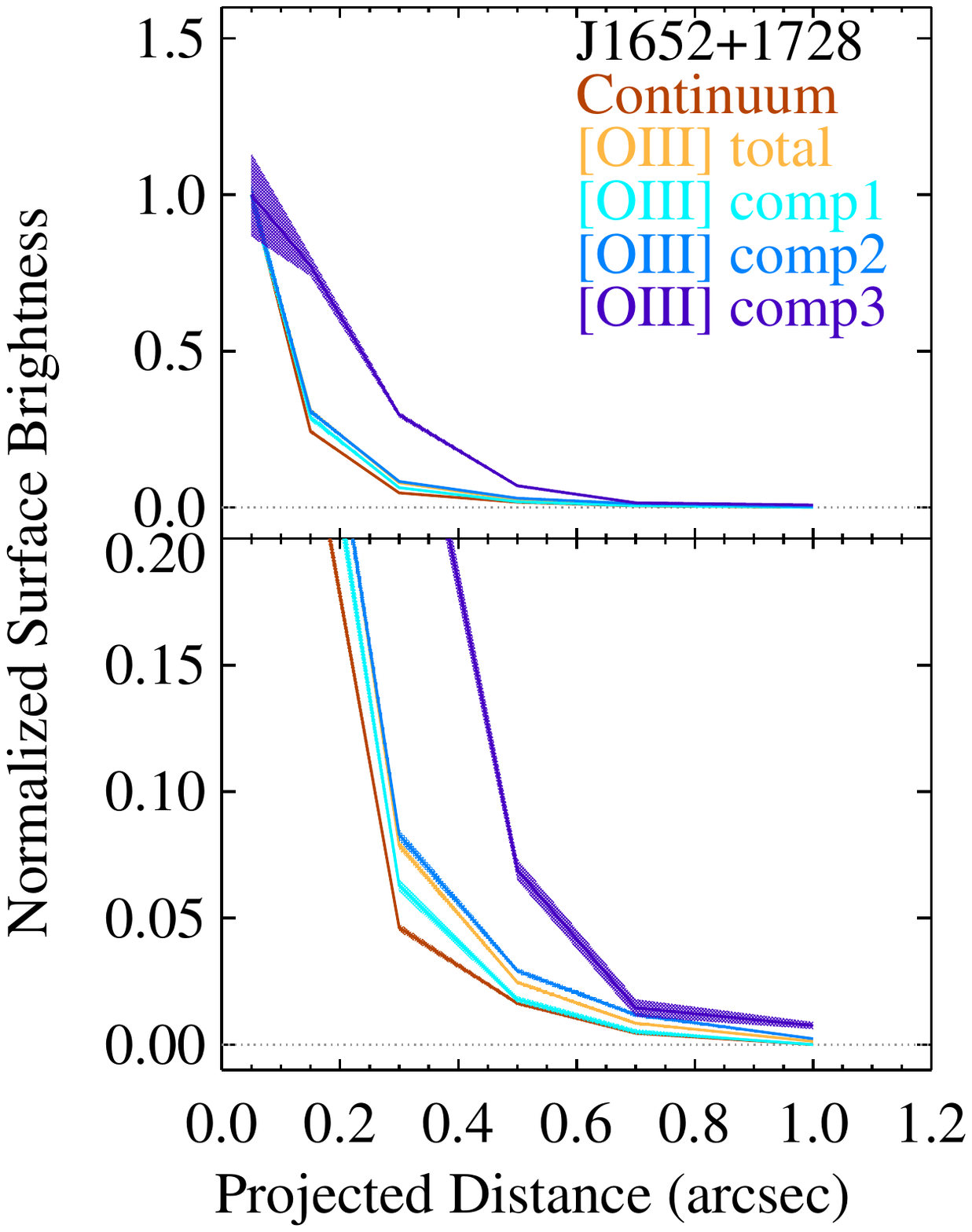}
\includegraphics[width=0.66\textwidth]{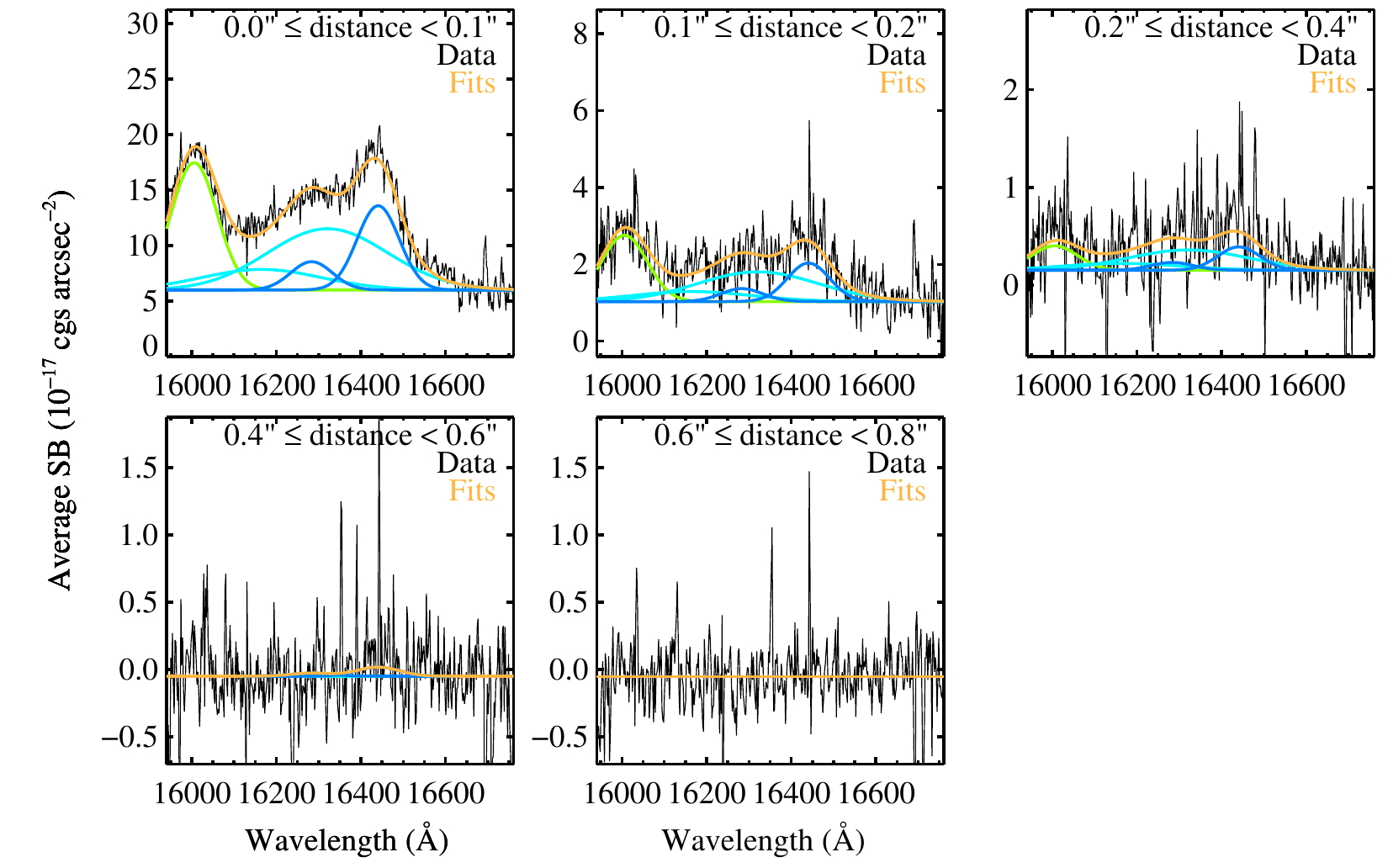}
\hspace{0.2in}
\includegraphics[width=0.30\textwidth]{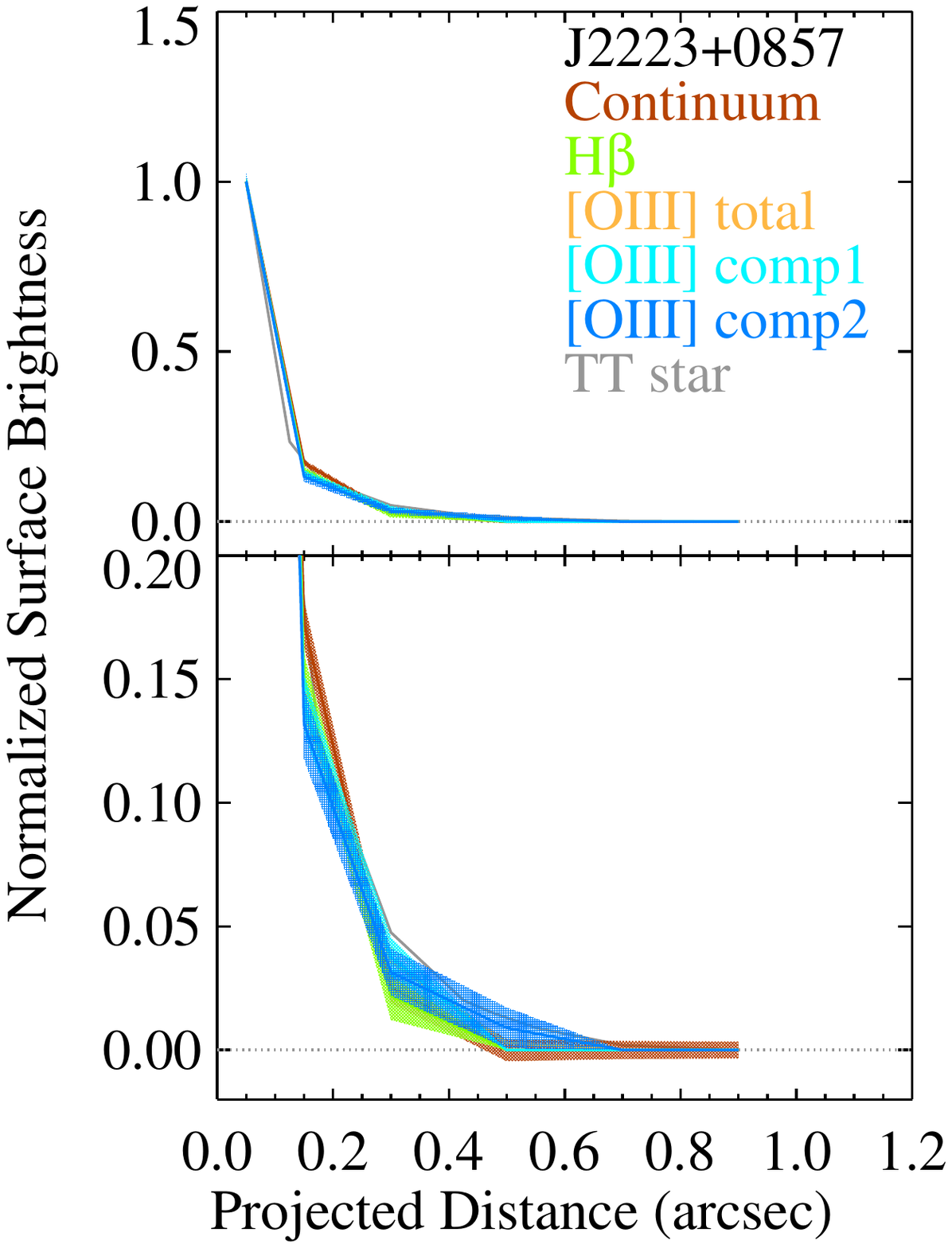}
\contcaption{}
\end{figure*}
\begin{figure*}
\includegraphics[width=0.66\textwidth]{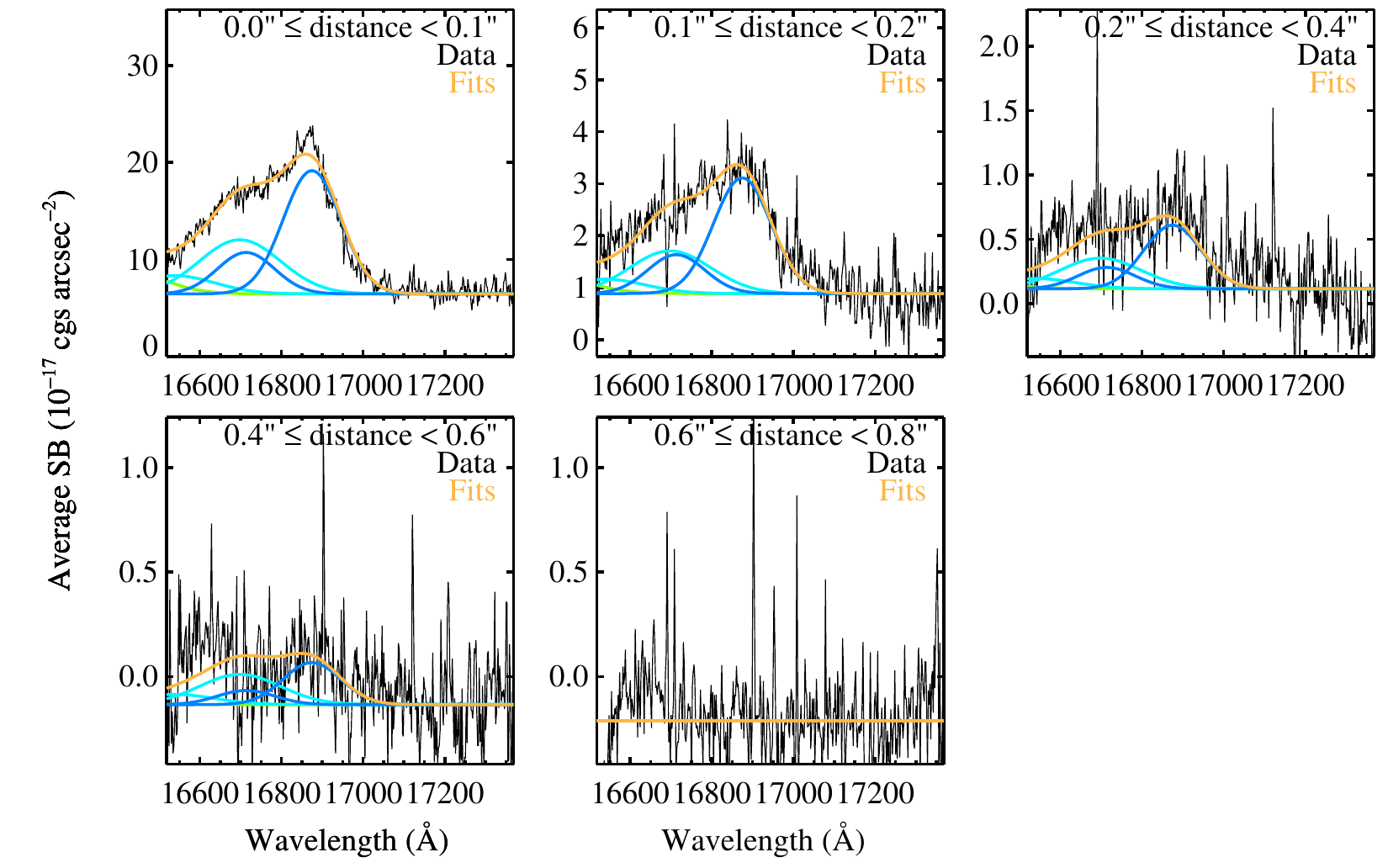}
\hspace{0.2in}
\includegraphics[width=0.30\textwidth]{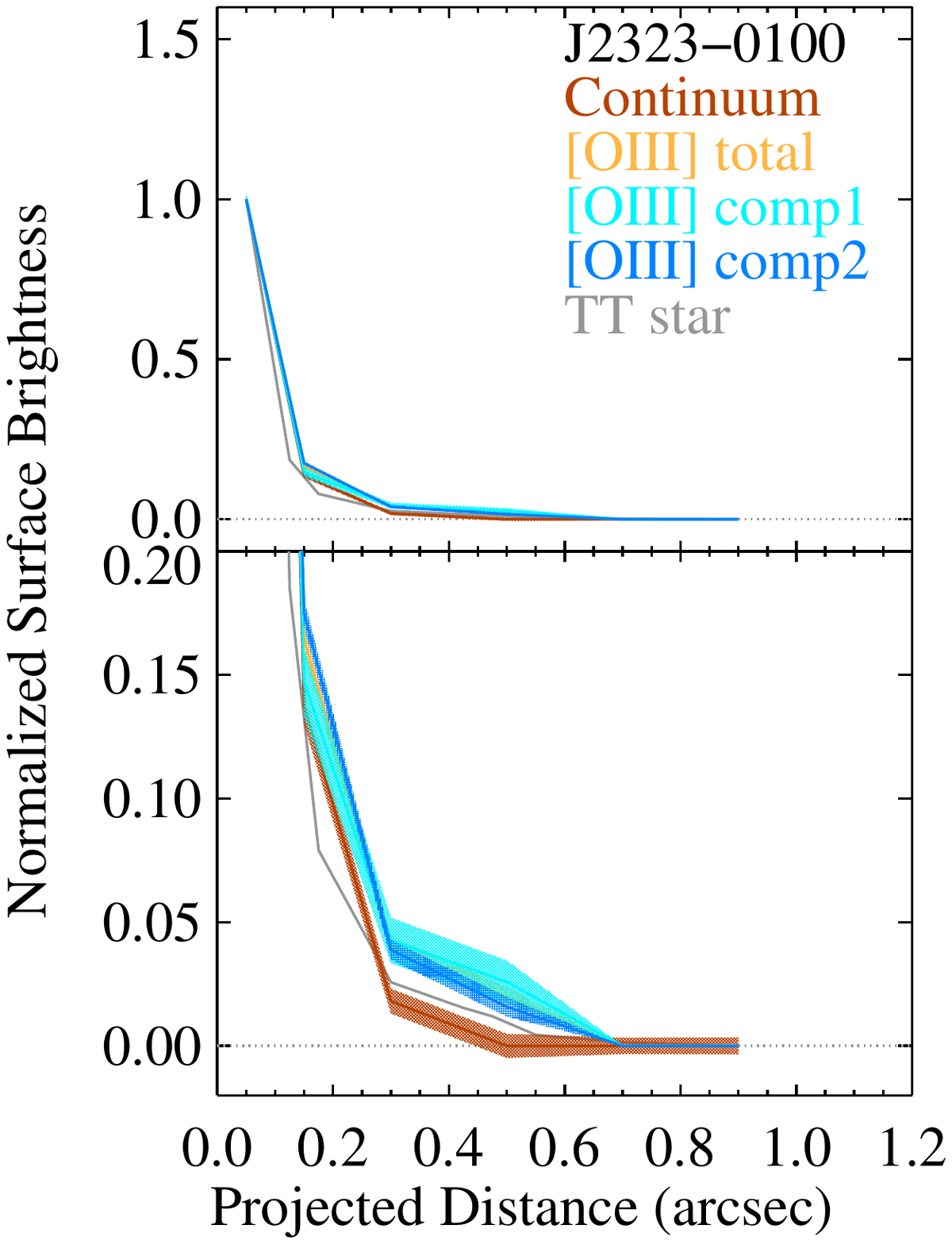}
\contcaption{}
\end{figure*}

\begin{figure*}
\includegraphics[width=0.33\textwidth]{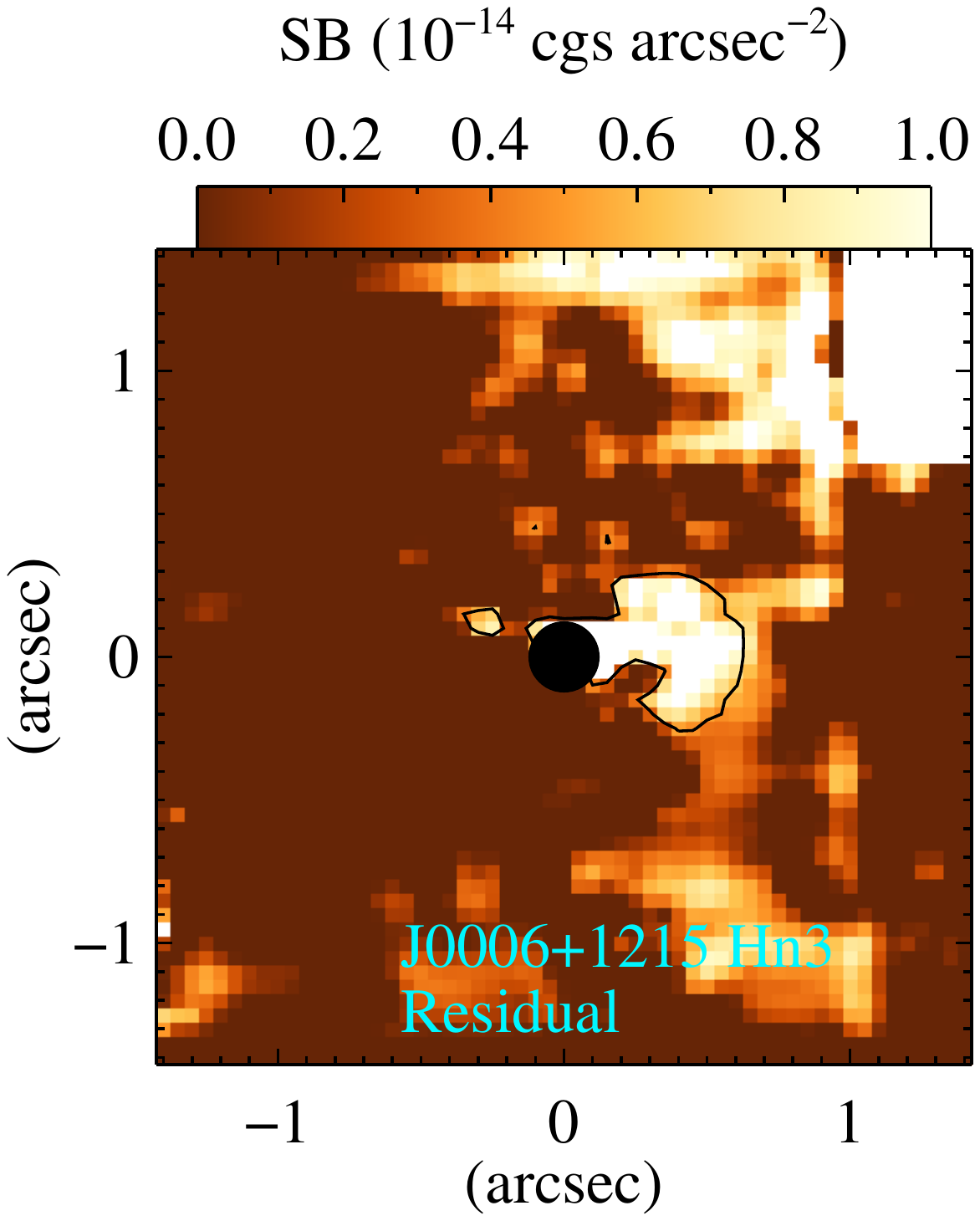}
\includegraphics[width=0.66\textwidth]{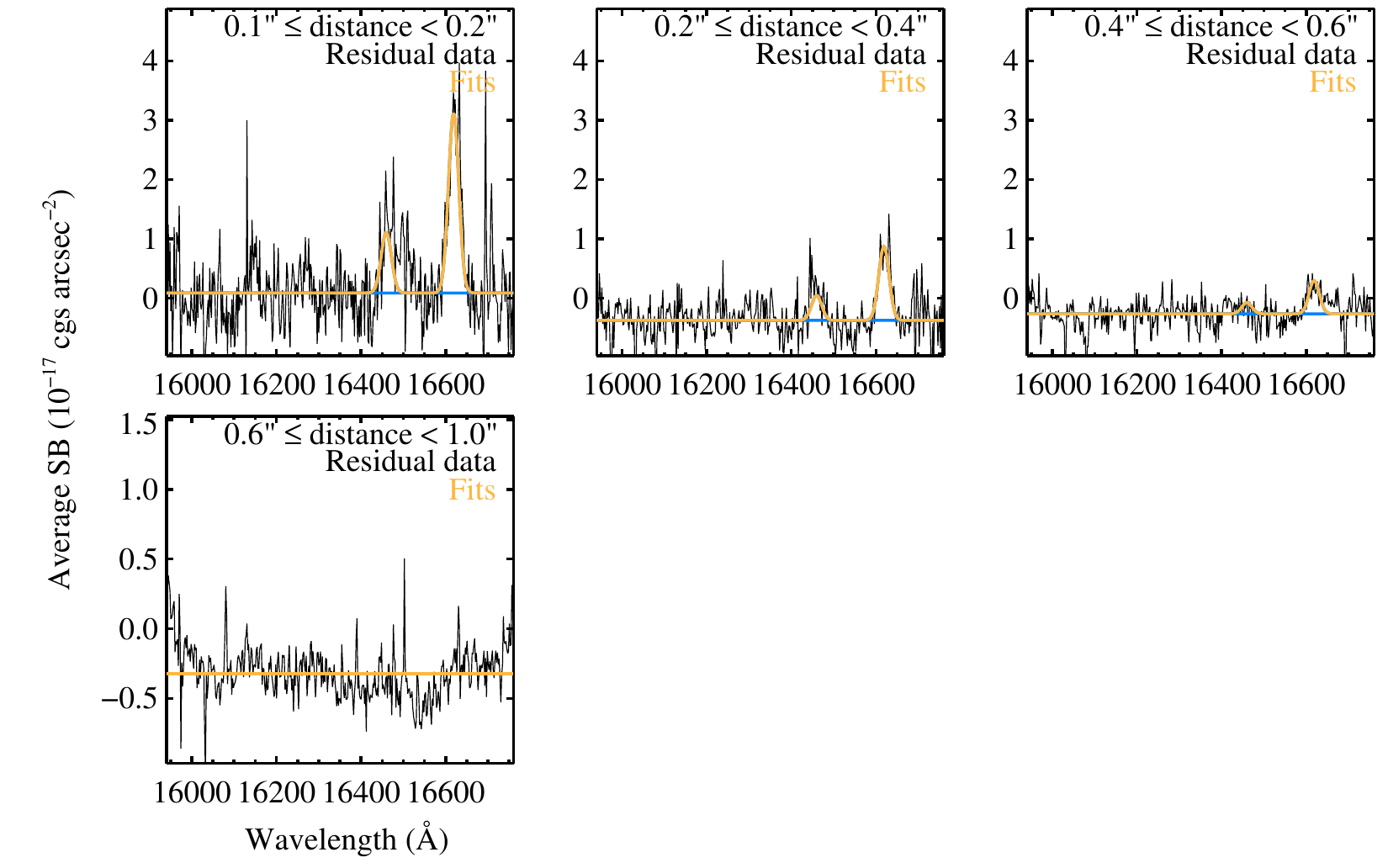}
\includegraphics[width=0.33\textwidth]{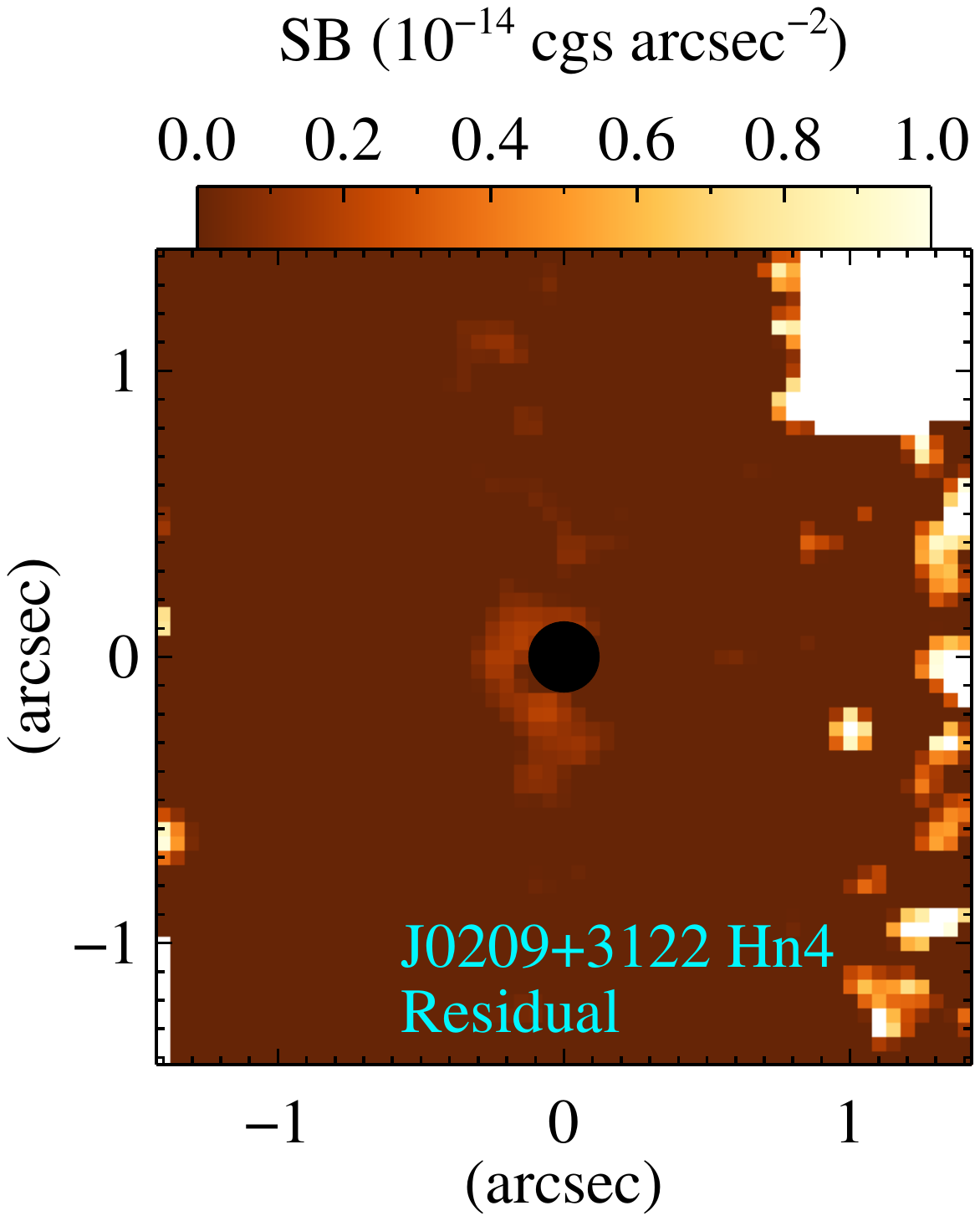}
\includegraphics[width=0.66\textwidth]{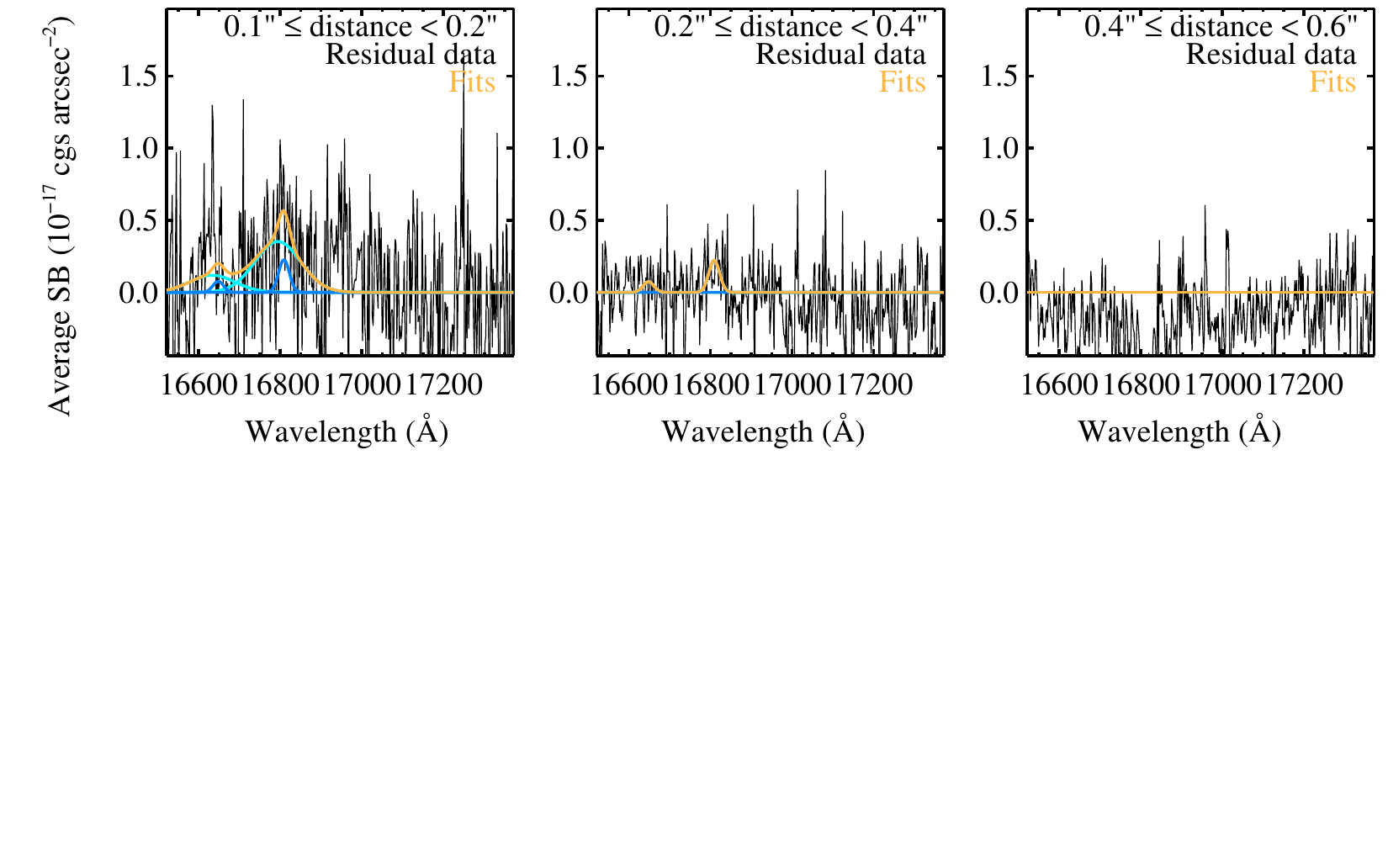}
\includegraphics[width=0.33\textwidth]{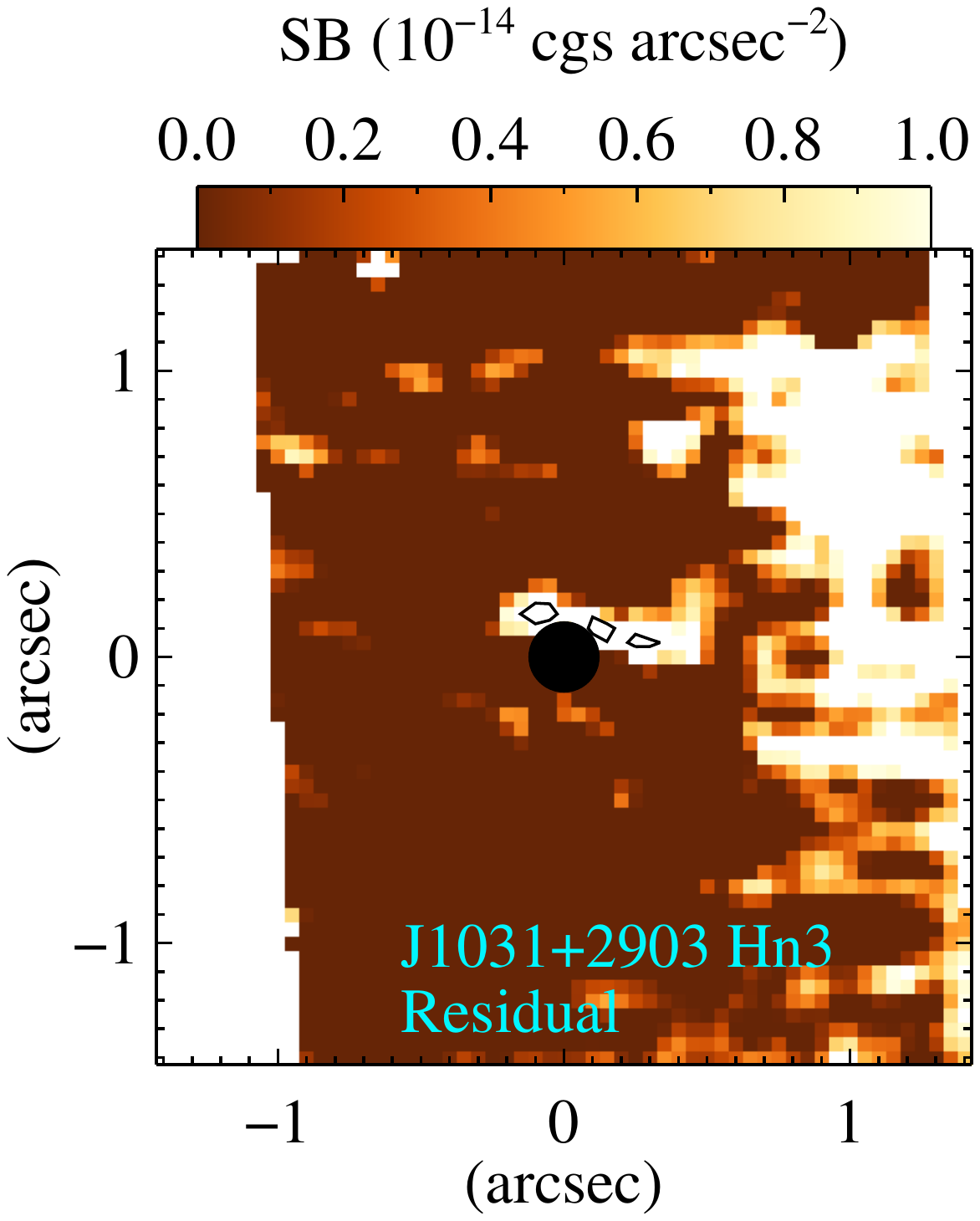}
\includegraphics[width=0.66\textwidth]{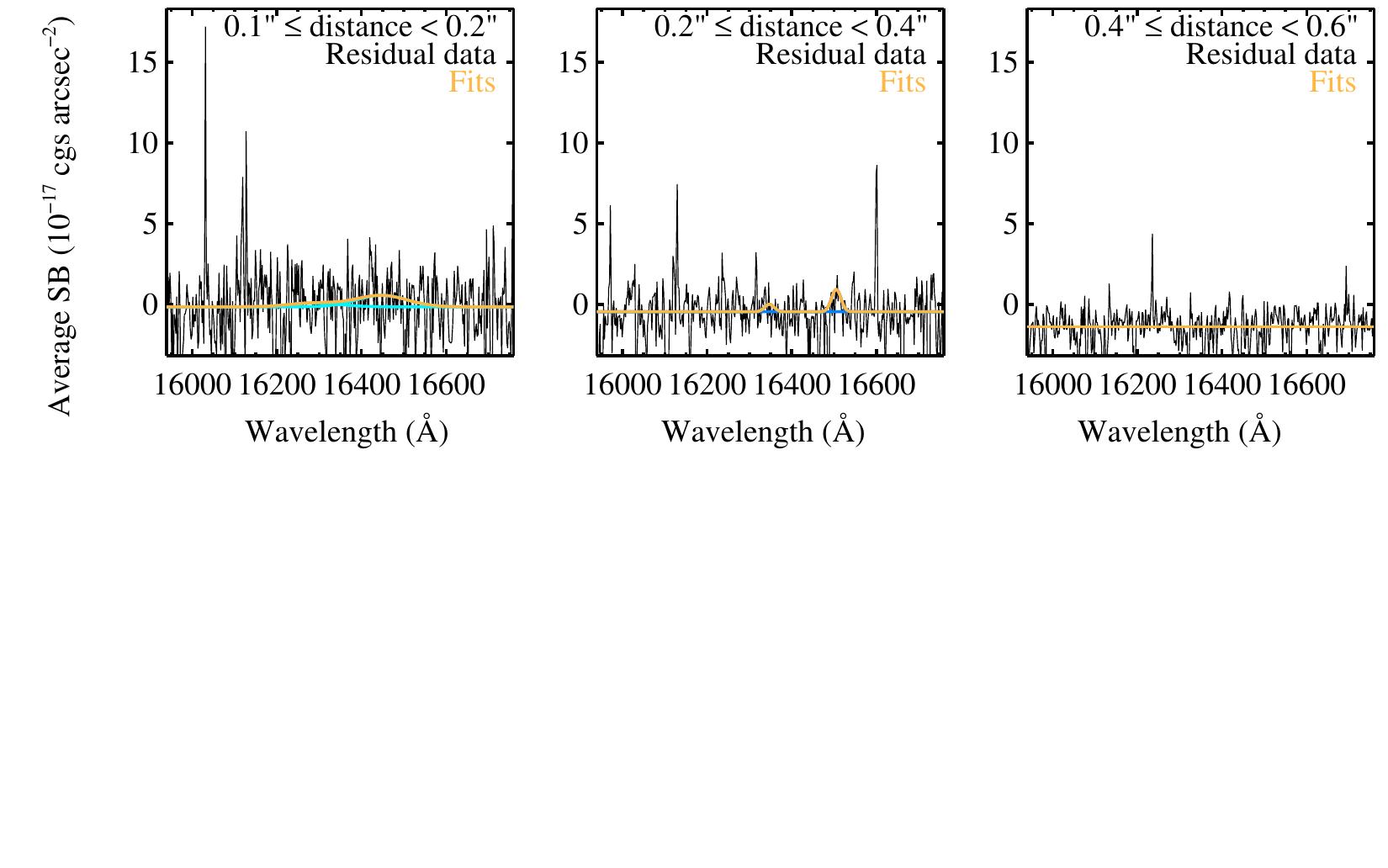}
\caption{The left-hand panels show PSF-subtracted residual surface brightness maps summing all 
channels and the right-hand panels show the nuclear and annular aperture spectra generated from 
the PSF-subtracted residuals of the data cubes. We overplot the total best-fits to the aperture 
spectra in orange. On each spectrum we overplot the [\ion{O}{III}] component 1 in cyan and the 
[\ion{O}{III}] component 2 in blue at the continuum emission level.}
\label{fig:psfres}
\end{figure*}
\begin{figure*}
\includegraphics[width=0.33\textwidth]{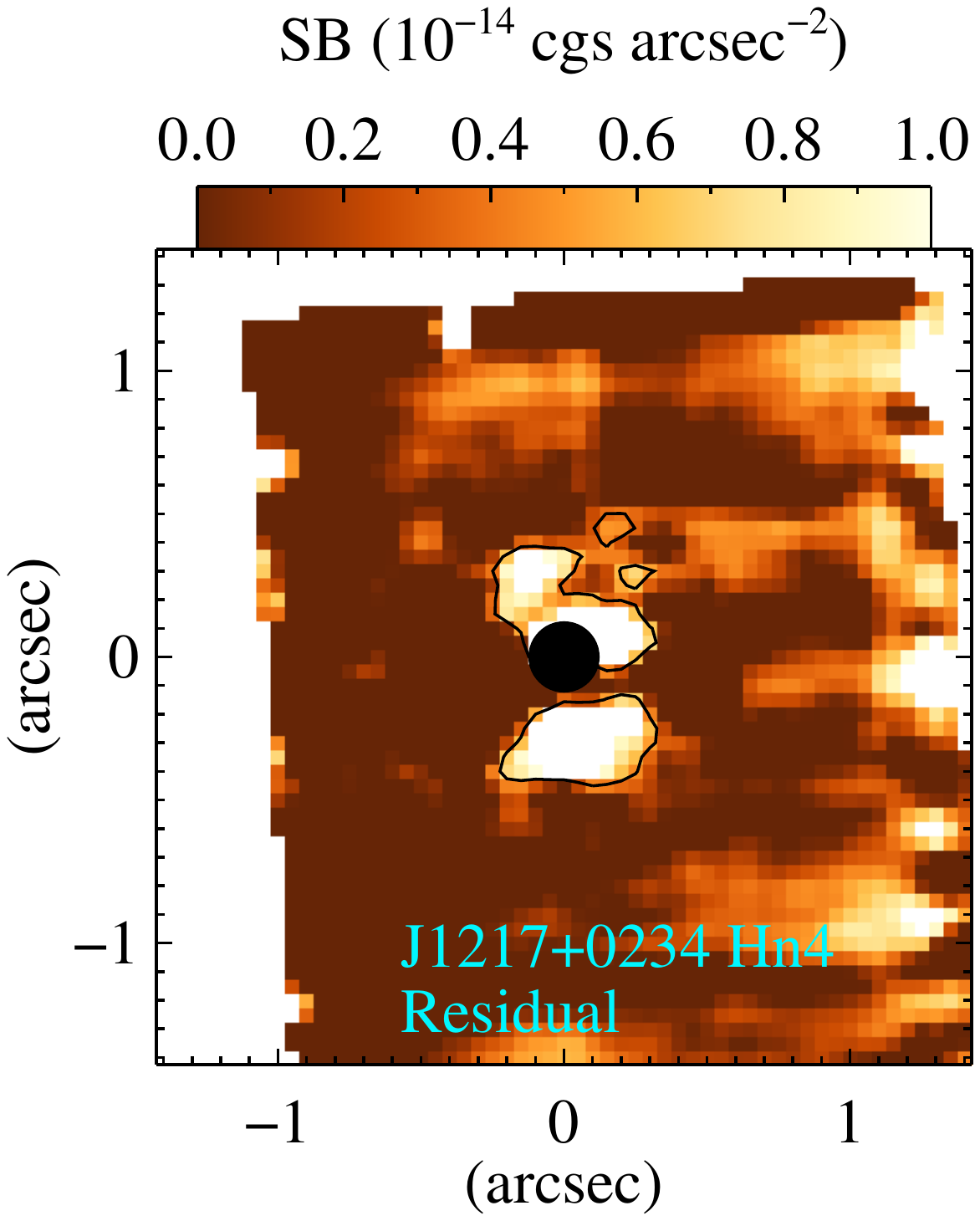}
\includegraphics[width=0.66\textwidth]{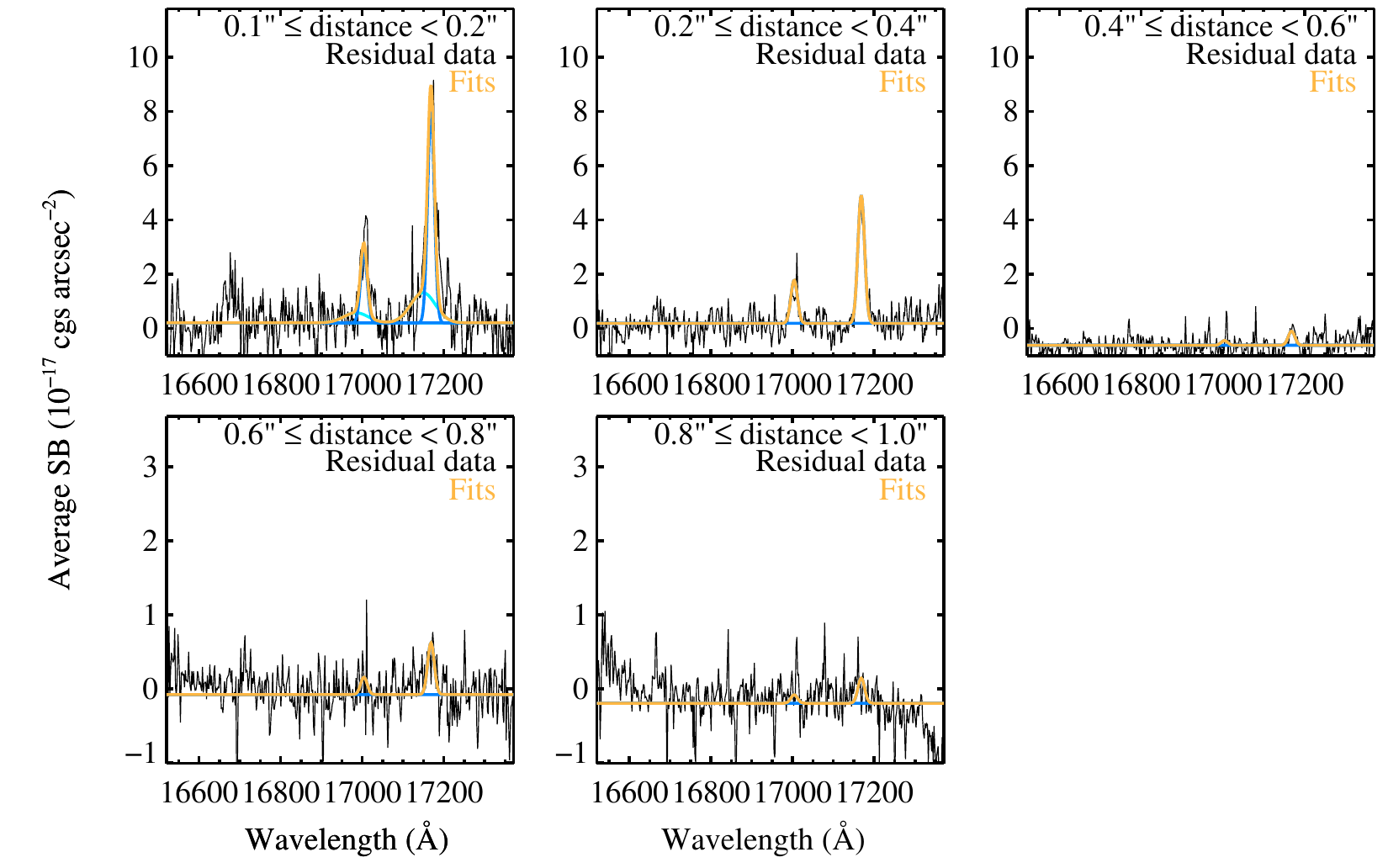}
\includegraphics[width=0.33\textwidth]{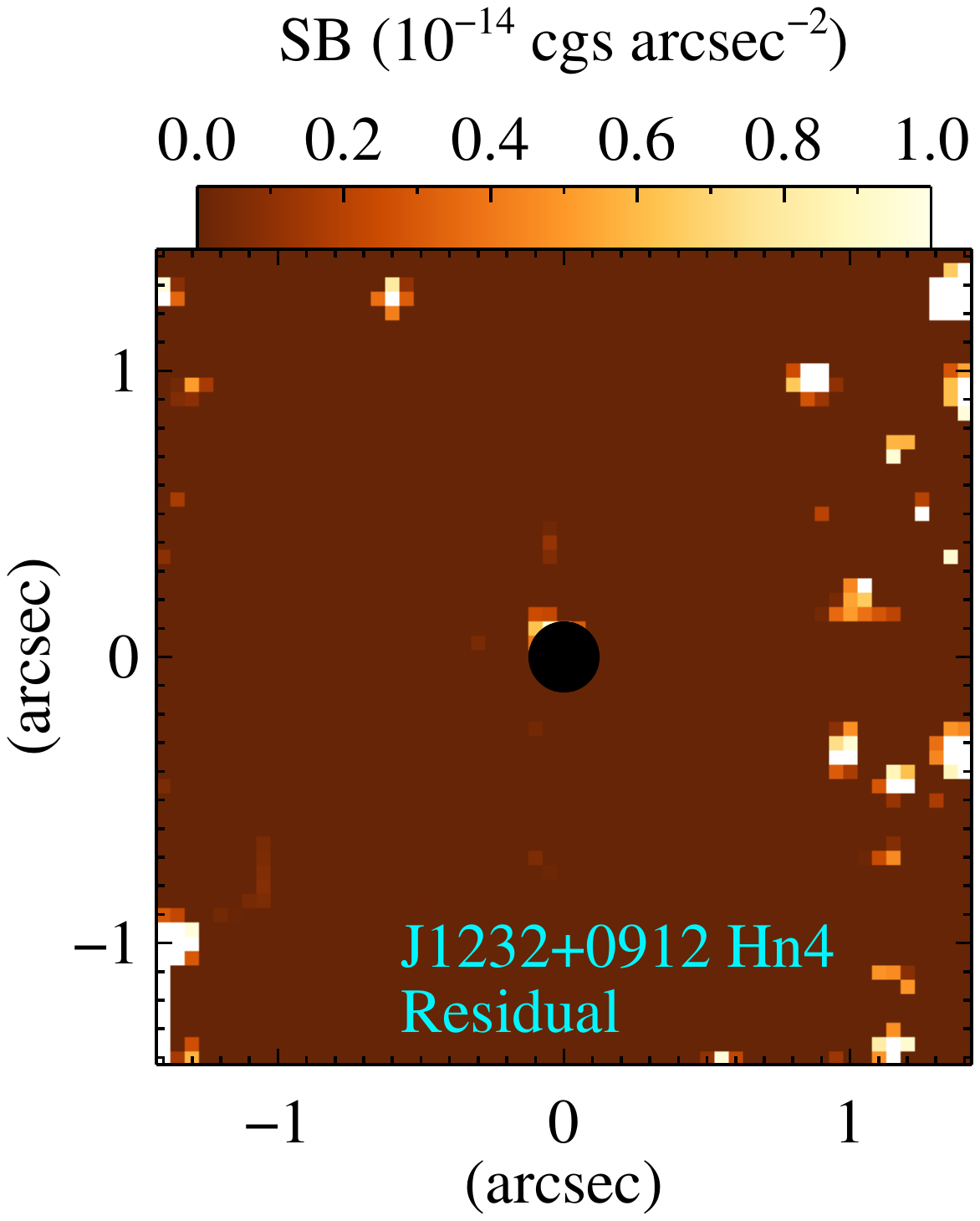}
\includegraphics[width=0.66\textwidth]{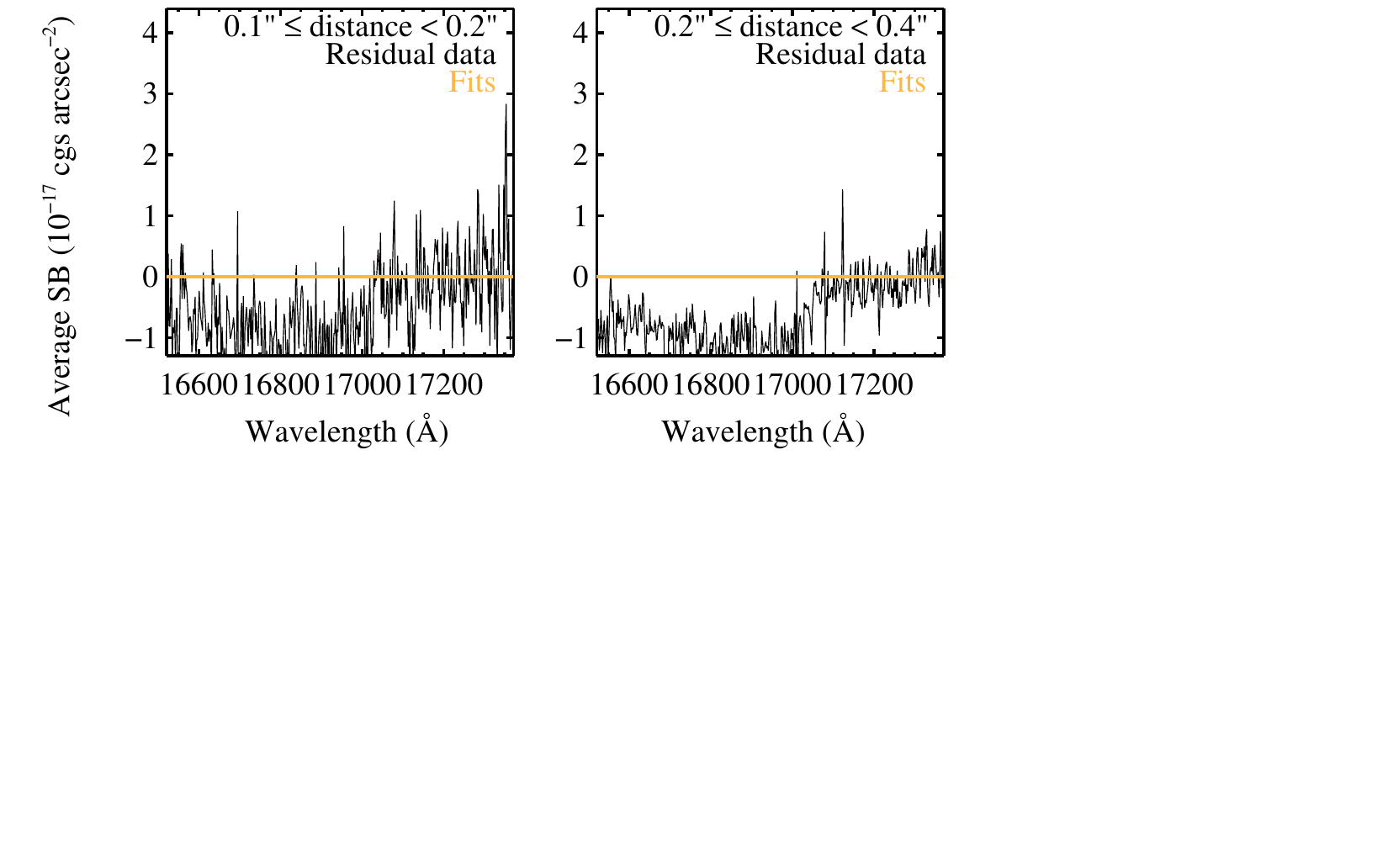}
\includegraphics[width=0.33\textwidth]{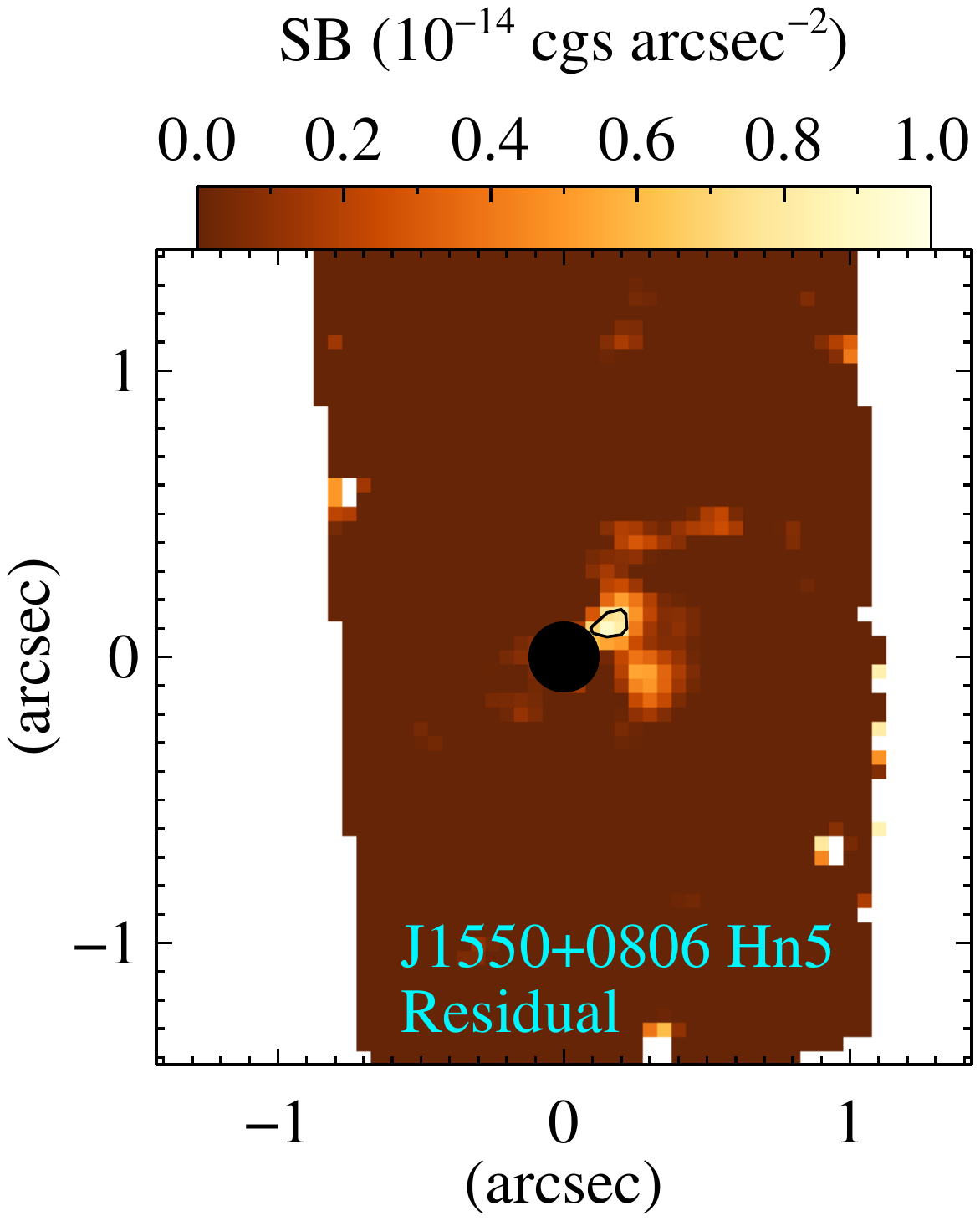}
\includegraphics[width=0.66\textwidth]{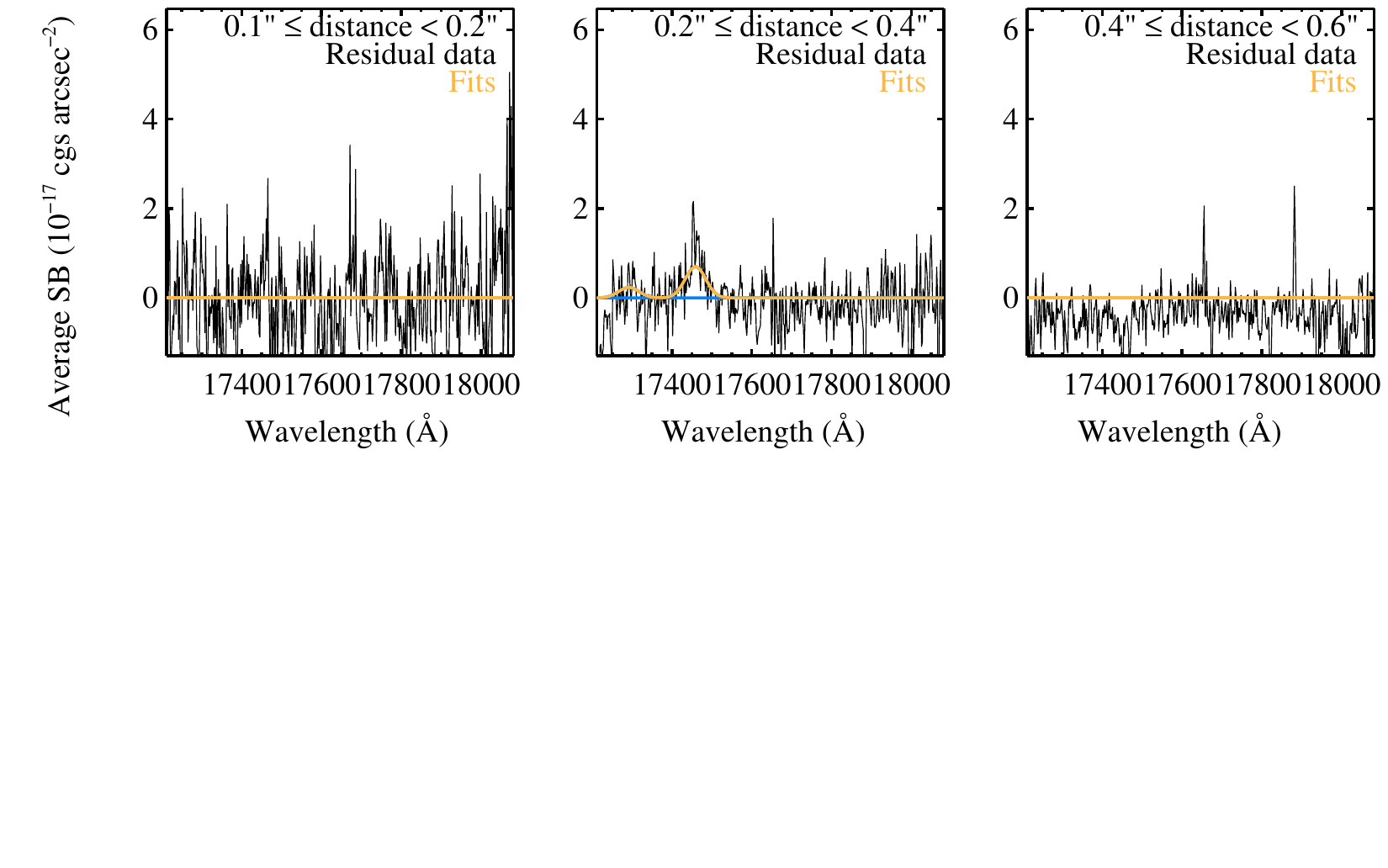}
\contcaption{}
\end{figure*}
\begin{figure*}
\includegraphics[width=0.33\textwidth]{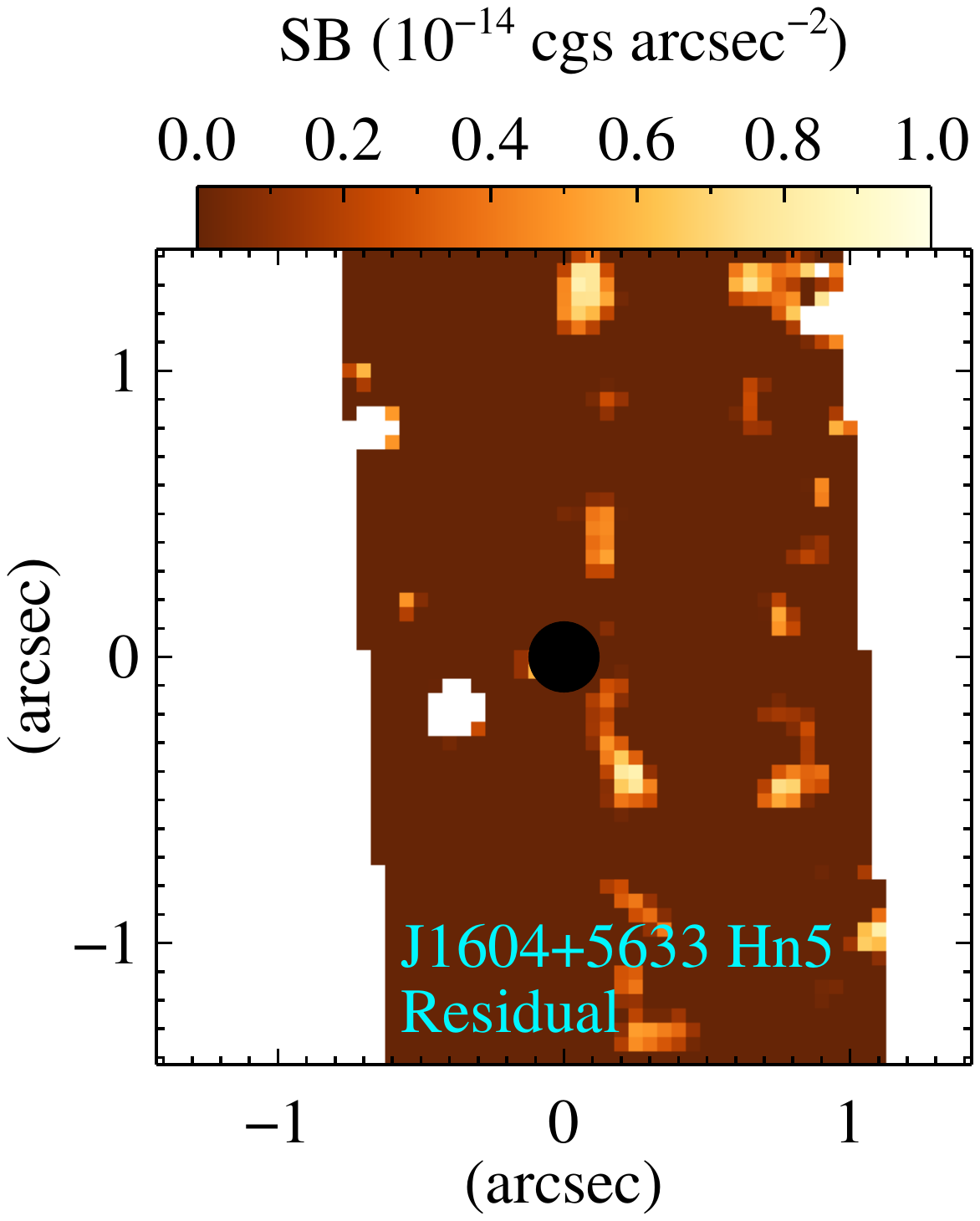}
\includegraphics[width=0.66\textwidth]{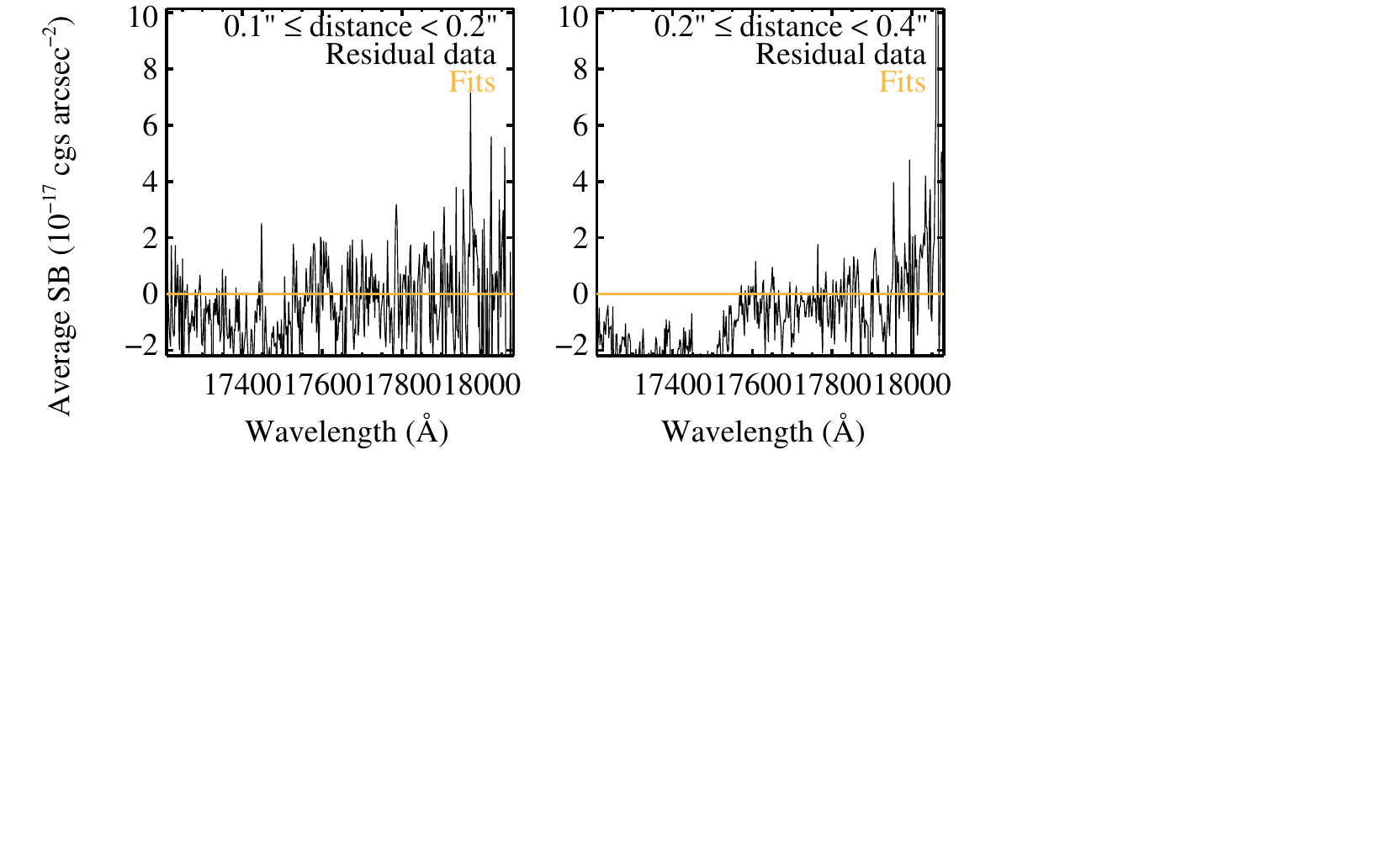}
\includegraphics[width=0.33\textwidth]{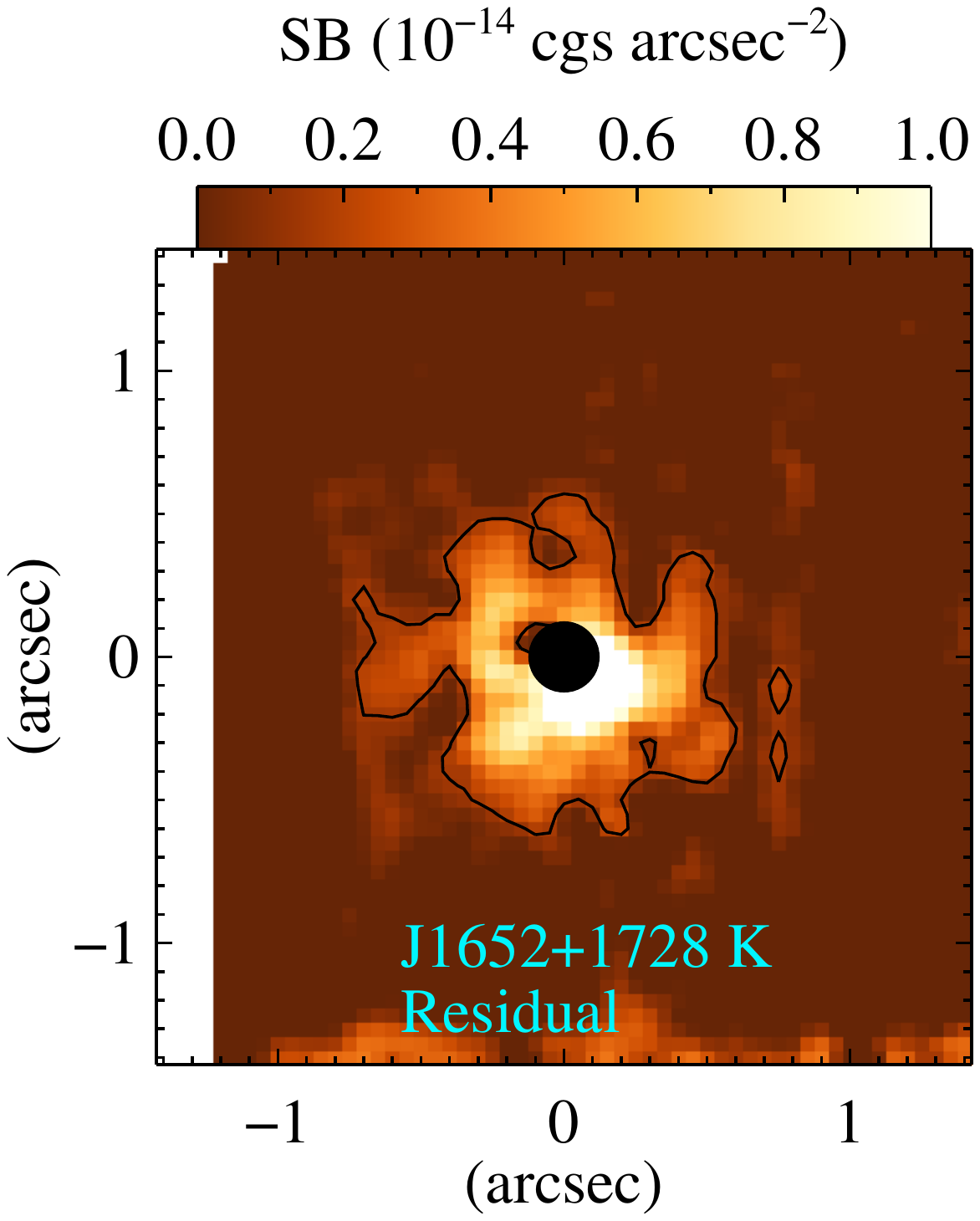}
\includegraphics[width=0.66\textwidth]{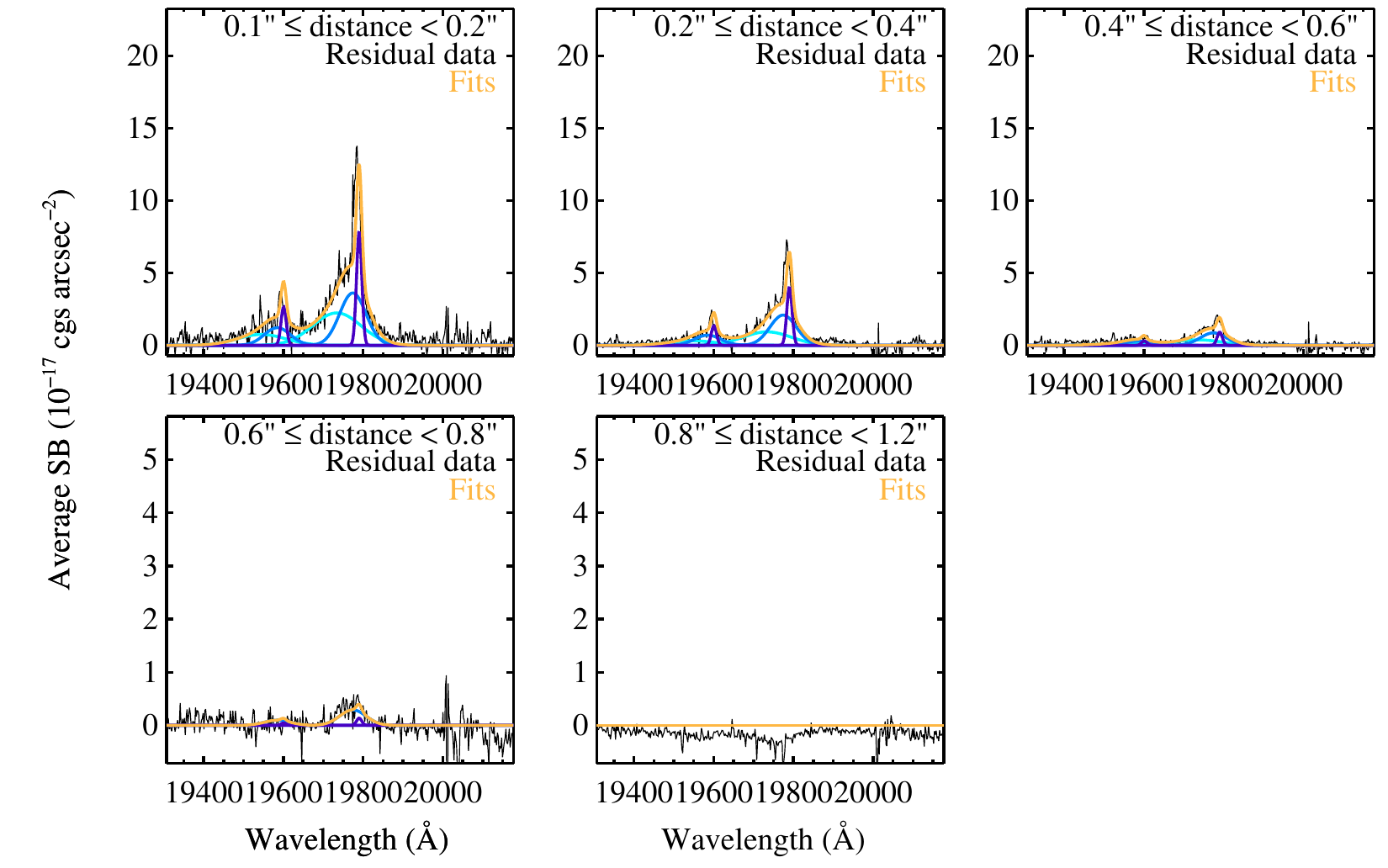}
\includegraphics[width=0.33\textwidth]{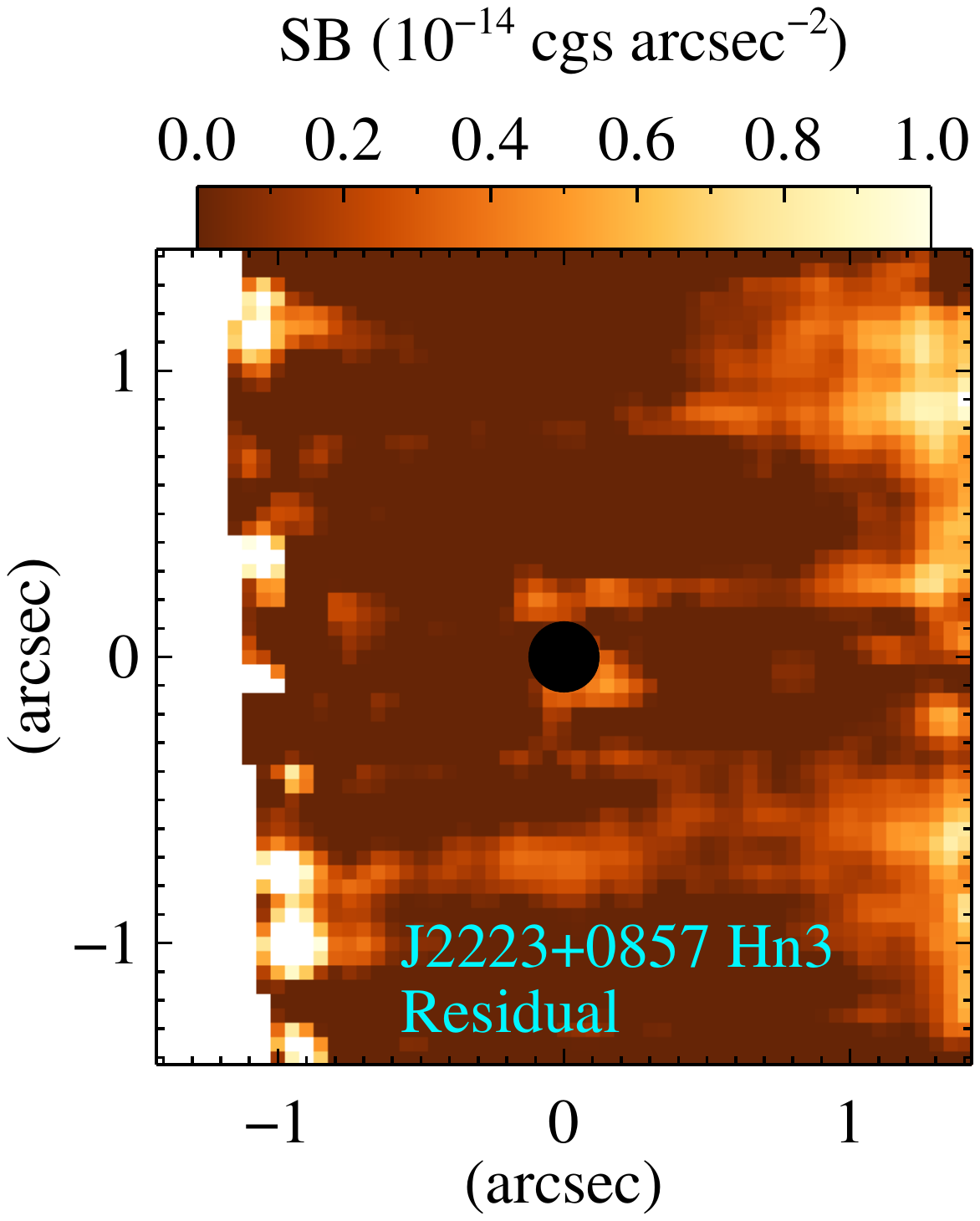}
\includegraphics[width=0.66\textwidth]{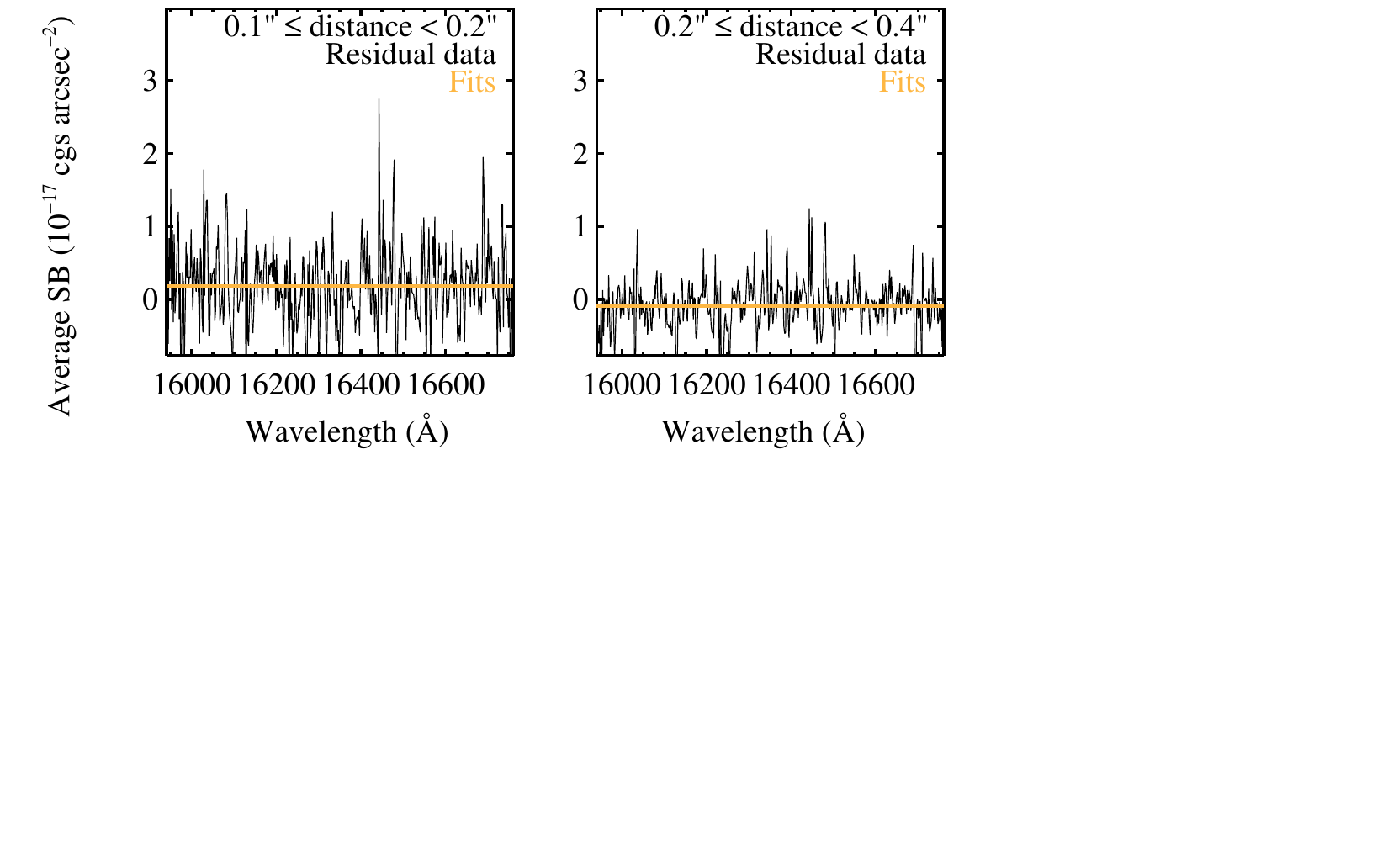}
\contcaption{}
\end{figure*}
\begin{figure*}
\includegraphics[width=0.33\textwidth]{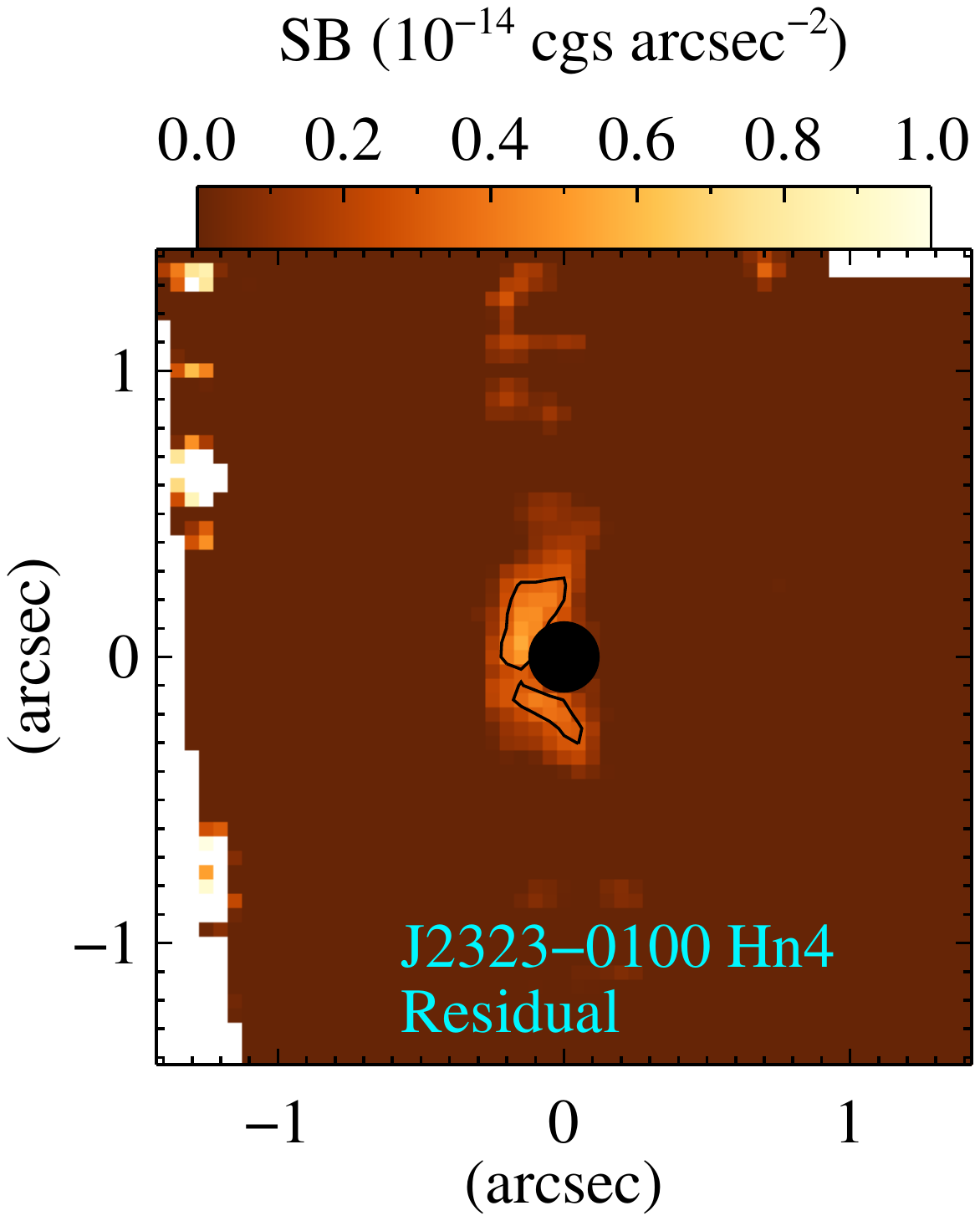}
\includegraphics[width=0.66\textwidth]{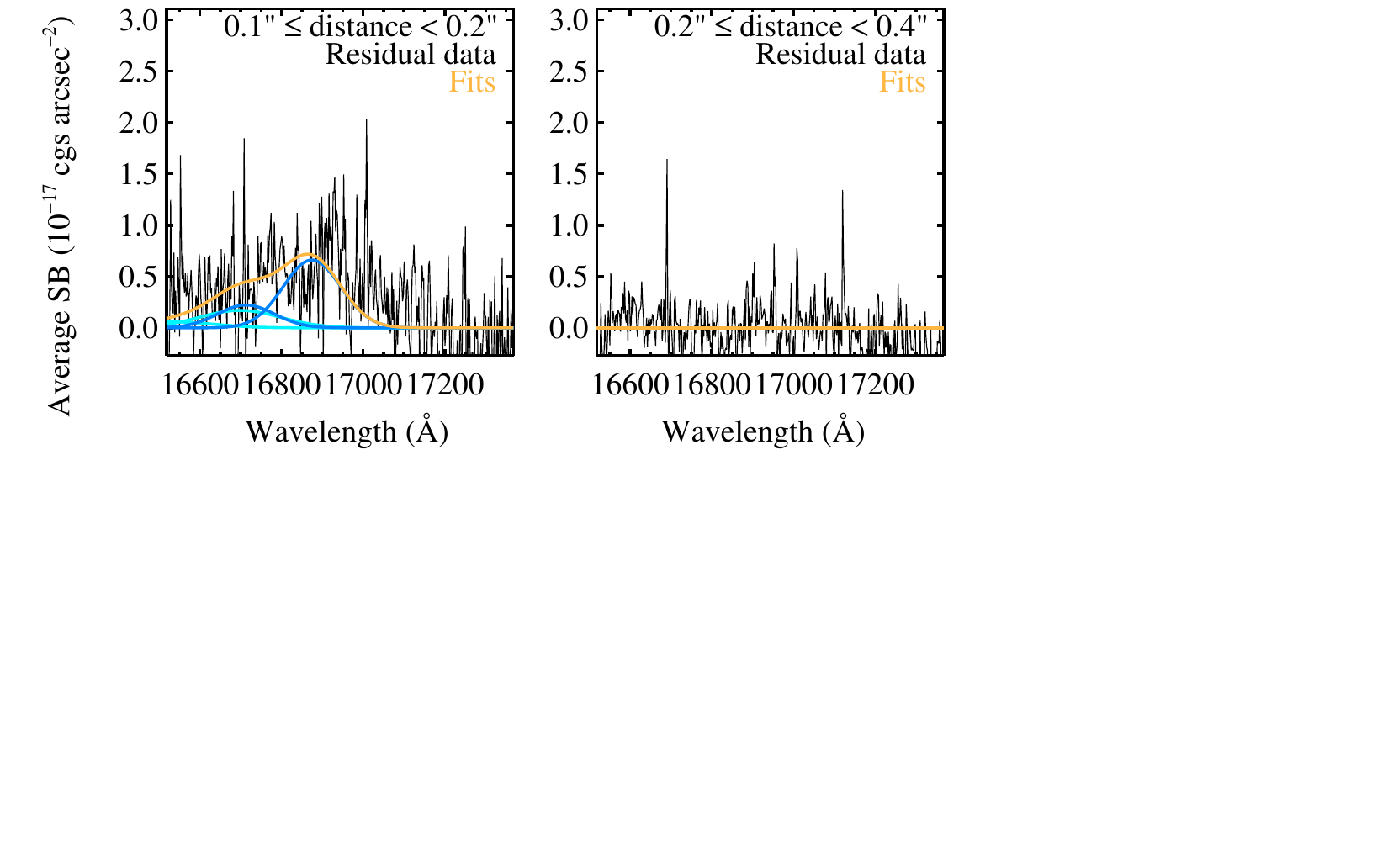}
\contcaption{}
\end{figure*}

\begin{table*}
\caption{Kinematic Parameters of Fitted [\ion{O}{III}] Components (fixed across all extracted apertures)}
\label{tab:kinematics}
\setlength{\tabcolsep}{0.1in}
\begin{tabular}{lcccccccc}
\hline
Name & [\ion{O}{III}] Comp1 & [\ion{O}{III}] Comp1\tablenotemark{a} & [\ion{O}{III}] Comp2 & [\ion{O}{III}] Comp2\tablenotemark{a} & [\ion{O}{III}] Comp3 & [\ion{O}{III}] Comp3\tablenotemark{a} \\
 & Velocity Centroid & Velocity Dispersion\tablenotemark{b} & Velocity Centroid & Velocity Dispersion\tablenotemark{b} & Velocity centroid & Velocity Dispersion\tablenotemark{b} \\
 & (km\,s$^{-1}$) & (km\,s$^{-1}$) & (km\,s$^{-1}$) & (km\,s$^{-1}$) & (km\,s$^{-1}$) & (km\,s$^{-1}$) \\
\hline
J0006$+$1215 & $-$2485$\pm$51 & 1997$\pm$103 & 0$\pm$30 & 221$\pm$41 & / & / \\
J0209$+$3122 & $-$554$\pm$19 & 998$\pm$103 & $-$286$\pm$18 & 237$\pm$18 & / & / \\
J1031$+$2903 & $-$1050$\pm$51 & 1100$\pm$57 & 0$\pm$21 & 201$\pm$29 & / & / \\
J1217$+$0234 & $-$340$\pm$33 & 530$\pm$18 & 0$\pm$17 & 142$\pm$29 & / & / \\
J1232$+$0912 & $-$3870$\pm$32 & 1872$\pm$18 & $-$1117$\pm$18 & 979$\pm$18 & / & / \\
J1550$+$0806 & $-$2565$\pm$47 & 1200$\pm$18 & $-$1952$\pm$17 & 443$\pm$17 & / & / \\
J1604$+$5633 & $-$1126$\pm$41 & 1174$\pm$18 & $+$55$\pm$17 & 525$\pm$17 & / & / \\
J1652$+$1728 & $-$1101$\pm$34 & 919$\pm$23 & $-$492$\pm$17 & 500$\pm$18 & $-$250$\pm$33 & 100$\pm$31 \\
J2223$+$0857 & $-$2876$\pm$58 & 2612$\pm$72 & $-$682$\pm$18 & 907$\pm$25 & / & / \\
J2323$-$0100 & $-$4142$\pm$105 & 1782$\pm$72 & $-$990$\pm$30 & 1253$\pm$29 & / & / \\
\hline
\end{tabular}
\tablenotetext{a}{\raggedright We order the components by their velocity dispersion values from high to low.}
\tablenotetext{b}{\raggedright We define the threshold for outflows that require AGN driving at velocity dispersion $\gtrsim$250\,km\,s$^{-1}$.}
\end{table*}

\begin{table*}
\caption{$w_{80}$ Velocity Widths of Total [\ion{O}{III}]\,$\lambda$5007 Line Profiles in Extracted Apertures}
\label{tab:annuw80}
\setlength{\tabcolsep}{0.1in}
\begin{tabular}{lcccccccc}
\hline
Name & (0.0\textendash0.1)\,arcsec & (0.1\textendash0.2)\,arcsec & (0.2\textendash0.4)\,arcsec & (0.4\textendash0.6)\,arcsec & Annulus 5\tablenotemark{a} & Annulus 6\tablenotemark{b} \\
 & (km\,s$^{-1}$) & (km\,s$^{-1}$) & (km\,s$^{-1}$) & (km\,s$^{-1}$) & (km\,s$^{-1}$) & (km\,s$^{-1}$) \\
\hline
J0006$+$1215 & 5170$\pm$130 & 5110$\pm$130 & 5060$\pm$180 & 4890$\pm$1070 & 570$\pm$110 & / \\
J0209$+$3122 & 2060$\pm$20 & 2110$\pm$70 & 2150$\pm$350 & 610$\pm$10 & 610$\pm$10 & 610$\pm$10 \\
J1031$+$2903 & 2790$\pm$90 & 2830$\pm$180 & 2550$\pm$280 & / & / & / \\
J1217$+$0234 & 1290$\pm$40 & 1170$\pm$50 & 980$\pm$170 & 1020$\pm$130 & 1080$\pm$260 & 360$\pm$70 \\
J1232$+$0912 & 5420$\pm$50 & 5330$\pm$130 & 5360$\pm$310 & 5230$\pm$3000 & / & / \\
J1550$+$0806 & 2550$\pm$120 & 2480$\pm$120 & 2240$\pm$200 & 2730$\pm$1080 & / & / \\
J1604$+$5633 & 2790$\pm$100 & 2640$\pm$290 & 1340$\pm$20 & / & / & / \\
J1652$+$1728 & 2000$\pm$30 & 1940$\pm$40 & 1840$\pm$30 & 1800$\pm$50 & 1710$\pm$80 & 1180$\pm$520 \\
J2223$+$0857 & 6090$\pm$200 & 6120$\pm$610 & 6180$\pm$1450 & 2330$\pm$60 & / & / \\
J2323$-$0100 & 5570$\pm$80 & 5450$\pm$470 & 5620$\pm$810 & 5780$\pm$1360 & / & / \\
\hline
\end{tabular}
\tablenotetext{a}{\raggedright (0.6\textendash1.0)\,arcsec for J0006+1215; (0.6\textendash0.8)\,arcsec for the rest.}
\tablenotetext{b}{\raggedright (1.0\textendash1.4)\,arcsec for J0006+1215; (0.8\textendash1.2)\,arcsec for J1232+0912 and J1652+1728; (0.8\textendash1.0)\,arcsec for the rest.}
\end{table*}

\begin{table*}
\caption{Size Measurements from Surface Brightness Radial Profiles}
\label{tab:COGmeasures}
\setlength{\tabcolsep}{0.05in}
\begin{tabular}{lcccccccccc}
\hline
Name & Cont. & Cont. & [\ion{O}{III}] Comp1 & [\ion{O}{III}] Comp1 & [\ion{O}{III}] Comp2 & [\ion{O}{III}] Comp2 & [\ion{O}{III}] Comp3 & [\ion{O}{III}] Comp3 & Total [\ion{O}{III}] & Total [\ion{O}{III}] \\
 & $R_{\rm half}$ & $R_{\rm max}$ & $R_{\rm half}$ & $R_{\rm max}$ & $R_{\rm half}$ & $R_{\rm max}$ & $R_{\rm half}$ & $R_{\rm max}$ & $R_{\rm half}$ & $R_{\rm max}$ \\
 & (kpc) & (kpc) & (kpc) & (kpc) & (kpc) & (kpc) & (kpc) & (kpc) & (kpc) & (kpc) \\
\hline
J0006$+$1215 & $>$2.8 & $>$8.9 & $<$1.0 & $<$3.0 & 2.8$\pm$0.5 & 5.8$\pm$2.2 & / & / & $<$1.0 & $<$3.0 \\
J0209$+$3122 & $<$1.1 & $<$4.6 & $<$1.1 & $<$4.6 & 1.7$\pm$0.5 & 4.6$\pm$2.3 & / & / & $<$1.1 & $<$4.6 \\
J1031$+$2903 & $<$1.1 & $<$4.5 & $<$1.1 & $<$4.5 & 2.1$\pm$0.5\tablenotemark{a} & $<$4.5 & / & / & $<$1.1 & $<$4.5 \\
J1217$+$0234 & 1.0$\pm$0.4 & $<$4.2 & $<$1.0 & $<$4.2 & $>$2.5 & $>$4.6 & / & / & 1.4$\pm$0.4 & $<$4.2 \\
J1232$+$0912 & $>$3.4 & $>$7.0 & $<$2.2 & $<$4.7 & $<$2.2 & $<$4.7 & / & / & $<$2.2 & $<$4.7 \\
J1550$+$0806 & $<$1.2 & $<$3.9 & $<$1.2 & $<$3.9 & 1.2$\pm$0.4 & $<$3.9 & / & / & $<$1.2 & $<$3.9 \\
J1604$+$5633 & 1.8$\pm$0.4 & $<$3.9 & $<$1.1 & $<$3.9 & $<$1.1 & $<$3.9 & / & / & $<$1.1 & $<$3.9 \\
J1652$+$1728 & $<$1.3 & $<$4.2 & $<$1.3 & $<$4.2 & 1.5$\pm$0.4 & 4.2$\pm$0.8 & $>$1.9 & $>$4.2 & $<$1.3 & $<$4.2 \\
J2223$+$0857 & $<$1.3 & $<$3.8 & $<$1.3 & $<$3.8 & $<$1.3 & $<$3.8 & / & / & $<$1.3 & $<$3.8 \\
J2323$-$0100 & $<$1.0 & $<$3.6\tablenotemark{b} & 1.2$\pm$0.4 & $<$3.6 & 1.0$\pm$0.4 & $<$3.6 & / & / & 1.0$\pm$0.4 & $<$3.6 \\
\hline
\end{tabular}
\tablenotetext{a}{\raggedright If the marginally detected component 2 is treated as non-detection in the nuclear region, its $R_{\rm half}$ becomes 2.2\,kpc with no change to other size measurements.}
\tablenotetext{b}{\raggedright A second nucleus is discernible at 1.3\,arcsec away in the north-northeast direction despite the ERQ host continuum being spatially unresolved.}
\end{table*}

\begin{table*}
\caption{Size Measurements from PSF-subtracted Residuals\tablenotemark{a}}
\label{tab:resmeasures}
\setlength{\tabcolsep}{0.1in}
\begin{tabular}{lcccccccc}
\hline
Name & Total [\ion{O}{III}] $R_{\rm half}$ & Total [\ion{O}{III}] $R_{\rm max}$ \\
 & (kpc) & (kpc) \\
\hline
J0006$+$1215 & 2.6$\pm$0.4 & 4.5$\pm$0.4 \\
J0209$+$3122 & 1.5$\pm$0.5 & $<$4.6 \\
J1031$+$2903 & 1.9$\pm$0.4 & $<$4.5 \\
J1217$+$0234 & 2.6$\pm$0.4 & 6.5$\pm$0.8 \\
J1232$+$0912 & $<$2.2 & $<$4.7 \\
J1550$+$0806 & 2.5$\pm$0.4 & $<$3.9 \\
J1604$+$5633 & $<$1.1 & $<$3.9 \\
J1652$+$1728 & 2.6$\pm$0.4 & 4.6$\pm$0.4 \\
J2223$+$0857 & $<$1.3 & $<$3.8 \\
J2323$-$0100 & 1.2$\pm$0.4 & $<$3.6 \\
\hline
\end{tabular}
\tablenotetext{a}{\raggedright Measured on the total [\ion{O}{III}] profile after subtracting the PSF. Further information on the derivation of these quantities is available in Section 3.3.}
\end{table*}

\section{Discussion}

We assess the implications of our measured [\ion{O}{III}] emitting region sizes and 
spatially-resolved kinematics of ERQs by comparing them with other non-jetted/radio-quiet and 
intrinsically luminous quasars at similar redshifts. The radio requirement on the control samples 
is to ensure comparing radiation-driven rather than jet-driven outflows. The luminosity 
requirement is to control for the quasar narrow-line region's size-luminosity relation 
\citep[e.g.][]{Liu+13,SunGreeneZakamska17}. Although the bolometric luminosities of ERQs are only 
estimates and uncertain by at least a factor of 2 \citep{Hamann+17}, our goal is not to look for 
luminosity trends but merely select comparably luminous quasars. The redshift requirement is to 
control for cosmic evolution. To minimize biases in comparing low surface brightness emissions, we 
only select control samples from (8\textendash10)\,m class adaptive-optics 
or JWST NIRSpec integral-field observations. Given that the outflow extent in J1652+1728 measured 
in NIFS data is much smaller than that measured with JWST, we measure sizes in JWST observations 
of the comparison sources only down to the typical surface brightness sensitivity of 
adaptive-optics observations which is about two orders of magnitude lower. 
From the \cite{Kakkad+20} adaptive-optics sample we select 11 well-measured normal blue quasars, 
and we select the \cite{Williams+17} adaptive-optics obsservations of a blue quasar, 
the \cite{Perna+23} JWST NIRSpec observations of a blue quasar, 
and the \cite{Brusa+16} adaptive-optics observations of a Type~II narrow-line quasar. 

\subsection{Compact ERQ-driven [\ion{O}{III}] Outflows}

\begin{figure}
\includegraphics[width=\columnwidth]{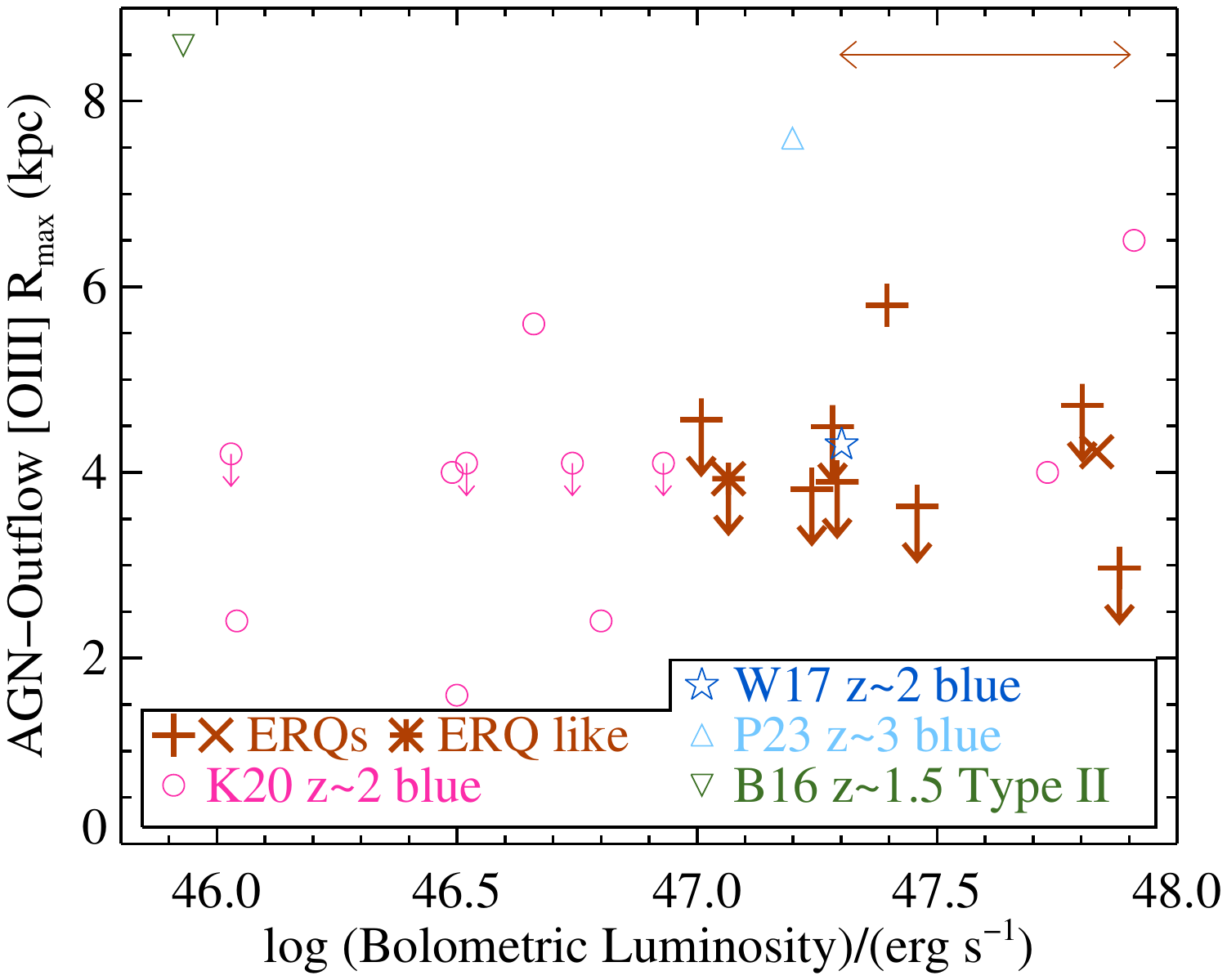}
\caption{The maximum radii of AGN-driven [\ion{O}{III}] outflows versus bolometric luminosities, 
of ERQs and other luminous quasars at cosmic noon in the literature. OSIRIS-measured ERQs are 
shown in brown plus symbols, the NIFS-measured ERQ from \citet{Vayner+21a} is shown in a brown 
cross symbol, and the red ERQ-like quasar is shown in a brown asterisk symbol. The typical 
uncertainty in the luminosities of the ERQs is indicated with a horizontal double arrow. Blue 
quasars at $z\sim2$ from the \citet{Kakkad+20} sample are shown in pink circles. One blue quasar 
at $z\sim2$ from \citet{Williams+17} is shown in a blue star. One blue quasar at $z\sim3$ from 
\citet{Perna+23} is shown in a light blue upward triangle. One Type~II quasar at $z\sim1.5$ from 
\citet{Brusa+16} is shown in a green downward triangle.}
\label{fig:RoutLbol}
\end{figure}

We compare the maximum radii of AGN-driven outflows in ERQs and other luminous quasars. For each 
ERQ, we calculate the maximum radius of AGN-driven outflows as the maximum radius of component 1 
or the maximum radius of the total of components 1 and 2 depending on whether one or both 
components meet the velocity dispersion $\gtrsim$250\,km\,s$^{-1}$ cut. 
The maximum radii of AGN outflows remain unchanged if we instead apply a corresponding 
$w_{80}\gtrsim$600\,km\,s$^{-1}$ cut on the total [\ion{O}{III}\,$\lambda$5007 line profile. 
For the comparison sources, the maximum radii are not always explicitly reported in the literature, 
we therefore measure the maximum radii from the published emission-line maps. Where the 
[\ion{O}{III}] emission is spatially resolved we measure the maximum radius of a S/N $>$ 2 spaxel 
that meets the same AGN-outflow cut, and where the emission is unresolved we adopt a 0.5\,arcsec 
upper limit informed by the typical adaptive optics correction performance. Fig.~\ref{fig:RoutLbol} 
presents the maximum radii of AGN-driven [\ion{O}{III}] outflows versus bolometric luminosities of 
ERQs and other radio-quiet, luminous, cosmic noon quasars in the literature. There is a hint that 
ERQ-driven outflows are more spatially compact than other luminous quasars. Using the ASURV 
software \citep{Feigelson+14} we perform a two-sample Peto-Prentice test and find the probability 
that the ERQ distribution and the comparison quasar distribution are drawn from the same parent 
population is only 0.04. Since individually detected spaxels are less sensitive than annular 
averages, the maximum radius measurements of the comparison sources made on the maps should be 
considered conservative underestimates relative to the measurements on the ERQs encompassing 90\% 
of the flux. While ERQs are on average extinct by three magnitudes in the rest-frame UV relative 
to a Type I spectral energy distribution \citep{Hamann+17}, ERQs are highly intrinsically luminous 
and their {\it W}3-band based bolometric luminosities are about an order of magnitude higher than 
the comparison samples. We therefore consider the size difference between ERQ-driven outflows and 
other AGN-driven outflows robust. 

\begin{figure}
\includegraphics[width=\columnwidth]{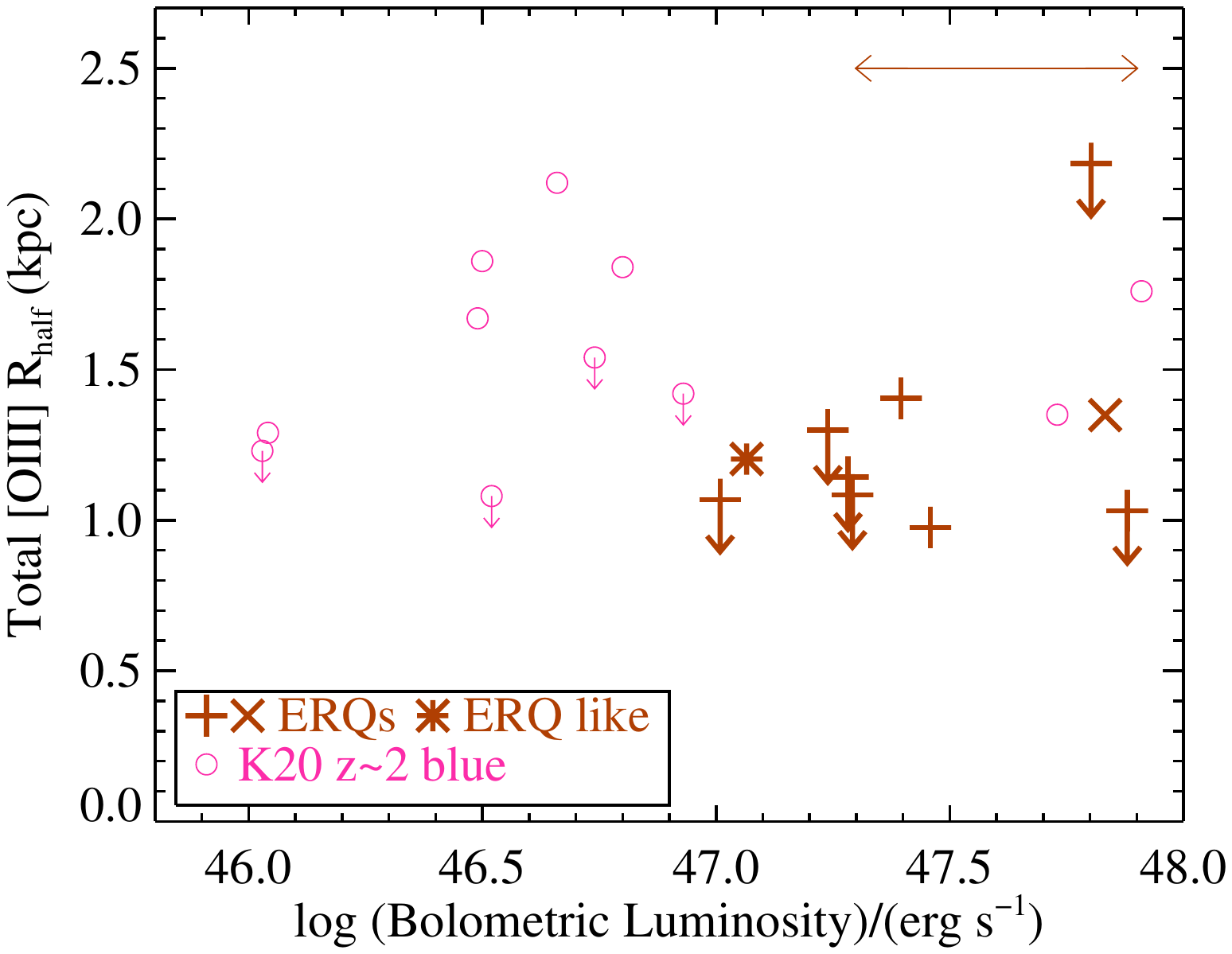}
\caption{The half-light radii of total [\ion{O}{III}] emissions versus bolometric luminosities, 
of ERQs and cosmic-noon luminous blue quasars from \citet{Kakkad+20}. The symbol scheme follows 
Fig.~\ref{fig:RoutLbol}.}
\label{fig:RtotLbol}
\end{figure}

We also compare the half-light radii of the total [\ion{O}{III}] emissions in ERQs versus the 
\cite{Kakkad+20} sample. The half-light radii of the total [\ion{O}{III}] line profiles are less 
measurement-dependent than the maximum radii of AGN-driven [\ion{O}{III}] outflows, and 
\cite{Kakkad+20} already have these measurements made and reported. Given the observed line 
profiles and the known relation between line velocity width and bolometric luminosity 
\citep[e.g.][]{ZakamskaGreene14,SunGreeneZakamska17}, the AGN-driven outflows dominate the line 
fluxes and the extents of the total [\ion{O}{III}] emissions largely trace extents of AGN-driven 
outflows. Fig.~\ref{fig:RtotLbol} presents the half-light radii of total [\ion{O}{III}] emissions 
versus bolometric luminosities of ERQs and the luminous blue quasars from \cite{Kakkad+20}. There 
is again a hint that the [\ion{O}{III}] emitting regions of ERQs are more compact. The 
Peto-Prentice test gives the probability that the ERQs and the comparison quasars follow the same  
distribution is only 0.03. 

Based on Figs.~\ref{fig:RoutLbol} and~\ref{fig:RtotLbol} we conclude that the extremely fast 
ERQ-driven ionized ouflows tend to be less extended than other AGN-driven ionized outflows 
controlled for luminosities and redshifts. As a low-density forbidden transition, 
[\ion{O}{III}]\,5007 cannot be produced on the scales of the broad-line region. The characteristic 
size of those spatially-unresolved [\ion{O}{III}]-emitting regions has a physical lower limit at 
$\sim$1\,kpc scales. Our results are consistent with the scenario that ERQs are in an earlier 
evolutionary stage than normal blue quasars and during this short-lived, transition blowout phase 
the outflows have not had enough time to further expand into their interstellar media. The 
difference between the extents of ERQ-driven outflows compared to those of Type~II quasars suggests 
the physical conditions in ERQs are unique and beyond orientation differences as 
in the AGN unified model \citep{Antonucci93,Netzer15}. Our results are consistent with models of 
quasar feedback where IR radiation pressure trapped in dusty environments can boost powerful, 
galactic-scale outflows during their more compact and younger stages \citep{Costa+18,
IshibashiFabianMaiolino18}. The post-starburst E+A galaxies, which are believed to 
be a transition stage from the active phase to the quenched phase of galaxy evolution, are found 
to have larger ionized outflow extents up to 17\,kpc \citep{Baron+18,Baron+24}. Their outflow 
extents are also consistent with the supposed evolution sequence. While the extended dust in ERQ 
host galaxies can affect the detectability of [\ion{O}{III}] emission, we note that the broad 
[\ion{O}{III}] component is compact even in ERQs where a narrow [\ion{O}{III}] component is 
detected and extended. 

\subsection{Small Total [\ion{O}{III}]-Emitting Regions}

\begin{figure}
\includegraphics[width=\columnwidth]{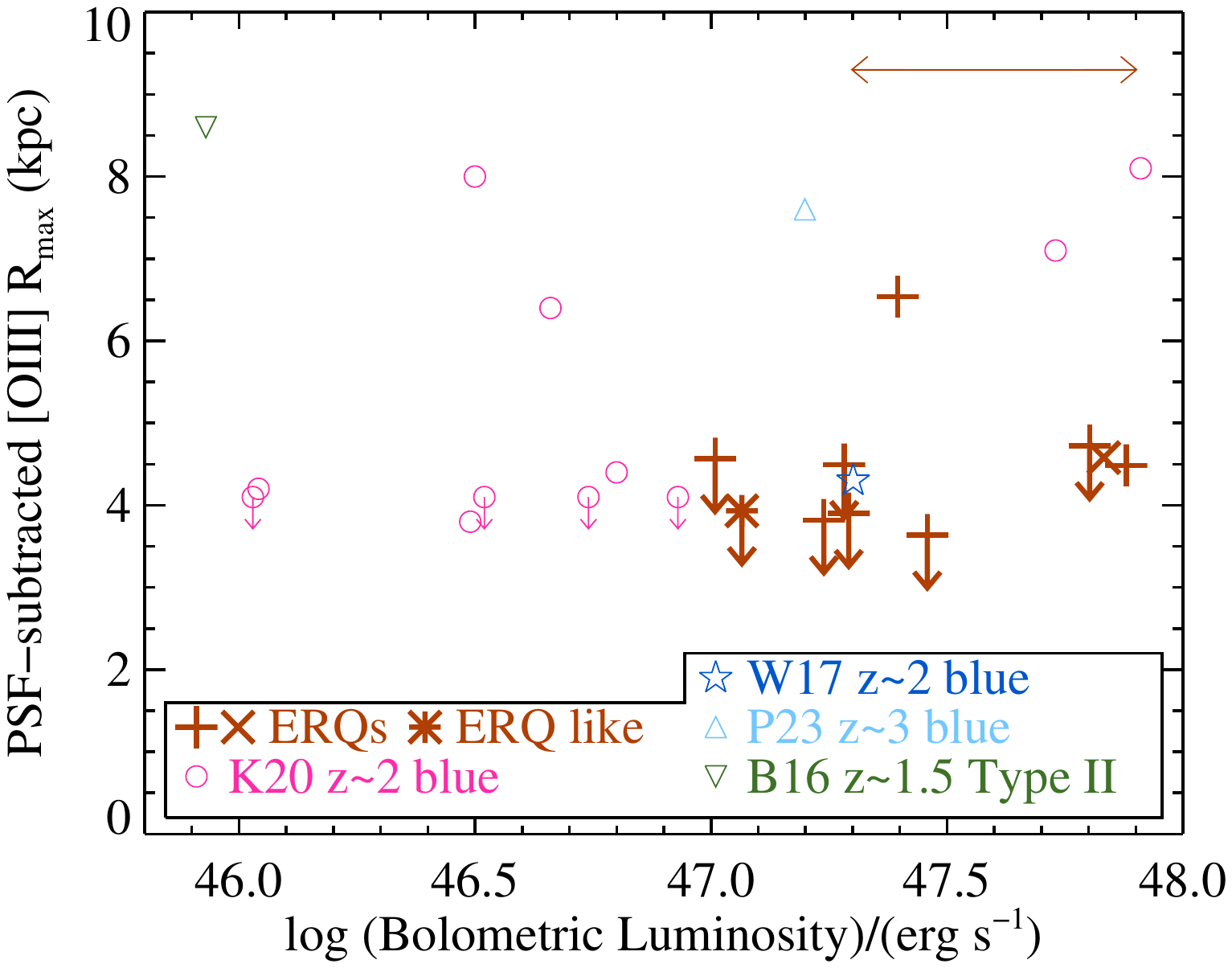}
\caption{The maximum radii of PSF-subtracted residual [\ion{O}{III}] emissions versus bolometric 
luminosities, of ERQs and other luminous quasars at cosmic noon in the literature. The symbol 
scheme follows Fig.~\ref{fig:RoutLbol}.}
\label{fig:RsubLbol}
\end{figure}

Given that the ERQ-driven outflows tend to be compact, the extents of the PSF-subtracted 
residual [\ion{O}{III}] emissions largely trace the extents of the quiescent ionized 
interstellar media. For the comparison sources, where the residual [\ion{O}{III}] is 
spatially resolved we measure the maximum radius of a S/N $>$ 2 spaxel, and where the residual is 
unresolved we adopt a 0.5\,arcsec upper limit. Fig.\ref{fig:RsubLbol} presents the maximum radii of 
PSF-subtracted residual [\ion{O}{III}] emissions versus bolometric luminosities of ERQs and 
other luminous quasars in the literature. The maximum radii of residual [\ion{O}{III}] are 
formally presented as upper limits for many ERQs, as the adaptive-optics-corrected PSF still has 
a weak seeing halo. We refer to the half-light radii of the [\ion{O}{III}] residuals in 
Table~\ref{tab:resmeasures} to determine that PSF-subtracted residuals are present in seven out of 
10 ERQs. The fraction of ERQs with PSF-subtracted residuals present is similar to that of the 
luminous blue quasar sample of \cite{Kakkad+20} despite being suppressed by typically three 
magnitudes along the lines-of-sight in the rest-frame UV relative to the spectral energy 
distribution of normal quasars \citep{Hamann+17}. This suggests that some ionizing radiation is 
able to escape and the [\ion{O}{III}] emission is not predominantly ionization-bounded, a 
conclusion that is also reached with comparing Ly$\alpha$ haloes of ERQs and normal blue quasars 
\citep{Lau+22,Gillette+23}. On a related note, the fraction of ERQs having a narrow [\ion{O}{III}] 
component, five out of 10, is also similar to the fraction of the \cite{Kakkad+20} luminous blue 
quasars having a narrow component. 

Although PSF-subtracted residuals are present in ERQs, they tend to be less extended than in 
luminous blue quasars and the Type~II quasar in comparison. The Pento-Prentice test gives the 
probability that the ERQs and the comparison quasars folow the same distribution is 0.11. 
This is consistent with ERQs having global and patchy obscuration patterns 
The global obscuration explains the small extents of the total ionized emitting regions 
independently of the outflow timescales, while the patchiness still allows channels for ionizing 
radiation to escape. Such geometry of quasar obscuration has been explored in \cite{Andonie+24} 
with submillimeter data, which find that dusty starburst galaxies can contribute substantial 
galactic-scale obscuration comparable to the nuclear-scale obscuration level 
\citep[see also][]{MunozElgueta+22,GonzalezLobos+23}. 

The ERQs J0006+1215, J1232+0912, J1652+1728, and J2323$-$0100 have their surrounding 
quasar-powered Ly$\alpha$ haloes measured with wide-field integral-field spectroscopy \citep{Lau+22,
Gillette+23}. We may combine analysis of their halo-scale Ly$\alpha$ emissions and galactic-scale 
[\ion{O}{III}] emissions to jointly assess the geometry of the quasar illumination. J0006+1215 
has moderately suppressed Ly$\alpha$ halo luminosity relative to luminous blue quasars or Type~II 
quasars, whereas the total [\ion{O}{III}]-emitting region has comparable extent as other quasar 
populations. J1232+0912 has moderately suppressed Ly$\alpha$ halo luminosity, whereas the 
[\ion{O}{III}]-emitting region has an unresolved size. J2323$-$0100 has heavily suppressed 
Ly$\alpha$ halo luminosity, whereas the [\ion{O}{III}]-emitting region is compact and resolvable 
only by the most excellent angular resolution and deep observations. J1652+1728 has a disturbed 
and luminous Ly$\alpha$ halo, whereas the total [\ion{O}{III}]-emitting region has comparable 
extent as other quasar populations. 
Overall, the halo-scale Ly$\alpha$ measurements and galactic-scale [\ion{O}{III}] measurements 
together present a picture that these ERQs are inconsistent with a simple Type~II geometry. 

Despite the aforementioned differences between ERQs and a simple Type~II geometry, the hint of the 
presence of ionization cones in 4 out of 10 ERQs suggests some of the obscuration patterns can be 
partially explained by the standard torus-based unified model. The idea of a patchy obscuring 
torus has been explored in \cite{Obied+16} to explain observations of scattered light in Type~II 
quasars. 


\section{Conclusions and Future Work}

ERQs are a non-radio-selected, intrinsically luminous population of quasars at cosmic noon 
redshifts selected by 
their extremely red colour from rest-frame UV to mid-IR. They are characterized by exceptionally 
broad and blueshifted [\ion{O}{III}]\,$\lambda\lambda$4959,5007 emission lines that carry outflow 
speeds reaching $>$6000\,km\,s$^{-1}$. We obtained new Keck/OSIRIS observations of nine ERQs and 
published Gemini/NIFS observations of one ERQ to form a sample of 10 ERQs with  
laser-guided adaptive-optics-assisted integral-field spectroscopy. The goal is to measure the 
sizes and spatially-resolved kinematics of the [\ion{O}{III}]-emitting regions. We reached 
angular resolutions of $\sim$0.2\,arcsec in Gaussian FWHM. We performed analysis on 
the [\ion{O}{III}] and continuum emissions of the ERQs and their reference empirical PSFs, using 
the surface brightness radial profile method and the PSF subtraction method. 

We identify extended, clumpy, disturbed continuum emission in the hosts of J0006+1215, J1217+0234, 
J1232+0912, and companion continuum emission around J2323$-$0100. We confirm signs of merger 
activities in ERQs although we do not quantify a merger fraction due to depth of the data. 

For each ERQ we fit two to three Gaussian components to the [\ion{O}{III}] emissions extracted 
from annular apertures, both before and after PSF subtraction. We consider a kinematic component 
to be tracing AGN-driven outflow if its velocity dispersion $\gtrsim$250\,km\,s$^{-1}$. For the 
overall sample, the 16 fitted [\ion{O}{III}] components that exceed the AGN-outflow threshold are 
spatially unresolved or compact with maximum radii $\lesssim$4\,kpc. 
There are five [\ion{O}{III}] components of the sample below the AGN-outflow cut and all are 
extended, possibly tracing the quiescent interstellar media on few kpc scales. 
Justified on their different spatial distributions and bimodality in the components' velocity 
distribution, we consider the broad versus narrow components physical distinct and having 
different dynamical mechanisms. 

Compared to other radio quiet, intrinsically luminous quasars at cosmic noon, the extremely fast 
ERQ-driven [\ion{O}{III}] outflows appear to be compact, at $\sim$1\,kpc scales. Fast and compact 
ionized outflows support the notion that ERQs are young quasars, representing a short-lived, 
transition, blowout phase in galaxy evolution, and their unique physical conditions cannot be 
explained by Type~I versus Type~II orientation differences. Our results are consistent with models 
of quasar feedback where IR radiation trapped by dust produces faster and more compact outflows. 

The fraction of total [\ion{O}{III}] emission being spatially resolved in ERQs is similar to 
that in other luminous quasars, suggesting that some of the ionizing radiation has channels to 
escape through the dust obscuration. The total [\ion{O}{III}]-emitting regions in ERQs, at few 
kpc scales, appear to be less extended compared to other luminous quasars. Resolved but small 
extents of the total [\ion{O}{III}] emissions support ERQs having global and patchy obscuration 
patterns that are inconsistent with a simple Type~II geometry. The hint of ionization cones in 
4 out of 10 ERQs suggests some of the obscuration patterns can also be partially explained by a 
patchy torus model. 

Finally, both the current generation of adaptive optics and JWST NIRSpec integral-field 
observations have PSF wings extending out to $\sim$1\,arcsec \citep{Law+18,
Veilleux+23}. How the ionized outflows traced by [\ion{O}{III}] compare to the unobscured star 
formation traced by H$\alpha$ in quasar-host galaxies remains an open question. Should there be 
evidence of suppression of star formation by quasar feedback manifested as spatial anticorrelation 
between [\ion{O}{III}] and H$\alpha$, new-generation instruments such as the VLT/ERIS and the Keck 
I/Liger are needed to detect it. 

\section*{Acknowledgements}

We acknowledge Gene C.\ K.\ Leung, Eilat Glikman, and Sanchit Sabhlok for 
discussions. We acknowledge Sherry Yeh, Jim Lyke, and Randy Campbell for supporting the 
observations. We acknowledge Devin Chu, Andrea Ghez, and Luke Finnerty for sharing their 
data. 
MWL, FH, and JG acknowledge support from the USA National Science Foundation grant AST-1911066. 
NLZ acknowledges support from NASA ADAP grant 80NSSC21K1569. This research was supported in part 
by the National Science Foundation under Grant No.\ NSF PHY-1748958.
This work is based on observations made at the W.\ M.\ Keck Observatory, which is operated as a
scientific partnership between the California Institute of Technology and the University of
California. It was made possible by the generous support of the W.\ M.\ Keck Foundation. The 
authors wish to recognize and acknowledge the very significant cultural role and reverence that 
the summit of Mauna Kea has always had within the indigenous Hawaiian community. We are most 
fortunate to have the opportunity to conduct observations from this mountain. 

\section*{Data Availability}

The data are available upon request. 

\bibliographystyle{mnras}
\bibliography{/Users/lwymarie/Documents/Bibli/allrefs}

\begin{thebibliography}{}
\makeatletter
\relax
\def\mn@urlcharsother{\let\do\@makeother \do\$\do\&\do\#\do\^\do\_\do\%\do\~}
\def\mn@doi{\begingroup\mn@urlcharsother \@ifnextchar [ {\mn@doi@}
  {\mn@doi@[]}}
\def\mn@doi@[#1]#2{\def\@tempa{#1}\ifx\@tempa\@empty \href
  {http://dx.doi.org/#2} {doi:#2}\else \href {http://dx.doi.org/#2} {#1}\fi
  \endgroup}
\def\mn@eprint#1#2{\mn@eprint@#1:#2::\@nil}
\def\mn@eprint@arXiv#1{\href {http://arxiv.org/abs/#1} {{\tt arXiv:#1}}}
\def\mn@eprint@dblp#1{\href {http://dblp.uni-trier.de/rec/bibtex/#1.xml}
  {dblp:#1}}
\def\mn@eprint@#1:#2:#3:#4\@nil{\def\@tempa {#1}\def\@tempb {#2}\def\@tempc
  {#3}\ifx \@tempc \@empty \let \@tempc \@tempb \let \@tempb \@tempa \fi \ifx
  \@tempb \@empty \def\@tempb {arXiv}\fi \@ifundefined
  {mn@eprint@\@tempb}{\@tempb:\@tempc}{\expandafter \expandafter \csname
  mn@eprint@\@tempb\endcsname \expandafter{\@tempc}}}

\bibitem[\protect\citeauthoryear{{Andonie} et~al.,}{{Andonie}
  et~al.}{2024}]{Andonie+24}
{Andonie} C.,  et~al., 2024, \mn@doi [\mnras] {10.1093/mnrasl/slad144}, \href
  {https://ui.adsabs.harvard.edu/abs/2024MNRAS.527L.144A} {527, L144}

\bibitem[\protect\citeauthoryear{{Antonucci}}{{Antonucci}}{1993}]{Antonucci93}
{Antonucci} R.,  1993, \mn@doi [\araa] {10.1146/annurev.aa.31.090193.002353},
  \href {http://adsabs.harvard.edu/abs/1993ARA\%26A..31..473A} {31, 473}

\bibitem[\protect\citeauthoryear{{Assef} et~al.,}{{Assef}
  et~al.}{2015}]{Assef+15}
{Assef} R.~J.,  et~al., 2015, \mn@doi [\apj] {10.1088/0004-637X/804/1/27},
  \href {https://ui.adsabs.harvard.edu/abs/2015ApJ...804...27A} {804, 27}

\bibitem[\protect\citeauthoryear{{Banerji}, {Alaghband-Zadeh}, {Hewett}  \&
  {McMahon}}{{Banerji} et~al.}{2015}]{Banerji+15}
{Banerji} M.,  {Alaghband-Zadeh} S.,  {Hewett} P.~C.,   {McMahon} R.~G.,  2015,
  \mn@doi [\mnras] {10.1093/mnras/stu2649}, \href
  {https://ui.adsabs.harvard.edu/abs/2015MNRAS.447.3368B} {447, 3368}

\bibitem[\protect\citeauthoryear{{Baron} et~al.,}{{Baron}
  et~al.}{2018}]{Baron+18}
{Baron} D.,  et~al., 2018, \mn@doi [\mnras] {10.1093/mnras/sty2113}, \href
  {https://ui.adsabs.harvard.edu/abs/2018MNRAS.480.3993B} {480, 3993}

\bibitem[\protect\citeauthoryear{{Baron}, {Netzer}, {Lutz}, {Davies}  \&
  {Prochaska}}{{Baron} et~al.}{2024}]{Baron+24}
{Baron} D.,  {Netzer} H.,  {Lutz} D.,  {Davies} R.~I.,   {Prochaska} J.~X.,
  2024, \mn@doi [arXiv e-prints] {10.48550/arXiv.2401.09576}, \href
  {https://ui.adsabs.harvard.edu/abs/2024arXiv240109576B} {p. arXiv:2401.09576}

\bibitem[\protect\citeauthoryear{{Barro} et~al.,}{{Barro}
  et~al.}{2023}]{Barro+23}
{Barro} G.,  et~al., 2023, \mn@doi [arXiv e-prints]
  {10.48550/arXiv.2305.14418}, \href
  {https://ui.adsabs.harvard.edu/abs/2023arXiv230514418B} {p. arXiv:2305.14418}

\bibitem[\protect\citeauthoryear{{Bischetti} et~al.,}{{Bischetti}
  et~al.}{2017}]{Bischetti+17}
{Bischetti} M.,  et~al., 2017, \mn@doi [\aap] {10.1051/0004-6361/201629301},
  \href {https://ui.adsabs.harvard.edu/abs/2017A&A...598A.122B} {598, A122}

\bibitem[\protect\citeauthoryear{{Brusa} et~al.,}{{Brusa}
  et~al.}{2016}]{Brusa+16}
{Brusa} M.,  et~al., 2016, \mn@doi [\aap] {10.1051/0004-6361/201527900}, \href
  {https://ui.adsabs.harvard.edu/abs/2016A&A...588A..58B} {588, A58}

\bibitem[\protect\citeauthoryear{{Calistro Rivera} et~al.,}{{Calistro Rivera}
  et~al.}{2021}]{CalistroRivera+21}
{Calistro Rivera} G.,  et~al., 2021, \mn@doi [\aap]
  {10.1051/0004-6361/202040214}, \href
  {https://ui.adsabs.harvard.edu/abs/2021A&A...649A.102C} {649, A102}

\bibitem[\protect\citeauthoryear{{Canalizo} \& {Stockton}}{{Canalizo} \&
  {Stockton}}{2001}]{CanalizoStockton01}
{Canalizo} G.,  {Stockton} A.,  2001, \mn@doi [\apj] {10.1086/321520}, \href
  {https://ui.adsabs.harvard.edu/abs/2001ApJ...555..719C} {555, 719}

\bibitem[\protect\citeauthoryear{{Chen} et~al.,}{{Chen} et~al.}{2017}]{Chen+17}
{Chen} C.-C.,  et~al., 2017, \mn@doi [\apj] {10.3847/1538-4357/aa863a}, \href
  {https://ui.adsabs.harvard.edu/abs/2017ApJ...846..108C} {846, 108}

\bibitem[\protect\citeauthoryear{{Costa}, {Rosdahl}, {Sijacki}  \&
  {Haehnelt}}{{Costa} et~al.}{2018}]{Costa+18}
{Costa} T.,  {Rosdahl} J.,  {Sijacki} D.,   {Haehnelt} M.~G.,  2018, \mn@doi
  [\mnras] {10.1093/mnras/sty1514}, \href
  {https://ui.adsabs.harvard.edu/abs/2018MNRAS.479.2079C} {479, 2079}

\bibitem[\protect\citeauthoryear{{Costa}, {Arrigoni Battaia}, {Farina},
  {Keating}, {Rosdahl}  \& {Kimm}}{{Costa} et~al.}{2022}]{Costa+22}
{Costa} T.,  {Arrigoni Battaia} F.,  {Farina} E.~P.,  {Keating} L.~C.,
  {Rosdahl} J.,   {Kimm} T.,  2022, \mn@doi [\mnras] {10.1093/mnras/stac2432},
  \href {https://ui.adsabs.harvard.edu/abs/2022MNRAS.517.1767C} {517, 1767}

\bibitem[\protect\citeauthoryear{{Cresci} et~al.,}{{Cresci}
  et~al.}{2015}]{Cresci+15}
{Cresci} G.,  et~al., 2015, \mn@doi [\apj] {10.1088/0004-637X/799/1/82}, \href
  {https://ui.adsabs.harvard.edu/abs/2015ApJ...799...82C} {799, 82}

\bibitem[\protect\citeauthoryear{{Cresci} et~al.,}{{Cresci}
  et~al.}{2023}]{Cresci+23}
{Cresci} G.,  et~al., 2023, \mn@doi [\aap] {10.1051/0004-6361/202346001}, \href
  {https://ui.adsabs.harvard.edu/abs/2023A&A...672A.128C} {672, A128}

\bibitem[\protect\citeauthoryear{{Croton} et~al.,}{{Croton}
  et~al.}{2006}]{Croton+06}
{Croton} D.~J.,  et~al., 2006, \mn@doi [\mnras]
  {10.1111/j.1365-2966.2005.09675.x}, \href
  {https://ui.adsabs.harvard.edu/abs/2006MNRAS.365...11C} {365, 11}

\bibitem[\protect\citeauthoryear{{Eisenstein} et~al.,}{{Eisenstein}
  et~al.}{2011}]{Eisenstein+11}
{Eisenstein} D.~J.,  et~al., 2011, \mn@doi [\aj] {10.1088/0004-6256/142/3/72},
  \href {https://ui.adsabs.harvard.edu/abs/2011AJ....142...72E} {142, 72}

\bibitem[\protect\citeauthoryear{{Feigelson}, {Nelson}, {Isobe}  \&
  {LaValley}}{{Feigelson} et~al.}{2014}]{Feigelson+14}
{Feigelson} E.~D.,  {Nelson} P.~I.,  {Isobe} T.,   {LaValley} M.,  2014,
  {ASURV: Astronomical SURVival Statistics}, Astrophysics Source Code Library,
  record ascl:1406.001 (\mn@eprint {ascl} {1406.001})

\bibitem[\protect\citeauthoryear{{Finnerty} et~al.,}{{Finnerty}
  et~al.}{2020}]{Finnerty+20}
{Finnerty} L.,  et~al., 2020, \mn@doi [\apj] {10.3847/1538-4357/abc3bf}, \href
  {https://ui.adsabs.harvard.edu/abs/2020ApJ...905...16F} {905, 16}

\bibitem[\protect\citeauthoryear{{Gebhardt} et~al.,}{{Gebhardt}
  et~al.}{2000}]{Gebhardt+00}
{Gebhardt} K.,  et~al., 2000, \mn@doi [\apjl] {10.1086/312840}, \href
  {http://adsabs.harvard.edu/abs/2000ApJ...539L..13G} {539, L13}

\bibitem[\protect\citeauthoryear{{Gillette}, {Lau}, {Hamann}, {Perrotta},
  {Rupke}, {Wylezalek}, {Zakamska}  \& {Vayner}}{{Gillette}
  et~al.}{2023}]{Gillette+23}
{Gillette} J.,  {Lau} M.~W.,  {Hamann} F.,  {Perrotta} S.,  {Rupke} D. S.~N.,
  {Wylezalek} D.,  {Zakamska} N.~L.,   {Vayner} A.,  2023, \mn@doi [\mnras]
  {10.1093/mnras/stad2923}, \href
  {https://ui.adsabs.harvard.edu/abs/2023MNRAS.526.2578G} {526, 2578}

\bibitem[\protect\citeauthoryear{{Gillette}, {Hamann}, {Lau}  \&
  {Perrotta}}{{Gillette} et~al.}{2024}]{Gillette+24}
{Gillette} J.,  {Hamann} F.,  {Lau} M.~W.,   {Perrotta} S.,  2024, \mn@doi
  [\mnras] {10.1093/mnras/stad2890}, \href
  {https://ui.adsabs.harvard.edu/abs/2024MNRAS.527..950G} {527, 950}

\bibitem[\protect\citeauthoryear{{Glikman} et~al.,}{{Glikman}
  et~al.}{2022}]{Glikman+22}
{Glikman} E.,  et~al., 2022, \mn@doi [\apj] {10.3847/1538-4357/ac6bee}, \href
  {https://ui.adsabs.harvard.edu/abs/2022ApJ...934..119G} {934, 119}

\bibitem[\protect\citeauthoryear{{Gonz{\'a}lez Lobos} et~al.,}{{Gonz{\'a}lez
  Lobos} et~al.}{2023}]{GonzalezLobos+23}
{Gonz{\'a}lez Lobos} V.,  et~al., 2023, \mn@doi [\aap]
  {10.1051/0004-6361/202346879}, \href
  {https://ui.adsabs.harvard.edu/abs/2023A&A...679A..41G} {679, A41}

\bibitem[\protect\citeauthoryear{{Hamann} et~al.,}{{Hamann}
  et~al.}{2017}]{Hamann+17}
{Hamann} F.,  et~al., 2017, \mn@doi [\mnras] {10.1093/mnras/stw2387}, \href
  {https://ui.adsabs.harvard.edu/abs/2017MNRAS.464.3431H} {464, 3431}

\bibitem[\protect\citeauthoryear{{Harrison} et~al.,}{{Harrison}
  et~al.}{2016}]{Harrison+16}
{Harrison} C.~M.,  et~al., 2016, \mn@doi [\mnras] {10.1093/mnras/stv2727},
  \href {https://ui.adsabs.harvard.edu/abs/2016MNRAS.456.1195H} {456, 1195}

\bibitem[\protect\citeauthoryear{{Hopkins} \& {Elvis}}{{Hopkins} \&
  {Elvis}}{2010}]{HopkinsElvis10}
{Hopkins} P.~F.,  {Elvis} M.,  2010, \mn@doi [\mnras]
  {10.1111/j.1365-2966.2009.15643.x}, \href
  {https://ui.adsabs.harvard.edu/abs/2010MNRAS.401....7H} {401, 7}

\bibitem[\protect\citeauthoryear{{Hopkins}, {Hernquist}, {Cox}, {Di Matteo},
  {Robertson}  \& {Springel}}{{Hopkins} et~al.}{2006}]{Hopkins+06}
{Hopkins} P.~F.,  {Hernquist} L.,  {Cox} T.~J.,  {Di Matteo} T.,  {Robertson}
  B.,   {Springel} V.,  2006, \mn@doi [\apjs] {10.1086/499298}, \href
  {http://adsabs.harvard.edu/abs/2006ApJS..163....1H} {163, 1}

\bibitem[\protect\citeauthoryear{{Hopkins}, {Hernquist}, {Cox}  \&
  {Kere\v{s}}}{{Hopkins} et~al.}{2008}]{Hopkins+08a}
{Hopkins} P.~F.,  {Hernquist} L.,  {Cox} T.~J.,   {Kere\v{s}} D.,  2008,
  \mn@doi [\apjs] {10.1086/524362}, \href
  {http://adsabs.harvard.edu/abs/2008ApJS..175..356H} {175, 356}

\bibitem[\protect\citeauthoryear{{Hopkins}, {Torrey}, {Faucher-Gigu{\`e}re},
  {Quataert}  \& {Murray}}{{Hopkins} et~al.}{2016}]{Hopkins+16}
{Hopkins} P.~F.,  {Torrey} P.,  {Faucher-Gigu{\`e}re} C.-A.,  {Quataert} E.,
  {Murray} N.,  2016, \mn@doi [\mnras] {10.1093/mnras/stw289}, \href
  {https://ui.adsabs.harvard.edu/abs/2016MNRAS.458..816H} {458, 816}

\bibitem[\protect\citeauthoryear{{Ishibashi}, {Fabian}  \&
  {Maiolino}}{{Ishibashi} et~al.}{2018}]{IshibashiFabianMaiolino18}
{Ishibashi} W.,  {Fabian} A.~C.,   {Maiolino} R.,  2018, \mn@doi [\mnras]
  {10.1093/mnras/sty236}, \href
  {https://ui.adsabs.harvard.edu/abs/2018MNRAS.476..512I} {476, 512}

\bibitem[\protect\citeauthoryear{{Kakkad} et~al.,}{{Kakkad}
  et~al.}{2020}]{Kakkad+20}
{Kakkad} D.,  et~al., 2020, \mn@doi [\aap] {10.1051/0004-6361/202038551}, \href
  {https://ui.adsabs.harvard.edu/abs/2020A&A...642A.147K} {642, A147}

\bibitem[\protect\citeauthoryear{{Kakkad} et~al.,}{{Kakkad}
  et~al.}{2023}]{Kakkad+23}
{Kakkad} D.,  et~al., 2023, \mn@doi [\mnras] {10.1093/mnras/stad439}, \href
  {https://ui.adsabs.harvard.edu/abs/2023MNRAS.520.5783K} {520, 5783}

\bibitem[\protect\citeauthoryear{{Lau}, {Hamann}, {Gillette}, {Perrotta},
  {Rupke}, {Wylezalek}  \& {Zakamska}}{{Lau} et~al.}{2022}]{Lau+22}
{Lau} M.~W.,  {Hamann} F.,  {Gillette} J.,  {Perrotta} S.,  {Rupke} D. S.~N.,
  {Wylezalek} D.,   {Zakamska} N.~L.,  2022, \mn@doi [\mnras]
  {10.1093/mnras/stac1823}, \href
  {https://ui.adsabs.harvard.edu/abs/2022MNRAS.515.1624L} {515, 1624}

\bibitem[\protect\citeauthoryear{{Law}, {Steidel}, {Chen}, {Strom}, {Rudie}  \&
  {Trainor}}{{Law} et~al.}{2018}]{Law+18}
{Law} D.~R.,  {Steidel} C.~C.,  {Chen} Y.,  {Strom} A.~L.,  {Rudie} G.~C.,
  {Trainor} R.~F.,  2018, \mn@doi [\apj] {10.3847/1538-4357/aae156}, \href
  {https://ui.adsabs.harvard.edu/abs/2018ApJ...866..119L} {866, 119}

\bibitem[\protect\citeauthoryear{{Liu}, {Zakamska}, {Greene}, {Nesvadba}  \&
  {Liu}}{{Liu} et~al.}{2013}]{Liu+13}
{Liu} G.,  {Zakamska} N.~L.,  {Greene} J.~E.,  {Nesvadba} N. P.~H.,   {Liu} X.,
   2013, \mn@doi [\mnras] {10.1093/mnras/stt1755}, \href
  {https://ui.adsabs.harvard.edu/abs/2013MNRAS.436.2576L} {436, 2576}

\bibitem[\protect\citeauthoryear{{Lockhart} et~al.,}{{Lockhart}
  et~al.}{2019}]{Lockhart+19}
{Lockhart} K.~E.,  et~al., 2019, \mn@doi [\aj] {10.3847/1538-3881/aaf64e},
  \href {https://ui.adsabs.harvard.edu/abs/2019AJ....157...75L} {157, 75}

\bibitem[\protect\citeauthoryear{{Lyke} et~al.,}{{Lyke} et~al.}{2017}]{Lyke+17}
{Lyke} J.,  et~al., 2017, {OSIRIS Toolbox: OH-Suppressing InfraRed Imaging
  Spectrograph pipeline}, Astrophysics Source Code Library, record
  ascl:1710.021 (\mn@eprint {ascl} {1710.021})

\bibitem[\protect\citeauthoryear{{Monadi} \& {Bird}}{{Monadi} \&
  {Bird}}{2022}]{MonadiBird22}
{Monadi} R.,  {Bird} S.,  2022, \mn@doi [\mnras] {10.1093/mnras/stac294}, \href
  {https://ui.adsabs.harvard.edu/abs/2022MNRAS.511.3501M} {511, 3501}

\bibitem[\protect\citeauthoryear{{Mu{\~n}oz-Elgueta}, {Arrigoni Battaia},
  {Kauffmann}, {De Breuck}, {Garc{\'\i}a-Vergara}, {Zanella}, {Farina}  \&
  {Decarli}}{{Mu{\~n}oz-Elgueta} et~al.}{2022}]{MunozElgueta+22}
{Mu{\~n}oz-Elgueta} N.,  {Arrigoni Battaia} F.,  {Kauffmann} G.,  {De Breuck}
  C.,  {Garc{\'\i}a-Vergara} C.,  {Zanella} A.,  {Farina} E.~P.,   {Decarli}
  R.,  2022, \mn@doi [\mnras] {10.1093/mnras/stac041}, \href
  {https://ui.adsabs.harvard.edu/abs/2022MNRAS.511.1462M} {511, 1462}

\bibitem[\protect\citeauthoryear{{Nelson} et~al.,}{{Nelson}
  et~al.}{2019}]{Nelson+19}
{Nelson} D.,  et~al., 2019, \mn@doi [\mnras] {10.1093/mnras/stz2306}, \href
  {https://ui.adsabs.harvard.edu/abs/2019MNRAS.490.3234N} {490, 3234}

\bibitem[\protect\citeauthoryear{{Nesvadba}, {De Breuck}, {Lehnert}, {Best}  \&
  {Collet}}{{Nesvadba} et~al.}{2017}]{Nesvadba+17}
{Nesvadba} N.~P.~H.,  {De Breuck} C.,  {Lehnert} M.~D.,  {Best} P.~N.,
  {Collet} C.,  2017, \mn@doi [\aap] {10.1051/0004-6361/201528040}, \href
  {https://ui.adsabs.harvard.edu/abs/2017A&A...599A.123N} {599, A123}

\bibitem[\protect\citeauthoryear{{Netzer}}{{Netzer}}{2015}]{Netzer15}
{Netzer} H.,  2015, \mn@doi [\araa] {10.1146/annurev-astro-082214-122302},
  \href {https://ui.adsabs.harvard.edu/abs/2015ARA&A..53..365N} {53, 365}

\bibitem[\protect\citeauthoryear{{Noboriguchi} et~al.,}{{Noboriguchi}
  et~al.}{2019}]{Noboriguchi+19}
{Noboriguchi} A.,  et~al., 2019, \mn@doi [\apj] {10.3847/1538-4357/ab1754},
  \href {https://ui.adsabs.harvard.edu/abs/2019ApJ...876..132N} {876, 132}

\bibitem[\protect\citeauthoryear{{Noboriguchi}, {Inoue}, {Nagao}, {Toba}  \&
  {Misawa}}{{Noboriguchi} et~al.}{2023}]{Noboriguchi+23}
{Noboriguchi} A.,  {Inoue} A.~K.,  {Nagao} T.,  {Toba} Y.,   {Misawa} T.,
  2023, \mn@doi [\apjl] {10.3847/2041-8213/ad0e00}, \href
  {https://ui.adsabs.harvard.edu/abs/2023ApJ...959L..14N} {959, L14}

\bibitem[\protect\citeauthoryear{{Obied}, {Zakamska}, {Wylezalek}  \&
  {Liu}}{{Obied} et~al.}{2016}]{Obied+16}
{Obied} G.,  {Zakamska} N.~L.,  {Wylezalek} D.,   {Liu} G.,  2016, \mn@doi
  [\mnras] {10.1093/mnras/stv2850}, \href
  {https://ui.adsabs.harvard.edu/abs/2016MNRAS.456.2861O} {456, 2861}

\bibitem[\protect\citeauthoryear{{Perna} et~al.,}{{Perna}
  et~al.}{2015}]{Perna+15}
{Perna} M.,  et~al., 2015, \mn@doi [\aap] {10.1051/0004-6361/201526907}, \href
  {https://ui.adsabs.harvard.edu/abs/2015A&A...583A..72P} {583, A72}

\bibitem[\protect\citeauthoryear{{Perna} et~al.,}{{Perna}
  et~al.}{2023}]{Perna+23}
{Perna} M.,  et~al., 2023, \mn@doi [\aap] {10.1051/0004-6361/202346649}, \href
  {https://ui.adsabs.harvard.edu/abs/2023A&A...679A..89P} {679, A89}

\bibitem[\protect\citeauthoryear{{Perrotta}, {Hamann}, {Zakamska},
  {Alexandroff}, {Rupke}  \& {Wylezalek}}{{Perrotta}
  et~al.}{2019}]{Perrotta+19}
{Perrotta} S.,  {Hamann} F.,  {Zakamska} N.~L.,  {Alexandroff} R.~M.,  {Rupke}
  D.,   {Wylezalek} D.,  2019, \mn@doi [\mnras] {10.1093/mnras/stz1993}, \href
  {https://ui.adsabs.harvard.edu/abs/2019MNRAS.488.4126P} {488, 4126}

\bibitem[\protect\citeauthoryear{{Ross} et~al.,}{{Ross} et~al.}{2015}]{Ross+15}
{Ross} N.~P.,  et~al., 2015, \mn@doi [\mnras] {10.1093/mnras/stv1710}, \href
  {https://ui.adsabs.harvard.edu/abs/2015MNRAS.453.3932R} {453, 3932}

\bibitem[\protect\citeauthoryear{{Rupke}}{{Rupke}}{2014}]{Rupke14a}
{Rupke} D. S.~N.,  2014, {IFSRED: Data Reduction for Integral Field
  Spectrographs} (\mn@eprint {ascl} {1409.004})

\bibitem[\protect\citeauthoryear{{Sanders}, {Soifer}, {Elias}, {Neugebauer}  \&
  {Matthews}}{{Sanders} et~al.}{1988}]{Sanders+88}
{Sanders} D.~B.,  {Soifer} B.~T.,  {Elias} J.~H.,  {Neugebauer} G.,
  {Matthews} K.,  1988, \mn@doi [\apjl] {10.1086/185155}, \href
  {https://ui.adsabs.harvard.edu/abs/1988ApJ...328L..35S} {328, L35}

\bibitem[\protect\citeauthoryear{{Shen}}{{Shen}}{2016}]{Shen16a}
{Shen} Y.,  2016, \mn@doi [\apj] {10.3847/0004-637X/817/1/55}, \href
  {https://ui.adsabs.harvard.edu/abs/2016ApJ...817...55S} {817, 55}

\bibitem[\protect\citeauthoryear{{Shen} et~al.,}{{Shen} et~al.}{2023}]{Shen+23}
{Shen} L.,  et~al., 2023, \mn@doi [Science Advances] {10.1126/sciadv.adg8287},
  \href {https://ui.adsabs.harvard.edu/abs/2023SciA....9G8287S} {9, eadg8287}

\bibitem[\protect\citeauthoryear{{Soliman} \& {Hopkins}}{{Soliman} \&
  {Hopkins}}{2023}]{SolimanHopkins23}
{Soliman} N.~H.,  {Hopkins} P.~F.,  2023, \mn@doi [\mnras]
  {10.1093/mnras/stad2460}, \href
  {https://ui.adsabs.harvard.edu/abs/2023MNRAS.525.2668S} {525, 2668}

\bibitem[\protect\citeauthoryear{{Sun}, {Greene}  \& {Zakamska}}{{Sun}
  et~al.}{2017}]{SunGreeneZakamska17}
{Sun} A.-L.,  {Greene} J.~E.,   {Zakamska} N.~L.,  2017, \mn@doi [\apj]
  {10.3847/1538-4357/835/2/222}, \href
  {https://ui.adsabs.harvard.edu/abs/2017ApJ...835..222S} {835, 222}

\bibitem[\protect\citeauthoryear{{Urrutia}, {Lacy}  \& {Becker}}{{Urrutia}
  et~al.}{2008}]{UrrutiaLacyBecker08}
{Urrutia} T.,  {Lacy} M.,   {Becker} R.~H.,  2008, \mn@doi [\apj]
  {10.1086/523959}, \href
  {https://ui.adsabs.harvard.edu/abs/2008ApJ...674...80U} {674, 80}

\bibitem[\protect\citeauthoryear{{Vayner} et~al.,}{{Vayner}
  et~al.}{2021a}]{Vayner+21a}
{Vayner} A.,  et~al., 2021a, \mn@doi [\mnras] {10.1093/mnras/stab1176}, \href
  {https://ui.adsabs.harvard.edu/abs/2021MNRAS.504.4445V} {504, 4445}

\bibitem[\protect\citeauthoryear{{Vayner} et~al.,}{{Vayner}
  et~al.}{2021b}]{Vayner+21b}
{Vayner} A.,  et~al., 2021b, \mn@doi [\apj] {10.3847/1538-4357/ac0f56}, \href
  {https://ui.adsabs.harvard.edu/abs/2021ApJ...919..122V} {919, 122}

\bibitem[\protect\citeauthoryear{{Vayner} et~al.,}{{Vayner}
  et~al.}{2023}]{Vayner+23}
{Vayner} A.,  et~al., 2023, \mn@doi [\apj] {10.3847/1538-4357/ace784}, \href
  {https://ui.adsabs.harvard.edu/abs/2023ApJ...955...92V} {955, 92}

\bibitem[\protect\citeauthoryear{{Vayner} et~al.,}{{Vayner}
  et~al.}{2024}]{Vayner+24}
{Vayner} A.,  et~al., 2024, \mn@doi [\apj] {10.3847/1538-4357/ad0be9}, \href
  {https://ui.adsabs.harvard.edu/abs/2024ApJ...960..126V} {960, 126}

\bibitem[\protect\citeauthoryear{{Veilleux} et~al.,}{{Veilleux}
  et~al.}{2023}]{Veilleux+23}
{Veilleux} S.,  et~al., 2023, \mn@doi [\apj] {10.3847/1538-4357/ace10f}, \href
  {https://ui.adsabs.harvard.edu/abs/2023ApJ...953...56V} {953, 56}

\bibitem[\protect\citeauthoryear{{Villforth}}{{Villforth}}{2023}]{Villforth23}
{Villforth} C.,  2023, \mn@doi [The Open Journal of Astrophysics]
  {10.21105/astro.2309.03276}, \href
  {https://ui.adsabs.harvard.edu/abs/2023OJAp....6E..34V} {6, 34}

\bibitem[\protect\citeauthoryear{{Williams}, {Maiolino}, {Krongold},
  {Carniani}, {Cresci}, {Mannucci}  \& {Marconi}}{{Williams}
  et~al.}{2017}]{Williams+17}
{Williams} R.~J.,  {Maiolino} R.,  {Krongold} Y.,  {Carniani} S.,  {Cresci} G.,
   {Mannucci} F.,   {Marconi} A.,  2017, \mn@doi [\mnras]
  {10.1093/mnras/stx311}, \href
  {https://ui.adsabs.harvard.edu/abs/2017MNRAS.467.3399W} {467, 3399}

\bibitem[\protect\citeauthoryear{{Wright}}{{Wright}}{2006}]{Wright06}
{Wright} E.~L.,  2006, \mn@doi [\pasp] {10.1086/510102}, \href
  {https://ui.adsabs.harvard.edu/abs/2006PASP..118.1711W} {118, 1711}

\bibitem[\protect\citeauthoryear{{Wright} et~al.,}{{Wright}
  et~al.}{2010}]{Wright+10}
{Wright} E.~L.,  et~al., 2010, \mn@doi [\aj] {10.1088/0004-6256/140/6/1868},
  \href {https://ui.adsabs.harvard.edu/abs/2010AJ....140.1868W} {140, 1868}

\bibitem[\protect\citeauthoryear{{Wylezalek} et~al.,}{{Wylezalek}
  et~al.}{2022}]{Wylezalek+22}
{Wylezalek} D.,  et~al., 2022, \mn@doi [\apjl] {10.3847/2041-8213/ac98c3},
  \href {https://ui.adsabs.harvard.edu/abs/2022ApJ...940L...7W} {940, L7}

\bibitem[\protect\citeauthoryear{{Zakamska} \& {Greene}}{{Zakamska} \&
  {Greene}}{2014}]{ZakamskaGreene14}
{Zakamska} N.~L.,  {Greene} J.~E.,  2014, \mn@doi [\mnras]
  {10.1093/mnras/stu842}, \href
  {https://ui.adsabs.harvard.edu/abs/2014MNRAS.442..784Z} {442, 784}

\bibitem[\protect\citeauthoryear{{Zakamska} et~al.,}{{Zakamska}
  et~al.}{2016}]{Zakamska+16}
{Zakamska} N.~L.,  et~al., 2016, \mn@doi [\mnras] {10.1093/mnras/stw718}, \href
  {https://ui.adsabs.harvard.edu/abs/2016MNRAS.459.3144Z} {459, 3144}

\bibitem[\protect\citeauthoryear{{Zakamska} et~al.,}{{Zakamska}
  et~al.}{2019}]{Zakamska+19}
{Zakamska} N.~L.,  et~al., 2019, \mn@doi [\mnras] {10.1093/mnras/stz2071},
  \href {https://ui.adsabs.harvard.edu/abs/2019MNRAS.489..497Z} {489, 497}

\bibitem[\protect\citeauthoryear{{Zubovas} \& {King}}{{Zubovas} \&
  {King}}{2012}]{ZubovasKing12}
{Zubovas} K.,  {King} A.,  2012, \mn@doi [\apjl] {10.1088/2041-8205/745/2/L34},
  \href {https://ui.adsabs.harvard.edu/abs/2012ApJ...745L..34Z} {745, L34}

\makeatother
\end{thebibliography}

\bsp    
\label{lastpage}
\end{document}